\documentclass{aa}

\usepackage[varg]{txfonts}
\usepackage{multirow}  
\usepackage{enumitem}  
\usepackage{subfigure}
\usepackage{epstopdf}  

\usepackage{nicefrac}  
\usepackage{color} 
\newcommand{\teff}{\ensuremath{{T_{\rm eff}}}}           
\newcommand{\tnom}{\ensuremath{{T_{\rm nom}}}}           
\newcommand{\logg}{\ensuremath{\log g}}                  
\newcommand{\feh}{\ensuremath{\mathrm{[Fe/H]}}}          
\def\micro{$\xi_{\rm micro}$}                            
\def\kms{$\mathrm {km~s}^{-1}$}

\def\liseven {$\mathrm{^{7}Li}$}
\def\lisix {$\mathrm{^{6}Li}$}
\def\iso{$\mathrm{^{6}Li/^{7}Li}$}
\def\1213C{$\mathrm{^{12}C/^{13}C}$}

\newcommand{\linfor}{{\sf Linfor3D}}
\newcommand{\nlte}{{\sf NLTE3D}}
\newcommand{\cobold}{CO5BOLD}
\newcommand{\lhdm}{{\sf LHD}}

\begin{document}

\title{Improving spectroscopic lithium abundances}
\subtitle{Fitting functions for 3D non-LTE corrections in
  FGK stars of different metallicity}

   \author{
		A. Mott\inst{1},
		M. Steffen\inst{1},
		E. Caffau\inst{2},
		K.G. Strassmeier\inst{1}
	}

   \institute{
		Leibniz-Institut f{\"u}r Astrophysik Potsdam, An der Sternwarte 16, 14482 Potsdam, Germany\\
		\email{msteffen@aip.de}
		\and
		GEPI, Observatoire de Paris, PSL Research University, CNRS, Place Jules Janssen, 92195 Meudon, France
	}

\authorrunning{A.~Mott et al.}

\titlerunning{Fitting functions for the lithium abundance in metal-poor stars}

\date{Received DD MM YY/\,Accepted DD MM YY}

\abstract
    {Accurate spectroscopic lithium abundances are essential in addressing a
      variety of open questions, such as the origin of a uniform lithium content
      in the atmospheres of metal-poor stars (Spite plateau) or the existence
      of a correlation between the properties of extrasolar planetary systems
      and the lithium abundance in the atmosphere of their host stars.}
    {We have developed a tool that allows the user to improve the accuracy of
      standard lithium abundance determinations based on 1D model atmospheres
      and the assumption of Local Thermodynamic Equilibrium (LTE) by applying
      corrections that account for hydrodynamic (3D) and non-LTE (NLTE) effects
      in FGK stars of different metallicity.}
    {Based on a grid of \cobold~3D models and associated 1D hydrostatic
      atmospheres, we computed three libraries of synthetic spectra of the
      lithium $\lambda\,670.8$\,nm line for a wide range of lithium abundances,
      accounting for detailed line formation in 3D\,NLTE, 1D\,NLTE,
      and 1D\,LTE, respectively. The resulting curves-of-growth were then
      used to derive 3D\,NLTE  and 1D\,NLTE lithium abundance
      corrections.}
    {For all metallicities, the largest corrections are found at the
      coolest effective temperature, \teff\,=\,$5000$\,K.
      They are mostly positive, up to $+0.2$\,dex, for
      the weakest lines (lithium abundance $A\rm(Li)_{1DLTE}=1.0$), whereas
      they become more negative towards lower metallicities, where they can
      reach $-0.4$\,dex for the strongest lines ($A\rm(Li)_{1DLTE}=3.0$) at
      [Fe/H]\,=\,$-2.0$. We demonstrate that 3D and NLTE effects are small
      for metal-poor stars on the Spite plateau, leading to errors of at most
      $\pm 0.05$\,dex if ignored.
      We present analytical functions evaluating the 3D\,NLTE
      and 1D\,NLTE corrections as a function of \teff\ [$5000\ldots 6500$\,K],
      \logg\ [$3.5\ldots 4.5$], and LTE lithium abundance $A$(Li)
      [$1.0\ldots 3.0$] for a fixed grid of metallicities \feh\
      [$-3.0\ldots 0.0$]. In addition, we also provide analytical
      fitting functions for directly converting a given lithium abundance into
      an equivalent width, or vice versa, a given equivalent width (EW) into a
      lithium abundance. For convenience, a Python script is made
      available that evaluates all fitting functions for given \teff, \logg,
      \feh, and $A$(Li) or EW. }
    {By means of the fitting functions developed in this work, the results of
      complex 3D and NLTE calculations are made readily accessible and quickly
      applicable to large samples of stars across a wide range of
      metallicities. Improving the accuracy of spectroscopic
      lithium abundance determinations will contribute to a better
      understanding of the open questions related to the lithium content in
      metal-poor and solar-like stellar atmospheres.}

\maketitle


\section{Introduction}
\label{Intro}

According to Standard Big Bang Nucleosynthesis (SBBN), \liseven\ together with
hydrogen and helium, is one of the few elements created in the primordial
fireball. To trace the primordial chemical composition of the Universe, one
of the best diagnostics is represented by old Population II stars,
characterized by effective temperatures $\teff$\,$\ga$\,$5900$ K.  Such
objects are metal-poor and their convection zones are shallow enough to
preserve their initial lithium content.  More than 30 years ago,
\cite{spite82} analyzed a sample of unevolved metal-poor (MP) stars
($-2.8\le\feh\le -2.0$) with temperatures in the range
$5700\le{\rm T}_{\rm eff}\le 6300$\,K and discovered that irrespective of
effective temperature, gravity, and metallicity, all analyzed stars had
essentially the same Li abundance of $A$(Li)\footnote{$A(\mathrm{X})$ =
$\log(N(\mathrm{X})/N(\mathrm{H})) + 12$, where X is the chemical element.}
$\approx$ 2.2 that defines the {Spite plateau}.  The invariable behavior
of lithium at low \feh\ is different from what occurs for other elements,
whose abundances generally drop with declining metallicity. The {Spite
  plateau} has been interpreted as the primordial lithium abundance, produced
during the SBBN, and preserved in the atmosphere of these stars due to their
shallow convection zones.  The fact that this constant abundance value is
three times smaller than the primordial Li abundance inferred from the
measurements of the Cosmic Microwave Background (CMB) by WMAP
\citep{spergel07} and \textit{Planck} \citep{plank14} poses a severe problem
for the interpretation of the {Spite plateau} as representative of the
primordial Li abundance. This puzzling discrepancy challenges our present
understanding of Galactic chemical evolution and stellar Li depletion.

Recent studies \citep[e.g.,][]{korn06,korn07,nordlander12} explain this
discrepancy by invoking a combination of diffusion processes and some form of
turbulent mixing operating at the bottom of the outer convective zone in
turn-off stars belonging to the {Spite plateau}, e.g.\ in globular cluster
NCG\,6397.  As discussed in \cite{richard05}, this extra ingredient is needed
to meet the constraints imposed by the observations of \liseven\ in old halo
stars: a thin and flat plateau can only be retained in the presence of
diffusion if extra mixing is introduced into the models. This process rapidly
transports the surface Li down to the hotter inner region, where it is readily
destroyed, resulting in a Li-depleted convection zone.  Since there is no
parameter-free physical description of the turbulent mixing, it is necessary
to introduce it into the stellar evolution models in an ad-hoc manner,
defining a turbulent diffusion coefficient $D_{\mathrm{T}}$, which is a
parametric function of density and the atomic diffusion coefficient of helium
at a reference temperature $T_0$
($D_{\mathrm{T}}=400D_{\mathrm{He}}(T_0)[\nicefrac{\rho}{\rho(T_0)}]^{-3}$;
see \citealt{richard05}). High-quality $A$(Li) measurements are fundamental
for understanding whether or not atomic diffusion is at work in metal-poor
stellar atmospheres and if so, to what extent. In their Figure\,5,
\cite{nordlander12} show the behavior of different diffusion models in
comparison with observed stars in globular cluster NCG\,6397.  Their measured
1D\,NLTE lithium abundances for the two metal-poor stars belonging to the
{Spite plateau} (at \teff\ $\approx$\,5900 and 6300\,K) can be fitted, within
error bars, with one of their fine-tuned diffusion models.  The predicted
primordial lithium abundance $A\rm (Li)=2.57\pm 0.1$ is compatible with the
primordial Li abundance inferred from \textit{Planck} data.  The same model,
however, fails to reproduce the red-giant branch (RGB) stars at lower
temperatures, probably due to a source of extra mixing that was not considered
in these calculations \citep{charbonnel95}.

The same globular cluster was also analyzed by \cite{gonzalezhernandez09},
who measured the 3D\,NLTE lithium abundance in a sample of subgiants (SG)
and main-sequence (MS) stars, aimed at investigating the cosmological Li
discrepancy in the framework of diffusion models by tracking the lithium
abundance trend with evolutionary phase.  As shown in their Figure\,2, they
found SG stars to have a higher lithium abundance than MS objects and in both
cases, a decreasing trend of $A$(Li) with \teff\ with similar slopes was
observed.  Although no conclusive explanation of such trends was provided,
they compared the results with different diffusion models but were not able to
fit satisfactorily the $A$(Li) versus \teff\ trend of their targets, starting
from an initial (primordial) Li abundance in agreement with the
\textit{Planck} observations.  The only model able to provide a decreasing
\teff\ trend, with abundances of SG stars larger than MS stars, failed
quantitatively because the predicted $A$(Li) for their warmest stars was about
0.05\,dex lower than observed, such that the overall slope of $A$(Li) with
\teff\ was not properly reproduced.  Similar difficulties were encountered in
the work by \cite{gruyters16}, again in the cluster NGC\,6397 while fitting
their abundances of lithium and other elements (Fe, Ca) as a function of
\teff\ with diffusion models.  The authors advise caution about the size of
the abundance trends predicted by the models, which is set by the relative
abundance difference between the hottest and the coolest stars of the target
sample.

These examples demonstrate how crucial it is to derive lithium abundances
with a precision of better than 0.1\,dex, needed to
distinguish between different diffusion models and to better understand the
underlying physics.  Furthermore, this would help in constraining the
efficiency of atomic diffusion and additional turbulent mixing processes,
which at the current stage represents a plausible scenario for explaining the
gap in lithium abundance as measured in unevolved MP stars and inferred from
the CMB measurements by WMAP and \textit{Planck}.
 
Another open issue is related to the question of whether the {Spite plateau}
presents some slope with the metallicity or if it is actually a
flat distribution of $A$(Li) \citep[see e.g.,][]
{thorburn94,norris94,ryan96,ryan99,asplund06,melendez10}.The
plateau, in fact, presents a flat trend with some dispersion in lithium
abundance between $\feh=-2.2$ and $-2.5$, as originally found by
\cite{spite82}, and the observed scatter can be as low as 0.033\,dex, as found
in \cite{ryan99}.  At metallicities lower than $\feh\leq -2.5$, instead, a
larger dispersion in the Li abundances is observed with a trend of decreasing
$A$(Li) with decreasing \feh\ \citep{bonifacio07,aoki09,sbordone2010}.  At
present, it is not clear from neither an observational nor a theoretical
perspective, why most stars in this metallicity range show lithium abundances
below the {Spite plateau}.  More recently, \cite{bonifacio18} revisited
this so-called ``meltdown'' effect of the {Spite plateau}, finding some
stars at $\feh<-3.0$ with $A$(Li) below the constant value around $\approx
2.2$\,dex \citep[as derived by][]{sbordone2010} but also two stars at
$\feh\approx-4.0$ with a lithium content that is fully compatible with the
usual flat distribution of lithium in MP stars (the two star symbols in
Figure\,2 of \citealt{bonifacio18} at $A\rm (Li)=2.2$ and $\feh\approx-4.0$).
The presence of these two stars is exceptional since, as pointed out by
\cite{matsuno17}, no star in the literature was known to have a Li abundance
comparable to the {Spite plateau} below \feh\ of $-3.5$, except for the
primary of the double-lined binary system CS 22876-032
\citep{gonzalezhernandez08}.  Very recently, an even more extreme case was
found: \citet{Aguado2019} discovered that SDSS\,J0023+0307, one of the most
iron-poor stars known ([Fe/H]\,$\la$\,6.1), lies on the Spite plateau with
$A$(Li)\,=\,$2.02\pm 0.05$.

In order to provide a better observational basis for meaningful comparisons
with stellar depletion models and to shed more light on the open questions
regarding the {Spite plateau}, precise Li abundances for a large sample of
MP stars are needed, accounting for both 3D and NLTE effects that may
affect significantly the strength of the lithium line in metal-poor stars
and, hence, their estimated lithium content.
Highly accurate lithium abundances are also required in statistical studies 
trying to detect significant differences in the lithium content of the 
atmospheres of planet host stars and stars without known planets
\citep[e.g.,][]{Ghezzi2009,Israelian2009,Baumann2010,Pavlenko2018}.

In this work, we present extended fitting functions to derive the Li abundance 
of solar-type stars of different metallicity, taking into account 
both hydrodynamical effects (3D) and departures from thermodynamical 
equilibrium (NLTE). These fitting functions allow to derive (i)
the 3D and 1D\,NLTE abundance correction; (ii) the 3D and 1D\,NLTE
equivalent width corresponding to a given Li abundance $A$(Li); (iii) the
NLTE lithium abundance $A$(Li)$_{\rm 3DNLTE}$ and $A$(Li)$_{\rm 1DNLTE}$
from measured equivalent widths. Our approach is similar to that of
\citet{harutyunyan18} who additionally provide 3D\,NLTE corrections for
the lithium isotopic ratio, albeit for a smaller range of stellar parameters.
The present work extends and supersedes the fitting
functions presented in \citet{sbordone2010}. 


\begin{figure*}[hbtp]
    \centering
	\includegraphics[angle=90,width=\textwidth]{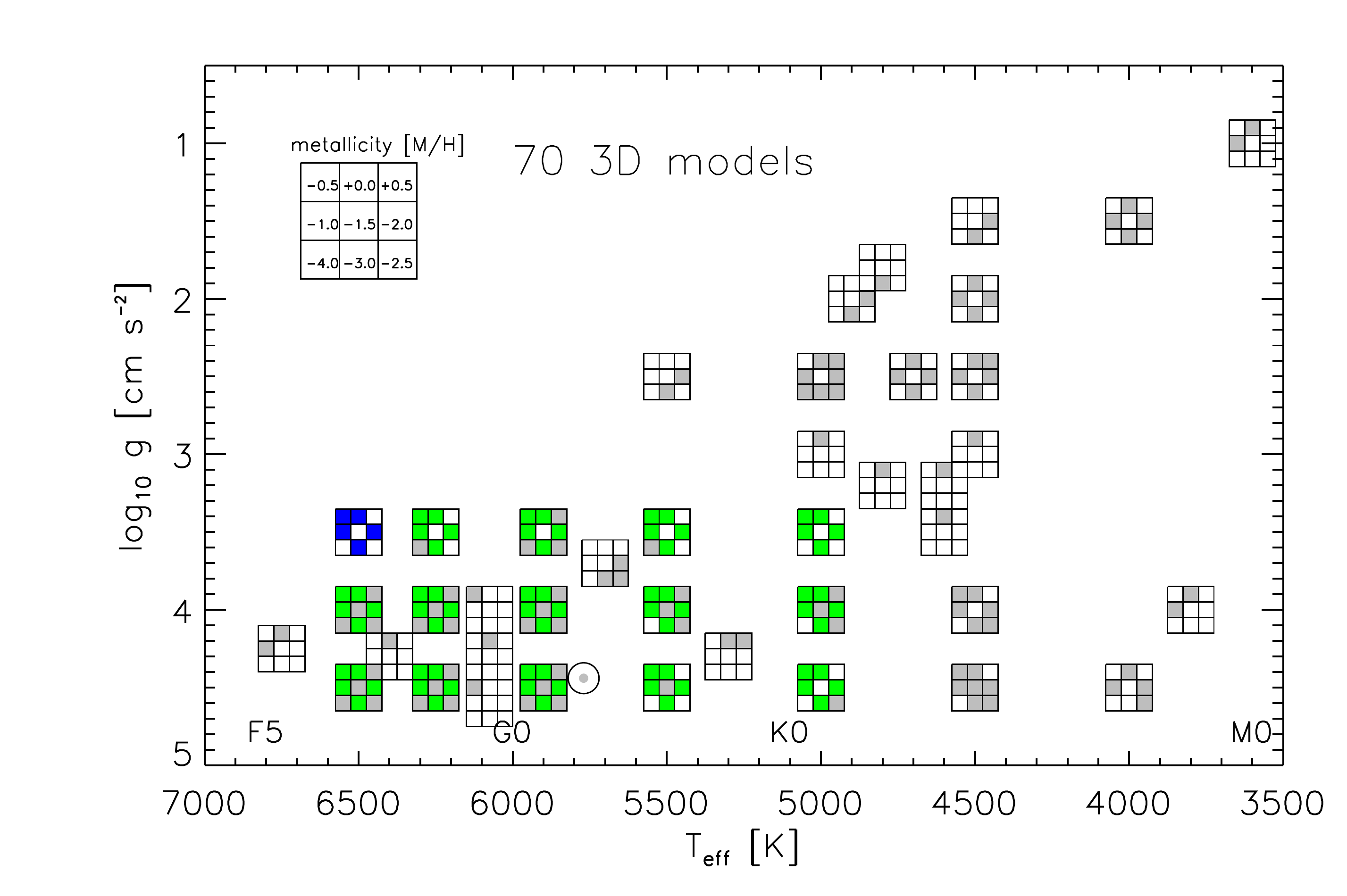} 
	\caption{Overview of the grid of \cobold~3D model atmospheres 
	adopted for the computation of the fitting functions. The green filled 
        squares mark the models that have been used in this work (five
        different metallicities, see legend at upper left), the grey squares 
        indicate existing models  that are part of the CIFIST grid but have 
        not been considered in the present context. The blue squares highlight 
        the missing models for which abundance corrections were 
        derived by extrapolation (see Section \ref{compCorr}). The solar
        3D~\cobold\ model is indicated by the Sun symbol.}
	\label{cifistFF}
\end{figure*}

\section{Stellar atmospheres and spectrum synthesis}\label{S4}
\subsection{3D~\cobold\ and 1D\,\lhdm\ model atmospheres}
\label{3d_1d_models}

For this work, we adopted a sub-set of the CIFIST grid of 3D hydro-dynamical
\cobold\ model atmospheres \citep{ludwig09} together with their corresponding
1D\,\lhdm\ models. These 1D models employ the same stellar parameters,
micro-physics, chemical composition, and radiative transfer scheme as a
\cobold\ model, facilitating an unbiased 3D-1D differential comparison
\citep{caffau07}.  Our model grid is shown in Figure \ref{cifistFF} in the
$\log g-\teff$ plane that can be seen as a pseudo Hertzsprung-Russell (HR)
diagram. The grid includes 70 3D models focusing on main-sequence and turn-off
cool stars, represented by the following stellar parameters:
\begin{itemize}[label={--}]
     \setlength{\itemsep}{0pt}
     \setlength{\parskip}{0pt}
     \setlength{\parsep}{0pt}
 \item \teff: \hskip 6mm 5000, 5500, 5900, 6300 and 6500 K, 
 \item \logg: \hskip 4mm 3.5, 4.0 and 4.5 [cm\,s$^{-2}$], 
 \item $\feh$: \hskip 1mm 0.00, -0.5, -1.0, -2.0 and -3.0 dex.
\end{itemize}
The region around $\teff=6500$ K and $\logg=3.5$ (blue filled squares in
Figure \ref{cifistFF}) is only scarcely populated in the (observed) H-R
diagram, and for this reason models with such parameters are not available in
the CIFIST grid.  However, since a rectangular \teff--\logg\ grid is more
practical for defining a homogeneous fitting function, we decided to provide
approximate results also for this missing combination of parameters by
extrapolation of the lithium abundance corrections derived from the available
atmospheres.

In Table\,\ref{tab:Models} (see Appendix \ref{appendixA}), the main properties
of the grid of selected models are listed. The effective
temperature reported in the fourth column of Table\,\ref{tab:Models} is the
average \teff\ over the sub-sample of $N_{\rm snap}$ selected snapshots (where
$N_{\rm snap}$ is given in the third column) for each \cobold\ simulation.
Note that these \teff\ deviate slightly from the nominal effective
temperatures \tnom\ defining the rectangular grid depicted in Figure
\ref{cifistFF} (where $\tnom=5000,\,5500,\,5900,\,6300$ and $6500$\, K for
all \logg\ and \feh). The fitting function will be developed on this fixed
temperature grid after interpolation of the relevant quantities from the true
\teff\ to \tnom.

For \feh\,=\,$0.0$ and $-2.0$, we have in addition computed
independent, refined grids of 1D\,\lhdm\ models, covering the same range of
gravities (\logg\,=\,$3.5$, $4.0$, $4.5$) and effective temperatures
($5000\le$\teff\,$\le 6500$\,K), but with a smaller and equidistant step of
$\Delta$\teff\,=\,$100$\,K. These auxiliary grids of 1D\,\lhdm\ models are
useful for estimating the interpolation errors of the fitting functions
developed in Section\,\ref{sfitCorr}.

\subsection{Library of 3D and 1D\,NLTE synthetic spectra}
\label{synthesis}
To ensure that our fitting functions cover a sufficiently wide range of 
lithium abundances, we consider eleven values of $A$(Li), from 1.0 to 3.0 
with a step of 0.2 dex. For each abundance, we calculate individual
NLTE departure coefficients for all snapshots of each 3D model and for their 
associated one-dimensional \lhdm\ counterparts, using our \nlte\ code 
\citep[see, e.g.,][]{steffen2015,mott17}. 
We adopted the same lithium model atom as described in \cite{mott17}, which is
more complete than the one used by \cite{sbordone2010}, who, moreover, 
calculated the NLTE departure coefficients only for one representative value 
of $A\rm(Li)=2.2$ and assumed them to be valid for all considered lithium
abundances in the range $0.9 \le$ $A$(Li) $\le 3.3$.

Our approach of calculating a separate set of departure coefficients for 
each abundance constitutes a more reliable treatment of the NLTE effects
over the full range of line strengths than the simplified approach adopted by
\cite{sbordone2010}. In this sense, their fitting functions are superseded 
by the updated and extended version provided in the present work.

Once the NLTE departure coefficients are available, they are fed into the 
NLTE spectral synthesis code 
\linfor\footnote{\url{https://www.aip.de/Members/msteffen/linfor3d}} 
to compute a library of synthetic lithium line profiles. 
Our library consists of 770 3D\,NLTE lithium line profiles 
in the wavelength range $\lambda=[670.69-670.87]\,\rm nm$
($14$ \logg--\teff\ combinations $\times 5$ 
metallicities $\times 11$ Li abundances) where each spectrum
is the average over the selected snapshots. For each of
these 3D\,NLTE spectra, we have computed corresponding 1D\,NLTE and 1D\,LTE 
spectra for comparison. In total, the library of synthetic spectra then
comprises $3\times 770=2310$ spectra. In addition, we have computed a
LTE curve-of growth (CoG) for each set of stellar parameters (\teff, \logg, 
\feh) which allows us to interpolate the 1D\,LTE equivalent width 
($\rm EW_{1DLTE}$) of the lithium line over a wide range of $A$(Li) between 
$0.0$ and $4.5$\,dex.
In exactly the same way, we calculated a total of $1056$ additional
1D\,NLTE line profiles ($48$ \logg--\teff\ combinations $\times 2$
metallicities $\times 11$ Li abundances) plus the corresponding LTE CoGs
from the auxiliary (refined) grids of 1D\,\lhdm\ models, available for
\feh\,=\,$0.0$ and $-2.0$ only.

\subsubsection{Microturbulence in 1D\,\lhdm\ models}
In the 1D models, the small-scale velocity fields in the stellar photosphere 
is represented by the so-called microturbulence parameter \micro. As
usual, \micro\ describes a depth-independent isotropic Gaussian velocity 
distribution with a dispersion of $v_{\rm rms}=\,$\micro$/\sqrt{2}$.

In the 3D models, on the other hand, there is no need to introduce such
artificial fudge parameter since the full velocity field is obtained 
naturally from the hydro-dynamical treatment of the convective motions 
in each \cobold\ model atmosphere.

For  a differential 3D-1D comparison, the microturbulence parameter of 
the 1D\,\lhdm\ model associated with a given 3D model should be adjusted
such that the resulting small-scale velocity field matches (as closely as
possible) the properties of the 3D hydrodynamical velocity field.
Such a calibration has been presented by \cite{dutraferreira16} who
provide the following (theoretical) relation for \micro\ as a 
function of effective temperature and surface gravity:
\begin{equation}
\label{microDutraFerreira}
\xi_{\mathrm{micro}}=0.998+3.16\times 10^{-4}X-0.253 Y -2.86\times 
10^{-4}XY + 0.165 Y^2,
\end{equation}
where $X=T_{\rm eff}-5500$ [K] and $Y=\log g-4.0$.

This formula is the result of an analysis of 30 \ion{Fe}{i} plus 7
\ion{Fe}{ii} lines for which synthetic spectra have been computed adopting
3D~\cobold\ and 1D\,\lhdm \ model atmospheres of solar metallicity (we refer
to this paper for more details). Equation (\ref{microDutraFerreira})
constitutes the most appropriate description of the \micro\ parameter
available in the context of 3D-1D comparisons. Even though formally valid for
solar metallicity only, we use this relation to assign \micro\ to all
\lhdm\ models used in the present study. This may be justified by empirical
evidence that \micro\ depends only weakly on \feh\ \citep{Sitnova2015,
  Mashonkina2017}.  The last column of Table\,\ref{tab:Models} lists the
values of \micro\ computed with Eq.\,(\ref{microDutraFerreira}).

\subsubsection{Line list}

The lithium resonance doublet at $\lambda$670.8 nm is represented by
12 hyper-fine structure components, six for each isotope, extracted 
from the original list of \cite{kurucz95}, as explained in \cite{mott17}.
For the spectrum synthesis we assume a solar system  isotopic ratio 
of \iso=8.2\% \citep[e.g.,][]{lodders2003}.

Although our grid includes models with solar metallicity for which
contributions to the equivalent width coming from some blend lines of
different elements may be present, we excluded such features in the spectral
synthesis, maintaining our focus on targets where these other atomic and
molecular species have a negligible effect on the equivalent width (hereafter
EW) of the lithium resonance doublet. See, however, Sect\,\ref{sec:blends},
where we discuss the applicability of our abundance corrections in the
presence of blend lines.

Once the grids of 3D\,NLTE theoretical spectra and 1D (both NLTE and LTE) CoGs
for the $\lambda$670.8 nm lithium line have been completed, we initiated the
procedure for deriving the abundance corrections for each model as a function
of $A$(Li) and their transformation into the functional fits. These are the
result of several steps that are described in the next sections.

\section{3D\,NLTE and 1D\,NLTE lithium abundance corrections}\label{FFcorr}
\label{sec:defcorr0}
\subsection{Definition}
\label{sec:defcorr1}
When dealing with the concept of abundance corrections, it is important to
define properly how such corrections are derived since they can be calculated
in multiple ways. Considering the 1D\,NLTE case, the corrections,
$\Delta_{\rm 1DNLTE-1DLTE}$, are usually defined as 
a function of the 1D\,LTE abundance, $A\rm(Li)_{1DLTE}$, such that:
\begin{equation}
A\mathrm{(Li)_{1DNLTE}}=A\mathrm{(Li)_{1DLTE}} + \Delta_{\rm 1DNLTE-1DLTE} \, .
\end{equation}
To derive this correction for a given $A\rm(Li)_{1DLTE}$
(which is usually the result of a standard 1D\,LTE analysis), one first has to
compute the 1D\,LTE equivalent width associated with this LTE lithium 
abundance, EW$_{\rm 1DLTE}$. The next step is the search for the 1D\,NLTE 
lithium abundance, $A\rm(Li)_{1DNLTE}$ that provides the same EW$_{\rm 1DLTE}$ 
value. Note that this step requires the computation of a 1D\,NLTE 
curve-of-growth.
The sought 1D\,NLTE abundance correction ($\Delta_{\rm 1DNLTE-1DLTE}$)
is then defined as the difference of the two abundances. These corrections 
can be applied to standard 1D\,LTE abundances to correct them for non-LTE 
effects.

\begin{figure}[hbtp]
    \centering
    \mbox{\includegraphics[trim=10 10 10 40, width=\linewidth]{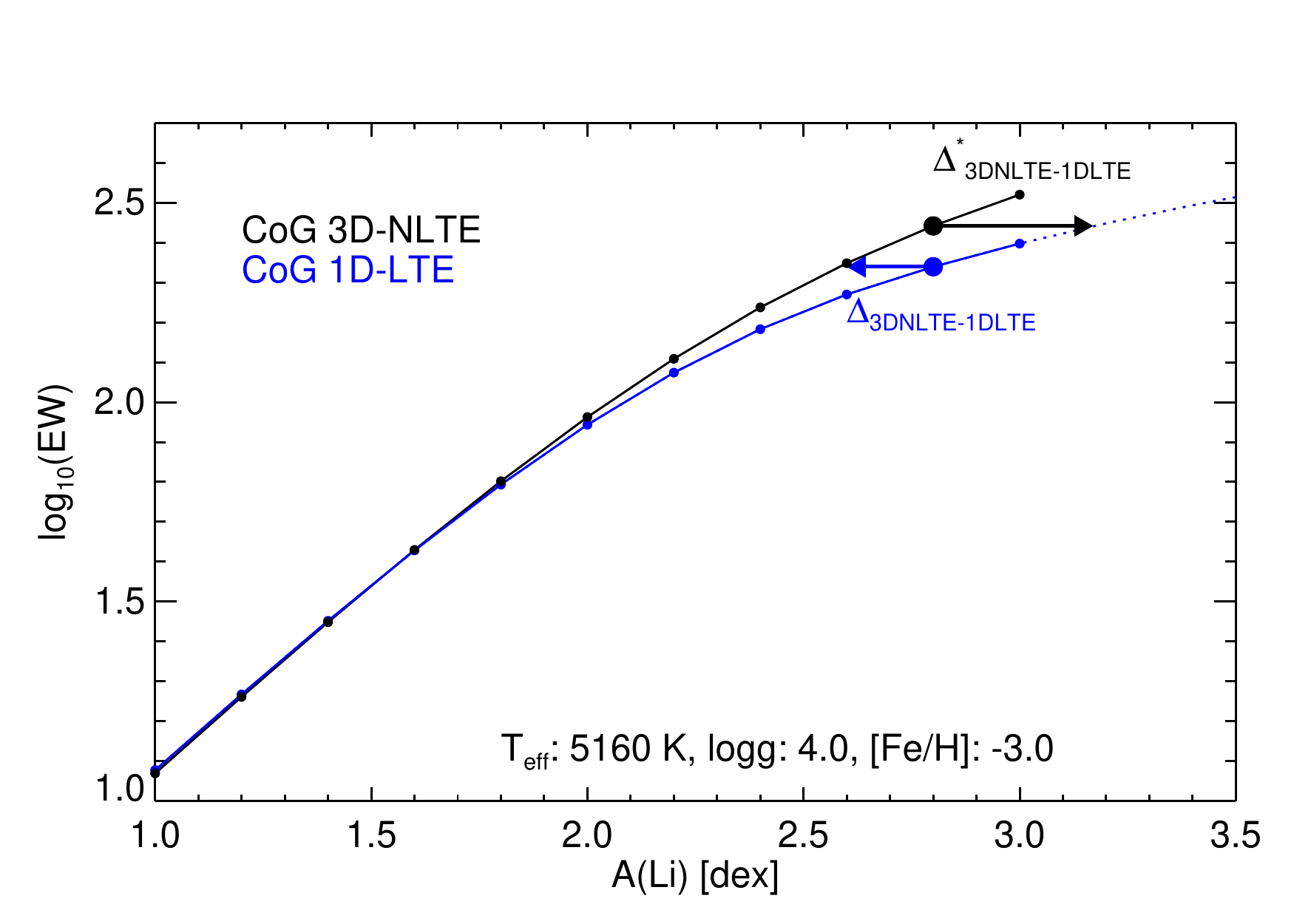}}
	\caption{Two different definitions of abundance corrections 
        for the example of a strong Li line on the  non-linear part of
        the CoG. The correction $\Delta^\ast_{\rm 3DNLTE-1DLTE}$ is indicated by 
        the black horizontal arrow starting on the 3D\,NLTE CoG (black)
        at $A$(Li)\,=\,$2.8$ and ending at a higher abundance given by the
        intersection with the 1D\,LTE CoG (blue).
	Conversely, the correction $\Delta_{\rm 3DNLTE-1DLTE}$ is given
        by the blue horizontal arrow starting on the 1D\,LTE CoG 
        (blue) at the same abundance $A$(Li)\,=\,$2.8$ and extending
        to the 3D\,NLTE CoG (black).}
	\label{defcorr}
\end{figure}

A completely analogous procedure could be applied to derive the 3D\,NLTE
corrections $\Delta_{\rm 3DNLTE-1DLTE}$, again as a function of the 1D\,LTE 
abundance  $A\rm(Li)_{1DLTE}$, allowing to correct the 1D\,LTE abundance for
the combined 3D plus NLTE effect:
\begin{equation}
A\mathrm{(Li)_{3DNLTE}}=A\mathrm{(Li)_{1DLTE}} + \Delta_{\rm 3DNLTE-1DLTE}\, .
\end{equation}  

The ``inverse'' abundance correction $\Delta^\ast_{\rm 3DNLTE-1DLTE}$ is defined
as the correction that has to be added to a given 3D\,NLTE abundance
in order to obtain the corresponding 1D\,LTE abundance. It has the opposite
sign but in general (for partly saturated lines) a different magnitude than
the commonly used ``forward'' correction $\Delta_{\rm 3DNLTE-1DLTE}$. While it is
technically more straightforward to derive the ``inverse'' corrections
$\Delta^\ast_{\rm 3DNLTE-1DLTE}$ from the theoretical models, they are of less
practical interest and we calculate them only as an intermediate
step. Eventually, we provide the ``forward'' corrections $\Delta_{\rm
  3DNLTE-1DLTE}$ as a function of the 1D\,LTE lithium abundance that is
typically derived by analyzing an observed stellar spectrum.

\begin{figure*}[!hbtp]
    \sidecaption
    \mbox{\includegraphics[trim=0 0 0 40,width=12cm]{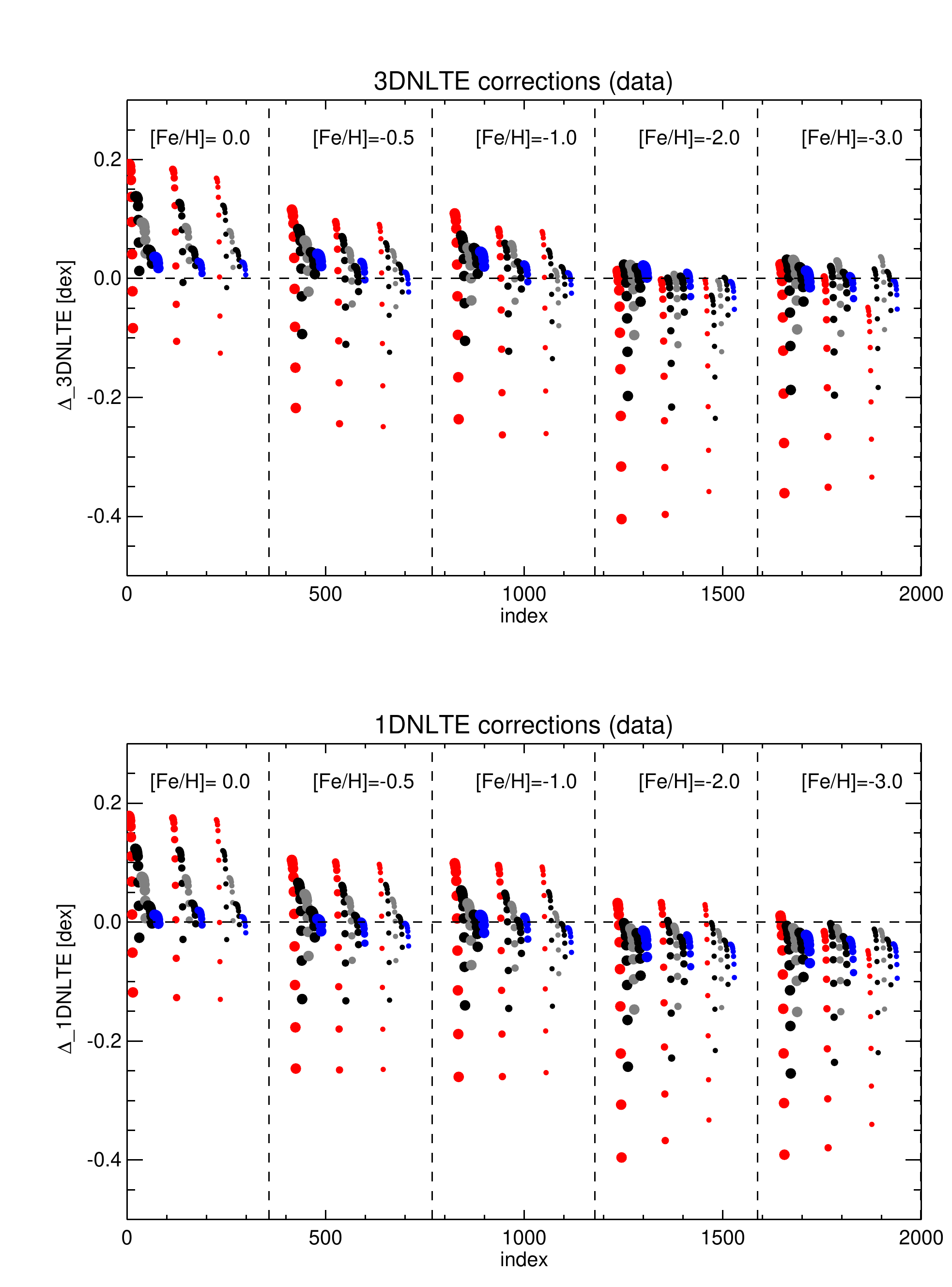}}
    \caption{Overview of the Li abundance corrections
      $\Delta_{\rm 3DNLTE-1DLTE}$ (top) and $\Delta_{\rm 1DNLTE-1DLTE}$
      (bottom) as evaluated on the complete 4D parameter grid.
      For each metallicity, a total of $165$ points indicates the
      corrections as a function of an arbitrary spectrum number
      that runs over $A$(Li), \teff, and \logg. Large (left group),
      intermediate (middle group), and small dots (right group) correspond to
      \logg\,=\,$3.5$, $4.0$, and $4.5$, respectively. For each gravity,
      \teff\ increases from left to right ($5000$\,K: red,
      $5500$\,K: black, $5900$\,K: gray, $6300$\,K: black, $6500$\,K: blue).
      Each effective temperature is represented by $11$ dots indicating the
      results for \mbox{$1.0 \le A$(Li)\,$\le 3.0$}.}
    \label{corr_overview}
\end{figure*}

\subsection{Computation of the lithium abundance corrections}\label{compCorr}
\label{sec:defcorr2}
The abundance corrections defined in Section\,\ref{sec:defcorr1}
are derived by the following procedure.

Given our fixed set of lithium abundances ($1.0\le A\rm (Li)\le 3.0$), 
we compute the equivalent width of the \ion{Li}{i} doublet from the
3D models in full 3D\,NLTE (EW$_{\rm 3DNLTE}$), and the LTE and NLTE EWs 
from the associated 1D LHD models (EW$_{\rm 1DLTE}$ and EW$_{\rm 1DNLTE}$,
respectively) by numerical integration of the synthetic lithium line
profiles over the wavelength range $\Delta\lambda$\,=\,$[670.69-670.87]$\,nm.
These data define three different curves-of-growth
for each set of stellar parameters, corresponding to 3D\,NLTE, 1D\,LTE,
and 1D\,NLTE treatment, respectively.

As an illustration, we show in Figure \ref{defcorr} the 3D\,NLTE CoG (black)
and the 1D\,LTE CoG (blue) for the models with \teff\,=\,$5160$\,K,
\logg\,=\,$4.0$ and \feh\,=\,$-3.0$.  Filled dots mark the computed EWs at the
prescribed grid of lithium abundances, continuous lines show the CoGs obtained
by piecewise cubic interpolation.  According to the above definition, the
3D\,NLTE abundance correction $\Delta_{\rm 3DNLTE-1DLTE}$ corresponds to the
horizontal distance between the two CoGs. For a given 1D\,LTE abundance
(e.g., $A\rm(Li)=2.8$), $\Delta_{\rm 3DNLTE-1DLTE}$ is negative and its
magnitude is indicated by the length of the blue horizontal arrow, which can
be easily obtained by interpolation.

For completeness, Figure \ref{defcorr} also shows the ``inverse'' abundance
correction $\Delta^\ast_{\rm 3DNLTE-1DLTE}$, indicated by the black horizontal
arrow. This is the correction that has to be applied to a given 3D\,NLTE
abundance in order to obtain the corresponding 1D\,LTE abundance. As mentioned
above, it is not considered here as it is not of practical interest.

We derived the 1D\,NLTE corrections $\Delta_{\rm 1DNLTE-1DLTE}$ following
the same method, replacing the $\rm EW_{3DNLTE}$ by the 1D\,NLTE
equivalent widths $\rm EW_{1DNLTE}$, computed with the microturbulence
parameter that is listed in Table\,\ref{tab:Models} for each model atmosphere.
The obtained grids of 3D\,NLTE and 1D\,NLTE abundance corrections have four
dimensions: \teff, \logg, \feh, $A$(Li). At this stage, however, the grids
need some further manipulation, before we are able to represent them by
analytical expressions.

The first issue is that the corrections are based on equivalent widths that are 
computed for the real effective temperatures that depart slightly from the
fixed nominal temperature grid  $\tnom$\,=\,$5000, 5500, 5900, 6300$,
and $6500$\,K. This means that the temperature grid is slightly different for
each (\logg, \feh) combination, preventing a straightforward comparison of
abundance corrections for models with different surface gravity or
metallicity at fixed \teff. This difficulty is eliminated by interpolating
the corrections from the actual \teff\ grid to the fixed \tnom\ grid,
separately for each (\logg, \feh) combination and for each of the 11
values of $A$(Li). Here we employ again a piecewise cubic interpolation,
in some cases involving some very slight (uncritical) extrapolation.
After this step, the grids of 3D\,NLTE and 1D\,NLTE abundance corrections
are represented on a four-dimensional rectangular grid. 
 
The second issue concerns the missing corrections for surface gravity
\logg\,=\,$3.5$ and \teff\,=\,$6500$\,K.  Due to the lack of the corresponding
model atmospheres in the CIFIST grid (blue squares in Figure \ref{cifistFF}),
we decided to find these values by means of extrapolation from the available
corrections. For given \feh\ and $A$(Li), we compute the missing corrections
as the average of the results obtained from two orthogonal linear
extrapolations: (i) using the available corrections at fixed \logg=3.5 and
\tnom=($5900,6300$)\,K, and (ii) using the corrections at fixed
\tnom=$6500$\,K and \logg=($4.0,4.5$).

Even if we could have generated the missing 1D LHD models and the
corresponding 1D\,LTE and NLTE spectra, we decided to compute the missing
1D\,NLTE corrections by extrapolation in exactly the same way as in the
3D case described above. Comparison with the independent results
from the auxiliary (refined) grid of 1D\,\lhdm\ models for \feh\,=\,$0.0$
and $-2.0$ (see Sect.\,\ref{3d_1d_models}) reveals that the extrapolation 
works perfectly for \logg\,=\,$4.5$, but may introduce an error of the order
of $-0.01$\,dex for \logg\,=\,$3.5$ and $4.0$.

After these preparatory steps, our rectangular grids of
$\Delta_{\rm 3DNLTE-1DLTE}$ and $\Delta_{\rm 1DNLTE-1DLTE}$ 
corrections are complete. An overview of the corrections as a function
of the grid parameters $A$(Li), \teff, \logg, and \feh\ is presented in
Figure\,\ref{corr_overview}. The dependence of the 3D and 1D corrections on
the stellar parameters and the lithium abundance is qualitatively similar. For
weak lines (low $A$(Li)), they are generally positive, while they tend to be
negative for the highest Li abundance. The corrections are most sensitive to
the line strength for the coolest models. They are most negative for the
lowest metallicities, reaching values of at least $-0.3$\,dex at
\feh\,$\le$\,$-2.0$, \teff\,=\,$5000$\,K, $A$(Li)\,=\,$3.0$.

\section{Analytical approximations}
\label{sfitCorr}
\subsection{Lithium abundance corrections}
\label{FFI}

In the following, we derive a first parametric equation (hereafter
\texttt{FFI$_{\rm 3DNLTE}$}, \texttt{FFI$_{\rm 1DNLTE}$}) that approximates the
$\Delta_{\rm 3DNLTE-1DLTE}$  and $\Delta_{\rm 1DNLTE-1DLTE}$ corrections,
respectively, as a function of \teff, \logg, \feh, and $A$(Li).

An effective way to achieve this is to perform, as a first step, a 2D-fit of
the corrections in the \tnom-$A$(Li) plane for each \feh\ and \logg, separately.
For this purpose, we utilized the IDL procedure \texttt{SFIT} which allows to
perform a surface fit to a set of 2-dimensional data (in our case
the $\Delta$ values given on a rectangular grid as a function of \tnom\ and
$A$(Li)). \texttt{SFIT} returns an array of coefficients, the number of which
depends on the chosen degree of the polynomial function used in the fitting.

Obviously, the goal is to achieve a reasonable fit of the corrections with a
relatively low number of coefficients.  After several tests we found that our
corrections were satisfactorily fitted by a polynomial function with a maximum
degree equal to 4, expressed as:
\begin{equation}
\label{temporary_eqcorr3DNLTE}
\Delta(y,z)=\sum_{j=0}^{4}y^j\sum_{k=0}^{4-j}z^k\cdot c_{jk}\, ,
\end{equation}
where $y=A\mathrm{(Li)}_{\rm 1DLTE}$ ($1.0\leq y\leq3.0$) and
\mbox{$z=(T_{\rm nom}/1000-5)$} ($0\leq z\leq1.5$).
Such normalized coordinates make the 15 coefficients $c_{jk}$ dimensionless
and of similar magnitude.

Lowering the degree of the polynomial function, we noticed a significant
degradation of the fitting quality for some metallicities or for some surface
gravities. As a consequence, we conclude that in order to fit homogeneously
the whole set of corrections with a unique polynomial form, the optimal choice
is a maximum degree equal to 4. 

\begin{figure*}[h!]
    \sidecaption
    \includegraphics[width=12cm]{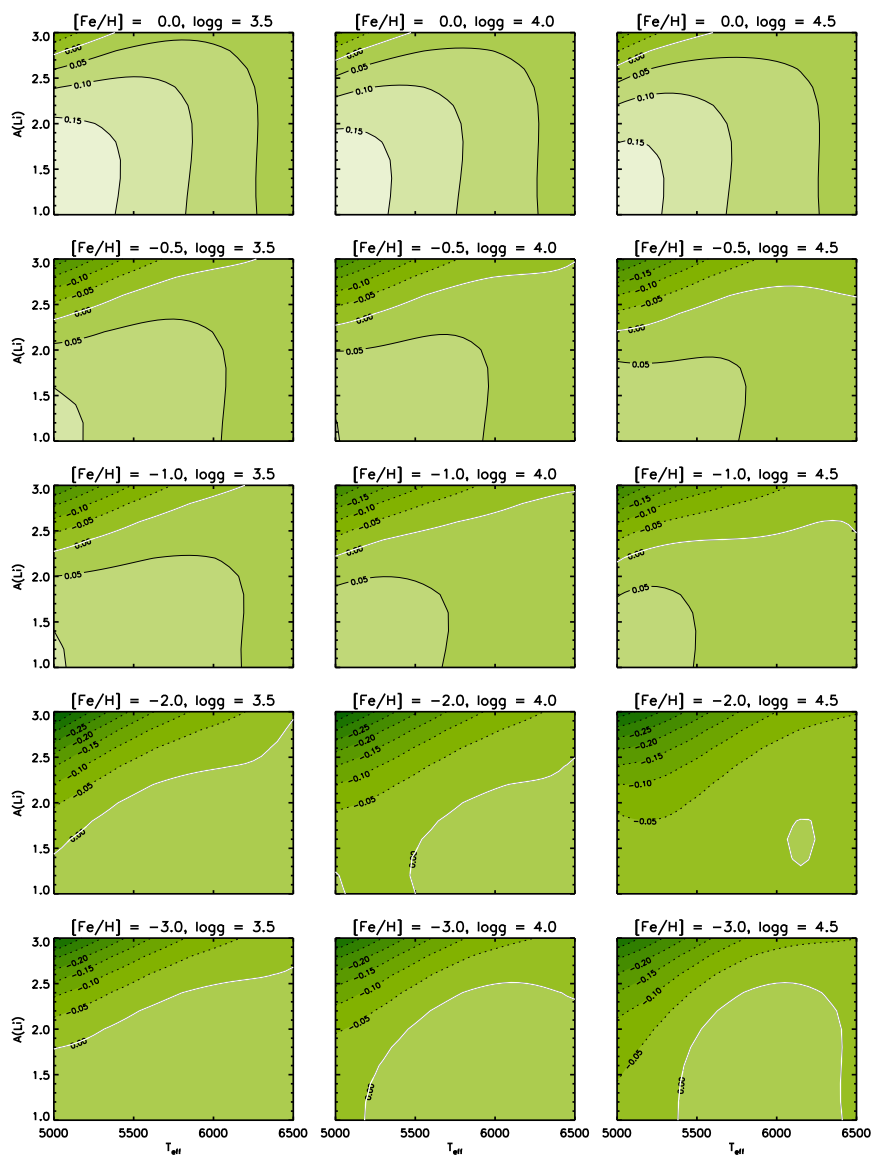}  
	\caption[Contour plots of the 3D\,NLTE abundance corrections]
                {Contour plots of the 3D\,NLTE abundance corrections in the
                 \teff--$A$(Li)$_{\rm1DLTE}$ plane. Each panel shows the
                 corrections computed with \texttt{FFI$_{\rm 3DNLTE}$}
                 (Eq.\,\ref{eqcorr3DNLTE}) for all combinations of surface
                 gravity and metallicity as given in the title of each
                 plot.}
        \label{contour_corr3D}
\end{figure*}

\begin{figure*}[h!]
    \sidecaption
    \includegraphics[width=12cm]{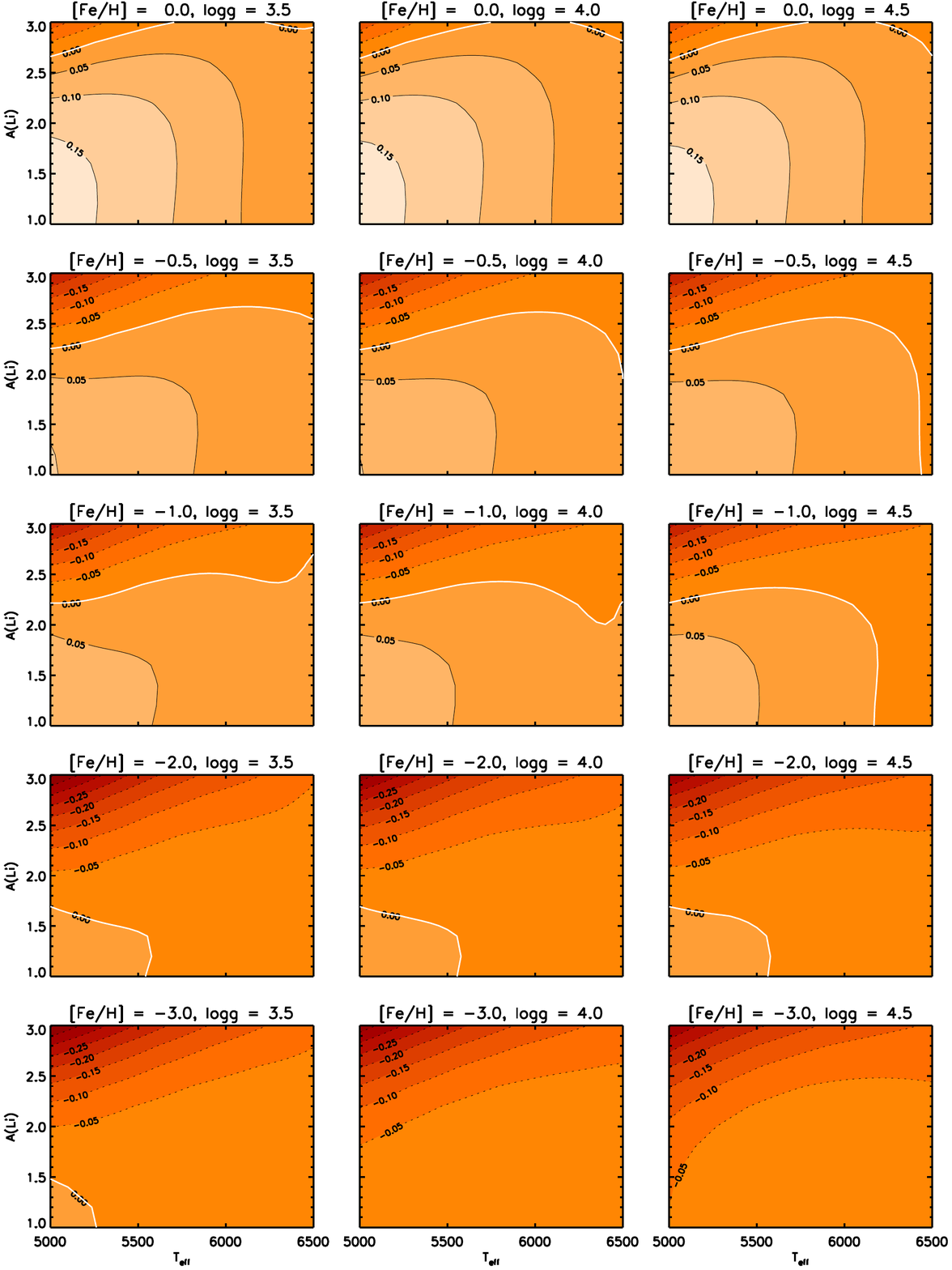}  
	    \caption[Contour plots for the 1D\,NLTE abundance corrections]
            {Contour plots of the 1D\,NLTE abundance corrections in the
                 \teff--$A$(Li)$_{\rm1DLTE}$ plane. Each panel shows the
                 corrections computed with \texttt{FFI$_{\rm 1DNLTE}$}
                 (Eq.\,\ref{eqcorr3DNLTE}) for all combinations of surface
                 gravity and metallicity as given in the title of each plot.}
            \label{contour_corr1D}
\end{figure*}

Following this strategy, we obtain a set of 15 numerical coefficients
for each given metallicity \feh\ and gravity \logg\ that defines the 
fitting surface of the corrections in the \tnom\--$A$(Li) plane.

Noticing that the gravity dependence of the corrections is rather weak, we
represent the gravity dependence of the coefficients $c_{jk}$ by a linear
function in the normalized surface gravity $x = \log g-3.5$ ($0\leq x\leq1$)
as:
\begin{equation}
\label{eq:cijk}
c_{jk}(x)=\sum_{i=0}^1x^i\cdot c_{ijk}\, .
\end{equation}
The coefficients $c_{ijk}$ are obtained from the $c_{jk}$ given at the three
gravities ($x=0.0,\,0.5,\,1.0$) by standard linear regression, separately for
each metallicity \feh.

For each particular metallicity value of our grid, we thus obtain the
following final fitting function that describes the $165$ NLTE lithium abundance
corrections as a function of \teff, \logg, and $A$(Li) through 30 numerical
coefficients:
\begin{equation}
\label{eqcorr3DNLTE}
\texttt{\rm FFI} = \Delta\left(\teff,\logg,A\mathrm{(Li)}\right)
                 =\sum_{i=0}^{1}x^i\sum_{j=0}^{4}y^j\sum_{k=0}^{4-j}z^k\cdot c_{ijk},
\end{equation}

The list of 30 coefficients $c_{ijk}$ are given, for each metallicity covered
by our grid ($\feh= 0.0, -0.5, -1.0, -2.0$ and $-$3.0 dex), in Tables
\ref{coeff_3Dcorr} ($\Delta_{\rm 3DNLTE-1DLTE}$) and \ref{coeff_1Dcorr} 
($\Delta_{\rm 1DNLTE-1DLTE}$) in Appendix \ref{appendixB}.

We decided not to include the metallicity as a fourth parameter of our fitting
function since the trend of the corrections as a function of \feh\
is not as simple as for the \logg\ dependence. Finding a global fitting
function able to fit our data for each combination of \teff, \logg, \feh,
and $A$(Li) simultaneously would have required the inclusion of an impractical
number of coefficients.
  
When applying this fitting function to metallicities other than those that are
tabulated, a further interpolation of the obtained $\Delta$ values to the
desired \feh\ is necessary. We do not expect that this final operation would
compromise the precision of the derived abundance corrections. 

To facilitate the practical application of the analytical fitting function,
we provide a Python script that computes the corrections as a function
of the four parameters  \teff, \logg, \feh, and $A$(Li), using the coefficients
listed in Tables \ref{coeff_3Dcorr} and \ref{coeff_1Dcorr}, respectively,
according to Eq.\,(\ref{eqcorr3DNLTE}), and subsequently performing a
piecewise cubic monotonic interpolation \citep[see][]{steffen90}
in \feh.
An overview of the lithium abundance corrections obtained with the fitting
functions \texttt{FFI$_{\rm 3DNLTE}$} and \texttt{FFI$_{\rm 1DNLTE}$} is shown
in Figures \ref{contour_corr3D} and \ref{contour_corr1D}, respectively,
as a series of contour plots in the \teff\,--\,$A$(Li) plane, separately for
each combination of \feh\ and \logg.

To check how well our fitting functions are able to represent the
$\Delta_{\rm 3DNLTE-1DLTE}$ and $\Delta_{\rm 1DNLTE-1DLTE}$ input data, we computed
for each combination of \feh, \logg, \teff, and $A$(Li) the
difference between the abundance correction predicted by the fitting
function and the value of the original input data. Figure\,\ref{ff1_errors}
shows these differences at the 4D grid points, both for the 3D\,NLTE and the
1D\,NLTE corrections.

For solar metallicity, the residuals $\delta_{\rm FFI}$ fall in the range
$|\delta_{\rm FFI}| \la 0.01$, with a rms mean of
$\langle\delta^2_{\rm FFI}\rangle^{1/2} < 0.004$.
For all subsolar metallicities, the residuals are slightly larger,
lying in the range $|\delta_{\rm FFI}| \la 0.02$, with a typical rms mean
of $0.005$. With very few exceptions, the maximum residuals occur at
\teff\,=\,$5000$\,K for the extreme lithium abundances, $A$(Li)\,=\,$1.0$ or
$3.0$.

In the 1D case, we can additionally evaluate the interpolation errors
of \texttt{FFI$_{\rm 1DNLTE}$} for \feh\,=\,$0.0$ and $-2.0$ by comparison
with the corrections derived independently from the auxiliary grid of
1D\,\lhdm\ models for a refined sample of effective temperatures. We find that
the interpolation errors at the intermediate temperatures have about the same
amplitude as the fitting function errors at the nodes of the original grid,
with $|\delta_{\rm FFI}| \la 0.016$\,dex in all cases, as shown in full detail
in Fig.\,\ref{ff1_errors_lhd}. At an arbitrary point in the multidimensional
parameter space (\teff, \logg, $A$(Li)), the interpolation error may be
somewhat larger, due to the additional (unknown) interpolation errors in
\logg\ and $A$(Li). However, we can assume that the latter two interpolation
errors are small compared to the interpolation error in \teff, due to the
weak gravity dependence of the corrections and the fine spacing of the
original $A$(Li) grid, respectively, such that the interpolation errors
shown in Fig.\,\ref{ff1_errors_lhd} can still be considered as a valid
order-of-magnitude estimate of the total interpolation error.

Taken together, the error analysis indicates an overall good fit of
tabulated abundance corrections by our fitting function
\texttt{FFI$_{\rm 1DNLTE}$}, and by implication \texttt{FFI$_{\rm 3DNLTE}$},
across the whole \teff--\logg--$A$(Li) space. As an additional benefit,
the fitting functions smooth out small numerical artifacts in the raw data.

\begin{figure*}[h!]
   \sidecaption
   \includegraphics[width=12cm]{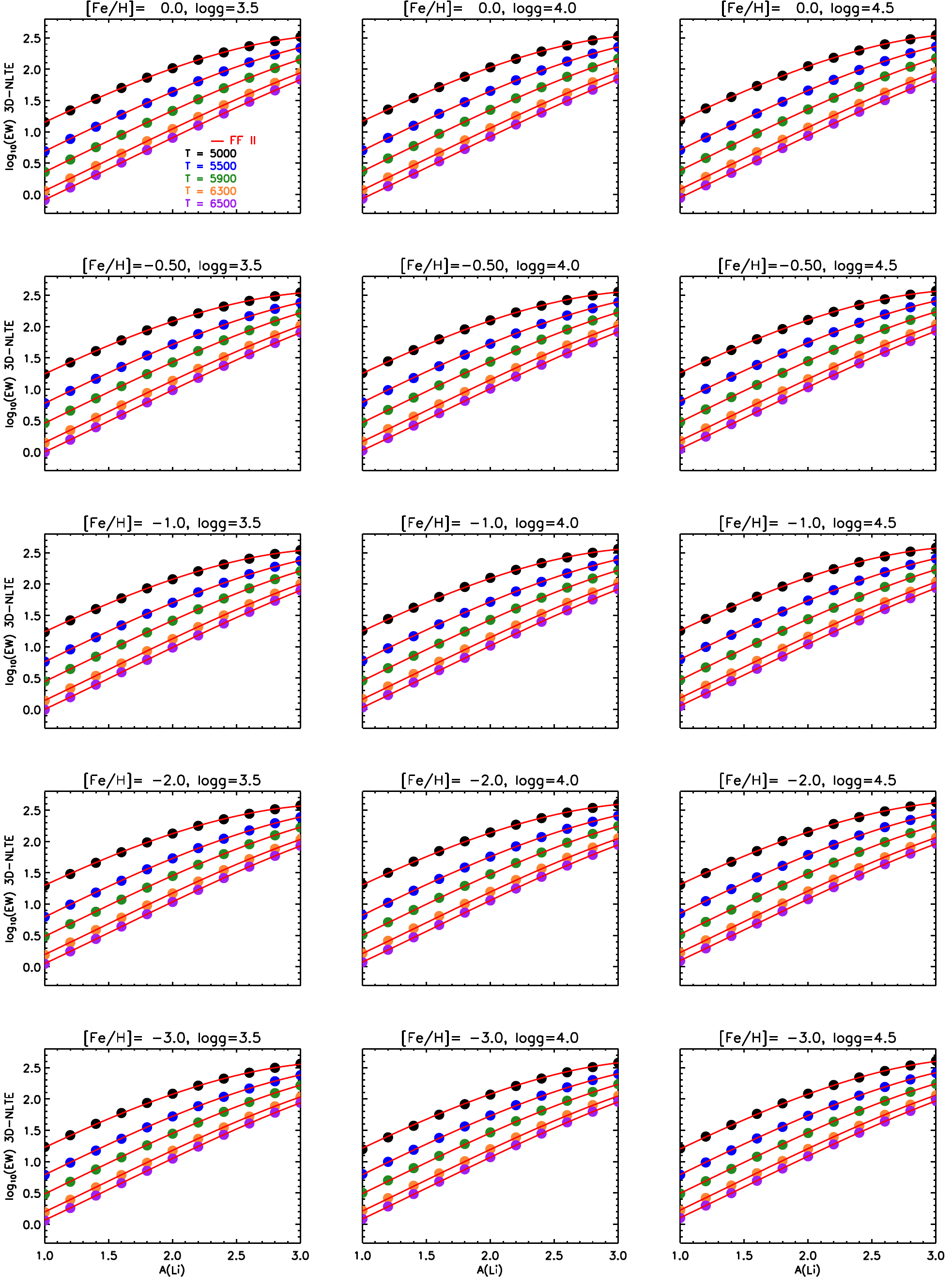}
   \caption[Comparison between EW data and \texttt{FFII$_{A\rm (Li)-EW}$}
     (3D\,NLTE)]{Comparison between the input data (filled dots, color-coded
     according to effective temperatures as shown in the legend of the upper
     left panel) and the resulting fitting polynomial
     \texttt{FFII$_{\rm 3DNLTE}$} (Eq.\,\ref{eqFFII}, red continuous line).
     The surface gravity and the metallicity values are shown in the title
     of each plot.}
\label{FFII_3DNLTE_comparison}
\end{figure*}

\subsection{From $A$(Li) to EW}
\label{FFII}
In this section we describe the derivation of a second parametric equation
(hereafter \texttt{FFII$_{\rm 3DNLTE}$}, \texttt{FFII$_{\rm 1DNLTE}$}) which
allows the computation of the 3D\,NLTE and 1D\,NLTE equivalent width values for
the input parameters \teff, \logg, \feh, and lithium abundance $A$(Li).

The principal ingredient is obviously represented by the equivalent width
values of the lithium doublet for each one of the 70 model atmospheres in our
grid as computed for each one of the 11 lithium abundances in the range
$A\rm(Li)=[1.0\,..\, 3.0]$. For the 1D case only, the microturbulence
parameter \micro\ of each model has been set to the value calculated by means
of Eq.\,(\ref{microDutraFerreira}).  Otherwise, the derivation of the
functional fit described in the following is the same for the 3D\,NLTE and the
1D\,NLTE case.

In a first step, the EWs obtained from the model atmospheres (at given \logg,
\feh, and $A$(Li)) were interpolated from the actual \teff\ values to the
fixed \tnom\ grid. In a second step, the resulting rectangular grid of
EW values is completed by estimating the missing EW values at $\tnom=6500$ K
and $\logg=3.5$ for each metallicity by means of linear extrapolation as
described in Section \ref{compCorr}.

For each \logg\,--\,\feh\ combination, we thus obtain a complete, rectangular
grid of EW values as a function of temperature and lithium abundance. The
derivation of the fitting function \texttt{FFII} is completely analogous to
the procedure to derive the fitting function \texttt{FFI} for the
abundance corrections as described in Sect.\,\ref{FFI}. First we use
the IDL procedure \texttt{SFIT} to compute the coefficients of the
two-dimensional polynomial function of maximum degree four providing the
best surface fit to the (logarithmic) EWs in the $y$\,--\,$z$ plane,
separately for each of the $15$ \logg\,--\,\feh\ sets. Then the
\logg-dependence of the coefficients is represented by a linear function
in $x= \log g-3.5$. In this way, we obtain for each of the five metallicities
a set of $30$ dimensionless coefficients $w_{ijk}$  that define the fitting
function \texttt{FFII} for the EWs. It has the same form as
Eq.\,(\ref{eqcorr3DNLTE}):

\begin{eqnarray}
\label{eqFFII}
\texttt{FFII} & = & \log_{10}(\mathrm{EW})\left(\teff,\logg,A\mathrm{(Li)}\right) \nonumber \\
& = &\sum_{i=0}^{1}x^i\sum_{j=0}^{4}y^j\sum_{k=0}^{4-j}z^k\cdot w_{ijk}\, ,
\end{eqnarray}
where EW is the equivalent width in m\AA, and as before $x$\,=\,$\log g-3.5$,
$y$\,=\,$A\mathrm{(Li)}$, \mbox{$z$\,=\,($T_{\rm nom}/1000-5$)}.
The coefficients $w_{ijk}$ are given in Tables \ref{FF_EW3D} and  
\ref{FF_EW1D} for the 3D\,NLTE and 1D\,NLTE case, respectively.

The comparison plot shown in Fig.\,\ref{FFII_3DNLTE_comparison}
demonstrates how well our function fits the 3D\,NLTE input data.
Clearly, the \texttt{FFII} fitting polynomial of maximum degree
four (continuous red lines) ensures a very good agreement between the input EW
(filled dots, color coded according to \tnom) across the full parameter space
covered by our grid. The fit of the 1D\,NLTE EWs is even slightly
better. For both 3D and 1D, the relative fitting function error is
$|\Delta$EW/EW|\,$\la$\,$2$\,\% with few exceptions, while the rms mean
fitting error on the 4D parameter grid is less than $1$\,\%. 
The maximum global fitting function error is $\Delta$EW/EW\,=\,$+3.9$\,\%
for \texttt{FFII}$_{\rm 3DNLTE}$ (\feh\,=\,$-3.0$, \logg\,=\,$4.0$,
\teff\,=\,$5000$\,K, $A$(Li)=$1.0$) and
$-3.4$\,\% for \texttt{FFII}$_{\rm 1DNLTE}$ (\feh\,=\,$-1.0$, \logg\,=\,$4.0$,
\teff\,=\,$5900$\,K, $A$(Li)=$1.0$). Interpolation errors are of similar size.
For further details see Figs.\,\ref{ff2_errors} and \ref{ff2_errors_lhd}.

For practical application of \texttt{FFII}, we provide a Python script
that computes the EWs as a function of the four parameters \teff, \logg, \feh,
and $A$(Li), using the coefficients listed in Tables \ref{FF_EW3D} and
\ref{FF_EW1D}, respectively, according to Eq.\,(\ref{eqFFII}), followed by a
piecewise cubic monotonic interpolation in \feh.

\subsection{From EW to $A$(Li)}\label{FFIII}
With the equivalent width fitting functions \texttt{FFII} presented in Section
\ref{FFII} and defined by Eq.\,\ref{eqFFII}, it is in principle already
possible to derive the 3D\,NLTE and 1D\,NLTE lithium abundances corresponding
to a given input equivalent width by numerical inversion.  Using the provided
\texttt{FFII} procedure, one could construct a curve-of-growth for an
appropriate range of abundances that maps to a range of equivalent widths
covering the ``observed'' EW.  The desired lithium abundance,
$A\rm(Li)_{3DNLTE}$ or $A\rm(Li)_{1DNLTE}$, for that particular EW may then be
easily derived through interpolation in the inverted curve-of-growth
EW$\,\rightarrow\,A$(Li).

As an alternative to this numerical procedure, we have developed a third
parametric fitting formula (hereafter \texttt{FFIII}) that represents
the inverse of \texttt{FFII}. As such, it provides the  3D\,NLTE or
1D\,NLTE value of $A$(Li) for a given input EW and stellar parameters \teff,
\logg, and \feh.
As an algebraic inversion of \texttt{FFII} is not feasible,
we construct the inverse fitting function  \texttt{FFIII} from
scratch, based on the raw data (EW as a function of $A$(Li)).

\begin{figure}[t!]
\includegraphics[angle=0,width=0.5\textwidth, clip]{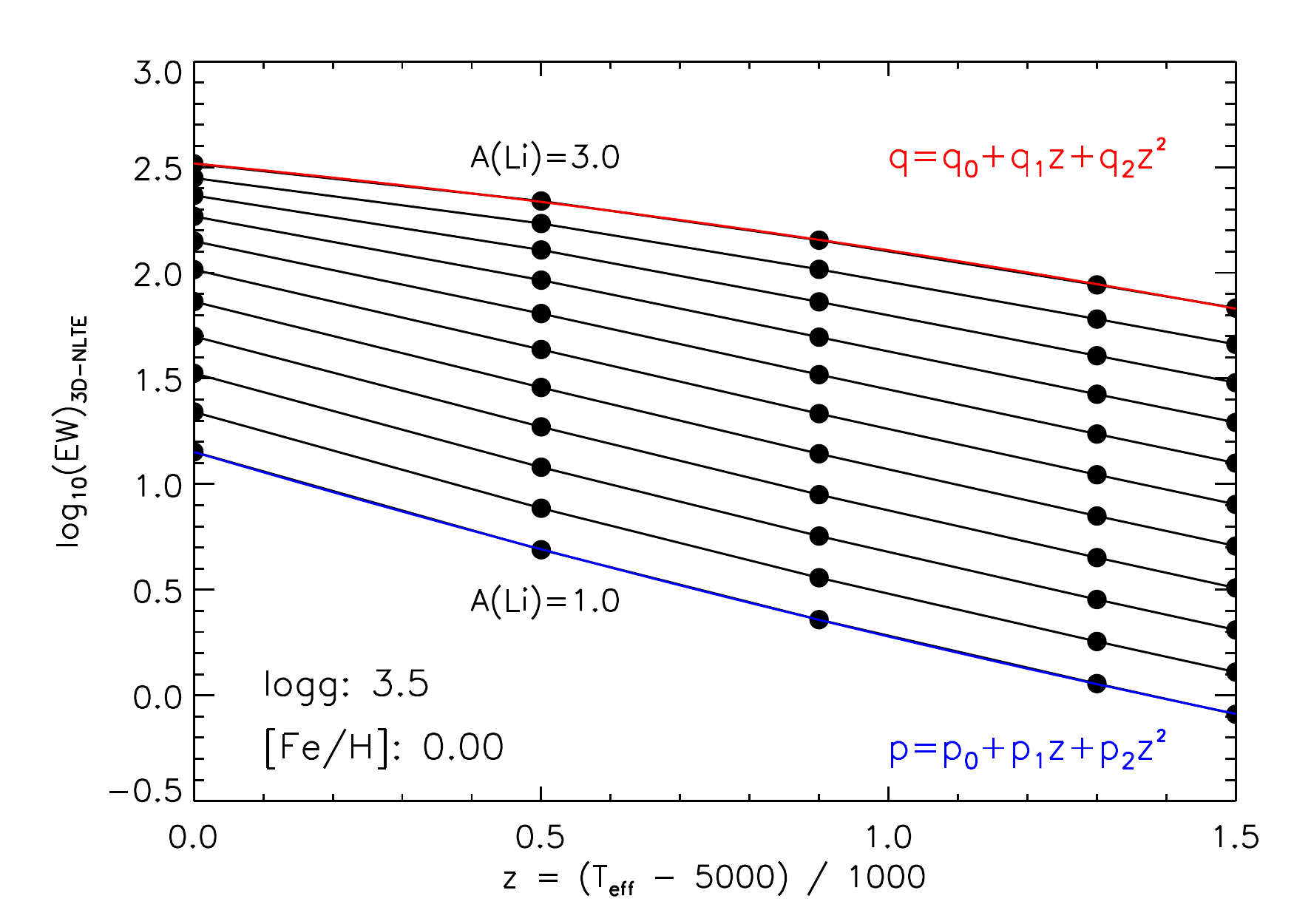}
\includegraphics[angle=0,width=0.5\textwidth, clip]{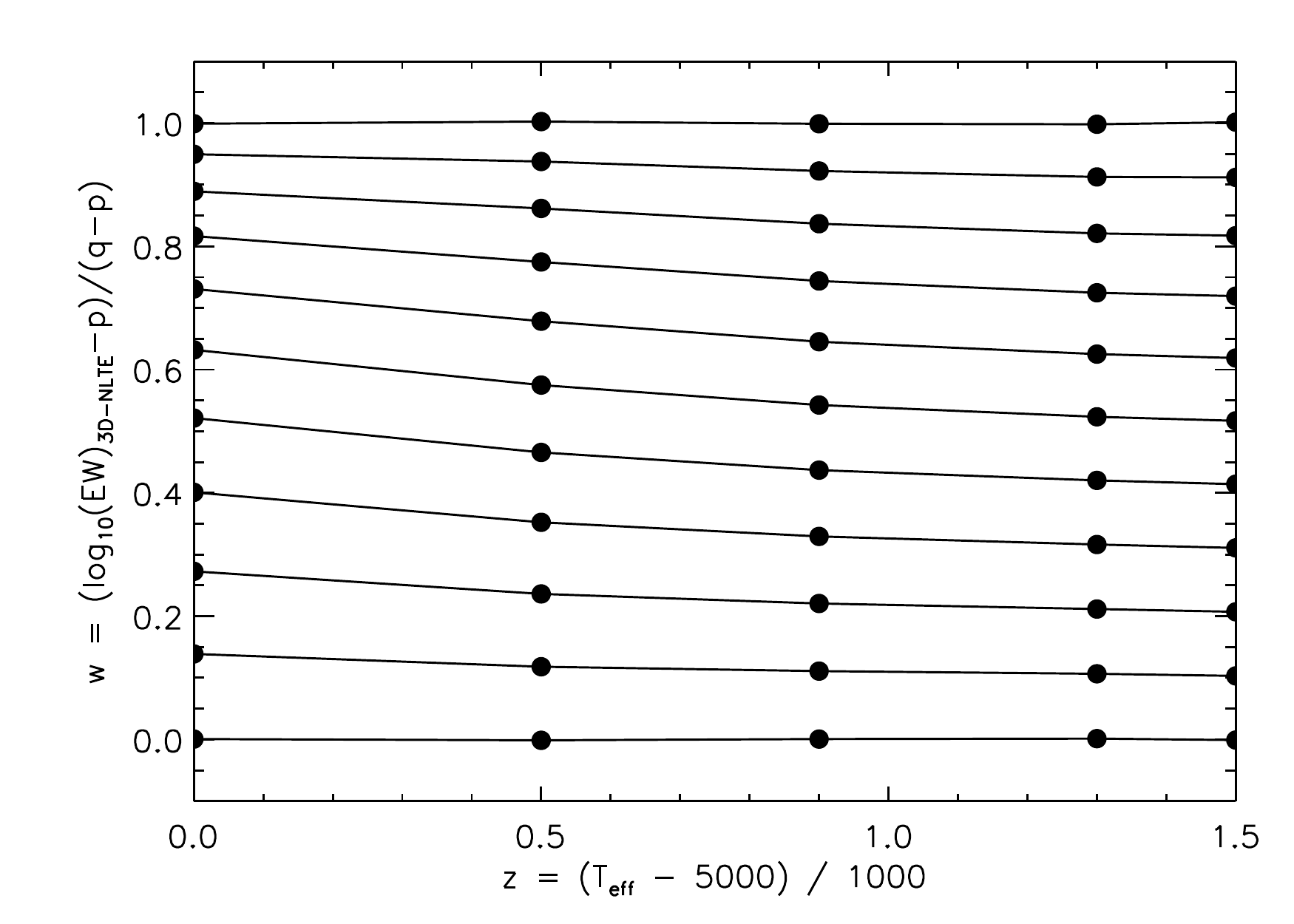}
\caption{
  Distribution of the 3D\,NLTE data points for \logg\,=\,$3.5$, \feh\,=\,$0.0$
  in the \teff\,--\,EW plane (upper panel), and in the \teff\,--\,$w$ plane
  (lower panel) where the systematic trend with temperature is eliminated by
  normalizing the EWs with the help of the parabolic boundary curves $p(z)$
  and $q(z)$ shown in the upper panel (see text for details).
}
\label{ConversionFFII}
\end{figure}

The EW grid we start with has already been interpolated to the nominal
temperature scale \tnom\ and completed by extrapolation where data are
missing, as described in Section \ref{FFII}.
In principle this grid would be ready for the fitting, but in this case
we encounter the additional difficulty that the data in the \tnom\,--\,EW
plane are not given on a rectangular grid, as shown in
Fig,\,\ref{ConversionFFII} (upper panel) for the 3D case 
\logg\,=\,$3.5$, \feh\,=\,$0.0$. While fitting data on an irregular
grid is not impossible, it becomes easier if the parameter space is
rectangular. Then it is possible to use the standard procedures to obtain
a good fit of the data with a polynomial function of relatively low degree.
However, it is not feasible to simply remap the data points to such a
rectangular grid, as this would require an unacceptable amount of
extrapolation. We also note that, from a physical point of view, a rectangular
\tnom\,--\,EW grid is not adapted to the systematic temperature dependence of
the line strength (see Fig,\,\ref{ConversionFFII}a).

In view of theses difficulties, we introduce the concept of a ``scaled
equivalent width'' $w$. As a first step, we define two auxiliary fitting
curves indicated in the lower panel of Figure \ref{ConversionFFII} as $p(z)$ and
$q(z)$.  These two quadratic curves approximate the EW values as a function of
\mbox{$z=(\tnom/1000-5)$} for the extreme abundances of our grid,
$A\rm(Li)=1.0$ ($p$, blue curve) and $A\rm(Li)=3.0$ ($q$, red curve),
respectively.  The scaled equivalent width is then defined by the following
temperature-dependent conversion:
\begin{equation}
\label{eqScalingEW}
w(z)=\frac{\log_{10}(\mathrm{EW})-p(z)}{q(z)-p(z)},
\end{equation} 
where $p(z)=p_0+p_1z+p_2z^2$ and $q(z)=q_0+q_1z+q_2z^2$ are the aforementioned
boundary fitting curves. We note that the quality of linear fits was found
to be unsatisfactory.

In principle, the coefficients $p_m, q_n$ depend on both \logg\ and \feh.
However, the fact that the EW values are only very weakly dependent on surface
gravity allows us to consider two ``global'' parabolas to scale all the
equivalent width values for a given metallicity. The coefficients $p_m(\feh)$
and $q_n(\feh)$ were obtained by averaging the respective individual
coefficients for the three surface gravities.  This approximation has the
advantage of decreasing the number of numerical coefficients defining our
inverse fitting function \texttt{FFIII}.  The averaged coefficients are
provided in Table \ref{coeffParabolas} of Appendix \ref{appendixB} as a
function of \feh\ for both the 3D\,NLTE and the 1D\,NLTE case.

The EW scaling largely removes the systematic trend with \teff, as shown in
the lower panel of Figure \ref{ConversionFFII}, where we now plot the
distribution of the data points in the $z-w$ plane. By construction,
$0 \le w \le 1$, but the grid is still not perfectly rectangular. However,
a rectangular grid can be easily created now by interpolation, without the
need for any extrapolation.

For this purpose, we selected 21 equidistant values between $w=0$ and $1$
with a step of $\Delta w=0.05$ that will constitute, together with the
temperature scale $z$, the rectangular parameter space on which our lithium
abundances are distributed.  The 21 $A$(Li) values that correspond to the
fixed $w$ grid are found by piecewise cubic, monotonic interpolation,
separately for each $z$, \logg, and \feh.

A this point the rectangular grid of lithium abundances can be fitted in the
rectangular $z-w$ parameter space. Also in this case, we choose to fit the
surface of the abundance values by a two-dimensional polynomial function of
maximum degree 4, following exactly the same method as for the abundance
corrections (\texttt{FFI}, Section \ref{FFI}) and the equivalent widths
(\texttt{FFII}, Section \ref{FFII}), again assuming a linear gravity
dependence of the coefficients.

The final analytic expression can be written as:
\begin{eqnarray}
\label{eqFFIII}
\texttt{FFIII} & = & A\mathrm{(Li)}\left(\teff,\logg,\mathrm{EW}\right) \nonumber \\
& =& \sum_{i=0}^{1}x^i\sum_{j=0}^{4}w^j\sum_{k=0}^{4-j}z^k\cdot a_{ijk}\, ,
\end{eqnarray}
with $w$\,=\,$(\log_{10}$(EW)$-p(z))/({q(z)-p(z))}$,
$p(z)=p_0+p_1z+p_2z^2$, $q(z)=q_0+q_1z+q_2z^2$, and
$x$\,=\,\logg\,$-3.5$, \mbox{$z=(\tnom/1000-5)$}.

The coefficients $a_{ijk}$ are presented as a function of \feh\ in
Appendix \ref{appendixB}, in Tables \ref{coeffFFIII3D} and \ref{coeffFFIII1D}
for the 3D\,NLTE and the 1D\,NLTE case, respectively, while the coefficients
$p_m$ and $q_n$ describing the transformation from EW to $w$ are provided in
Table \ref{coeffParabolas}.

Employing Eq.\,(\ref{eqFFIII}), optionally using the provided
Python script that in addition performs interpolation in
\feh, the user obtains  directly the 3D\,NLTE or the 1D\,NLTE lithium
abundance for given stellar parameters and the measured EW, avoiding
intermediate steps such as the aforementioned inversion that is required
if one uses the forward fitting function \texttt{FFII} for this purpose.

We have checked that our inverse fitting function \texttt{FFIII} works
very well. The fitting function error is evaluated by taking the equivalent
widths EW(\teff, \logg, \feh, $A$(Li)) on the grid points of our full 4D
parameter space as input and comparing the output of \texttt{FFIII},
$A$(Li)$_{\rm FFIII}$, with the true (input) lithium abundance $A$(Li)$_{\rm inp}$.
As shown in Fig.\,\ref{ff3_errors}, we find a maximum fitting
function error of $A$(Li)$_{\rm FFIII}-A$(Li)$_{\rm inp}=-0.033$\,dex, both for
\texttt{FFII}$_{\rm 3DNLTE}$ and \texttt{FFII}$_{\rm 1DNLTE}$,
occurring at \feh=$-0.5$, \logg=$4.0$, \teff=$5000$\,K, $A$(Li)=$3.0$,
where the equivalent with of the lithium doublet is $\ga 360$\,m\AA.
Apart from these extreme conditions, \texttt{FFIII} performs much
better, with a rms mean fitting error of $\approx 0.006$\,dex in both cases.
The interpolation error is typically confined in the range
$|\delta_{\rm FFIII}|\la 0.02$ (see Fig.\,\ref{ff3_errors_lhd}).

\section{Comparison with other authors}
In this section, we compare the results of our fitting functions with some of
the existing work in the literature that is focused on lithium abundance
corrections and parametric equations for deriving NLTE lithium abundances from
EW measurements.  The availability of such independent analysis performed on a
common range of stellar parameters is used to check the consistency of our
results and to discuss potential differences in the adopted methods.

\subsection{$A$(Li) to EW: comparison with \cite{sbordone2010}}
The work presented in the previous sections has some important features in
common with the work published by \cite{sbordone2010}.  Both investigations
are based on 3D~\cobold\ model atmospheres and the related 1D\,\lhdm \ models,
and use the same codes to compute the NLTE departure coefficients (\nlte) and
the line formation of the lithium doublet (\linfor). The immediate advantage
of this comparison is that systematic differences in the results due to
different input model atmospheres can be excluded.

\begin{figure*}[hbtp]
    \sidecaption
    \includegraphics[width=12cm,trim=2cm 1cm 2cm 1cm,clip]{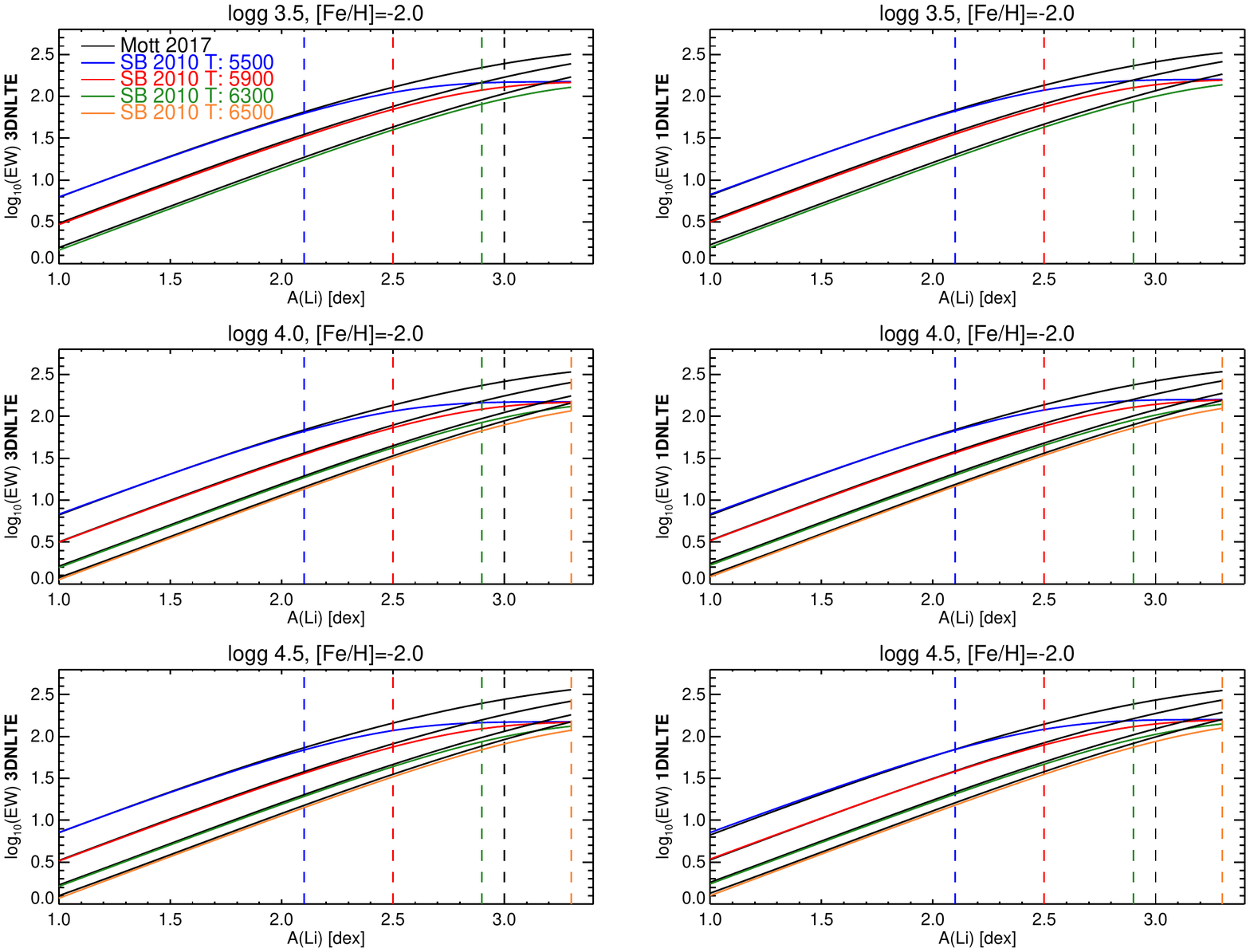}
        \caption{Comparison between the Li\,670.8\,nm curves-of-growth
          computed with the fitting function of \cite{sbordone2010} for
          \feh\,=\,$-2.0$ (colored continuous lines) and with our function 
	  \texttt{FFII} (black solid lines) for the 3D\,NLTE (left) and 1D\,NLTE
          (right) cases, respectively, and for three values of surface gravity
          (\logg\,=\,$3.5$, $4.0$, and $4.5$). The vertical colored dashed
          lines indicate the upper limits of the lithium abundance range
          for which the relation of \cite{sbordone2010} is well defined
          (beyond these limits the fitting function is evaluated in
          extrapolation). Different colors denote the different effective
          temperatures at which the relations are evaluated (see legend in
          upper left panel). The black dashed line represents the upper
          abundance limit up to which our fitting function is defined
          ($A$(Li)\,=\,$3.0$ for all temperatures).}
	\label{SB10_mm20}
\end{figure*}

\begin{figure*}[hbtp]
    \sidecaption
    \includegraphics[width=12cm,trim=2cm 8cm 2cm 0cm,clip]{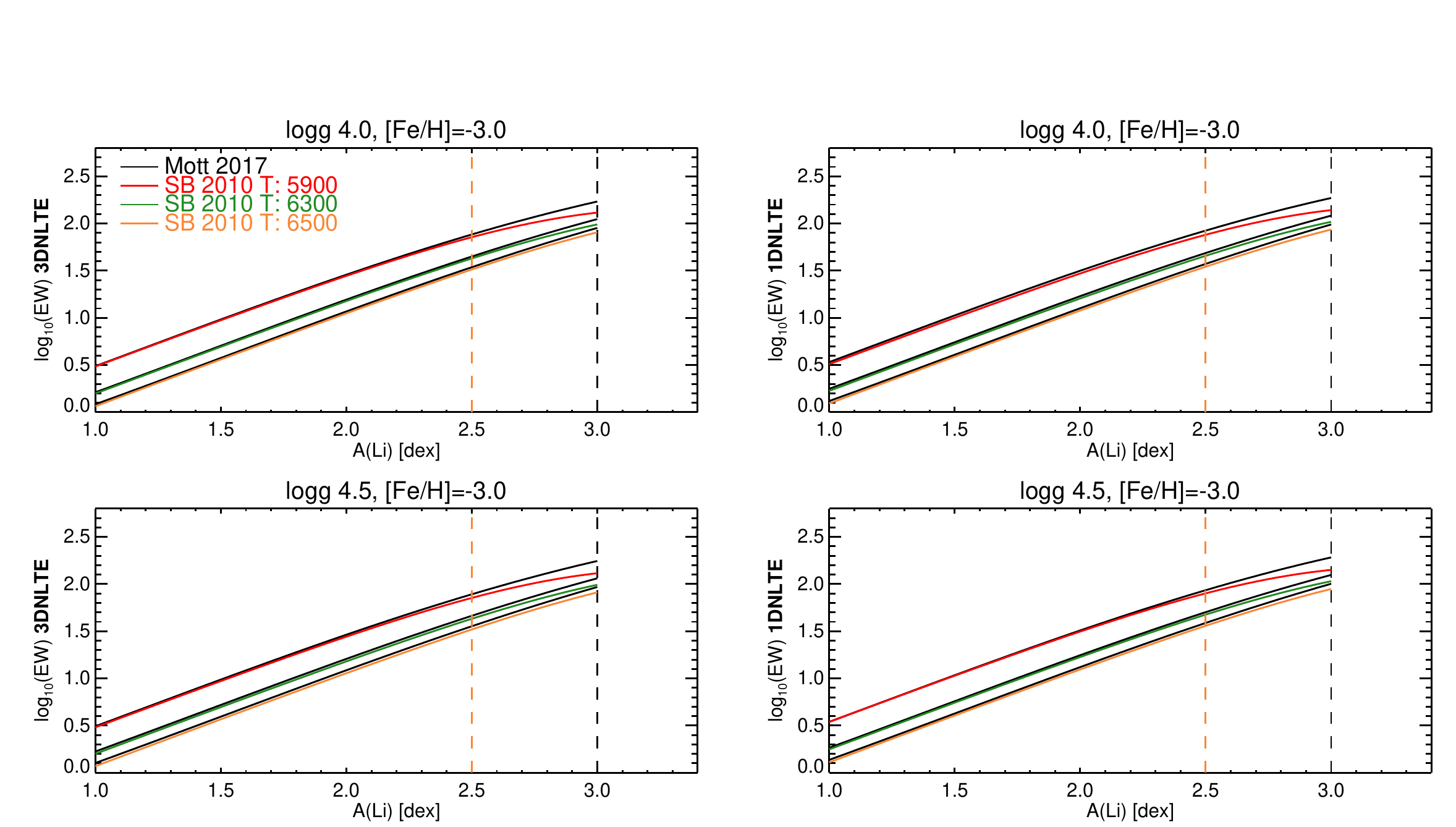}
\caption{Same as Figure \ref{SB10_mm20} but for \feh\,=\,$-3.0$, 
	  \logg\,=\,$4.0$ and $4.5$. In this case, the upper limits to which
          the fitting function by \cite{sbordone2010}
	  is safely applicable coincides with $A \rm (Li)=2.5$ for all
          temperatures (vertical dashed orange line). Our fitting function is
          valid up to $A$(Li)\,=\,$3.0$ (vertical dashed black line).}
	\label{SB10_mm30}
\end{figure*}

In Figures\,\ref{SB10_mm20} and \ref{SB10_mm30}, we show a comparison of the
results, in terms of CoGs obtained by applying both \cite{sbordone2010}'s
fitting function in the form EW(\teff,\logg,\feh,$A$(Li)) and our
\texttt{FFII}(\teff, \logg, $A$(Li); \feh) (Eq. \ref{eqFFII}) to the lithium
abundance range $A\rm(Li)=[1.0, 3.3]$.  Although our functions are valid over
a wider range of metallicity, we limit the comparison to the stellar
parameters that both works have in common, namely \teff\,=\,$5500, 5900, 6300,
6500$\,K, \logg\,=\,$3.5, 4.0, 4.5$ and \feh\,=\,$-2.0$
(\mbox{Figure\,\ref{SB10_mm20}}) and $-3.0$ (\mbox{Figure \ref{SB10_mm30}}).

It is worth noting that the lithium abundance range, specifically the upper
boundary below which the function of Sbordone et al.\ is applicable, depends
on the stellar parameters. This is shown in Figures\,\ref{SB10_mm20} and
\ref{SB10_mm30} by means of colored vertical dashed lines. Beyond these upper
limits, their equation is not trustworthy, since it operates in complete
extrapolation. On the other hand, our fitting function \texttt{FFII} is valid
across the whole abundance range from $A$(Li)\,=\,$[1.0$ to $3.0]$,
independently of the stellar parameters.

If we consider the abundance range where both fitting functions are well
defined, we can conclude that the results are in a very good agreement.
Only very subtle differences in the derived CoGs can be appreciated towards
hotter models and at larger lithium abundances.  This could be related to
the different lithium model atom or to the choice of \cite{sbordone2010} of
using a single set of NLTE departure coefficients, computed for each model
atmosphere assuming $A\rm(Li)=2.2$.  This approximation can only be considered
reasonable as long as the lithium line is rather weak. In the case of a
stronger lithium line (larger $A$(Li)), the NLTE departure coefficients are no
longer independent of the lithium abundance, thus introducing additional
non-linearities in the partly saturated region of the curve-of-growth. In the
present work, we therefore compute consistent departure coefficients for each
individual $A$(Li).

We have verified that, on average, the departure coefficients tend to
be driven closer to unity (i.e.,\ the level populations are driven towards
LTE) as the lines become optically thick. This implies that the Sbordone
at al.\ approximation produces synthetic lithium lines that tend to be too
weak for $A$(Li)\,$>$\,$2.2$. In case of the 3D model with
\teff\,=\,$5500$\,K, \logg\,=\,$4.0$, \feh\,=\,$-2.0$, and $A$(Li)\,\,$=3.0$,
for instance, we find EW\,=\,$226$\,m\AA\ when using the departure
coefficients computed for $A$(Li)\,=\,$2.20$, and  EW\,=\,$263$\,m\AA\ when
using the consistent departure coefficients computed for $A$(Li)\,=\,$3.00$,
thus $\Delta\log(\mathrm{EW})$\,$\approx$\,$0.07$\,dex. While going in the
right direction, this effect is too small to explain the discrepancies
of the order of $0.3$\,dex seen in Fig.\,\ref{SB10_mm20} at $A$(Li)\,\,$=3.0$.

We conclude that these large differences are mainly due to extrapolation errors
arising when the \cite{sbordone2010} formula is applied outside its range of
validity. Such errors can now be avoided by using the extended fitting function
\texttt{FFII} which correctly tracks the impact of NLTE effects across
the abundance grid for each given model atmosphere. If $A$(Li) falls in the
range where both approximations are valid, \texttt{FFII} and
\cite{sbordone2010} are in very close agreement, as expected.
Since the ``inverse'' fitting function of Sbordone et al., EW to $A$(Li),
is the algebraic inversion of their ``forward'' fitting function, a
comparison with our \texttt{FFIII} would not allow for any further conclusions.

\begin{figure*}[hbtp]
    \sidecaption
    \includegraphics[width=12cm,trim=1cm 1cm 1cm 1cm,clip]{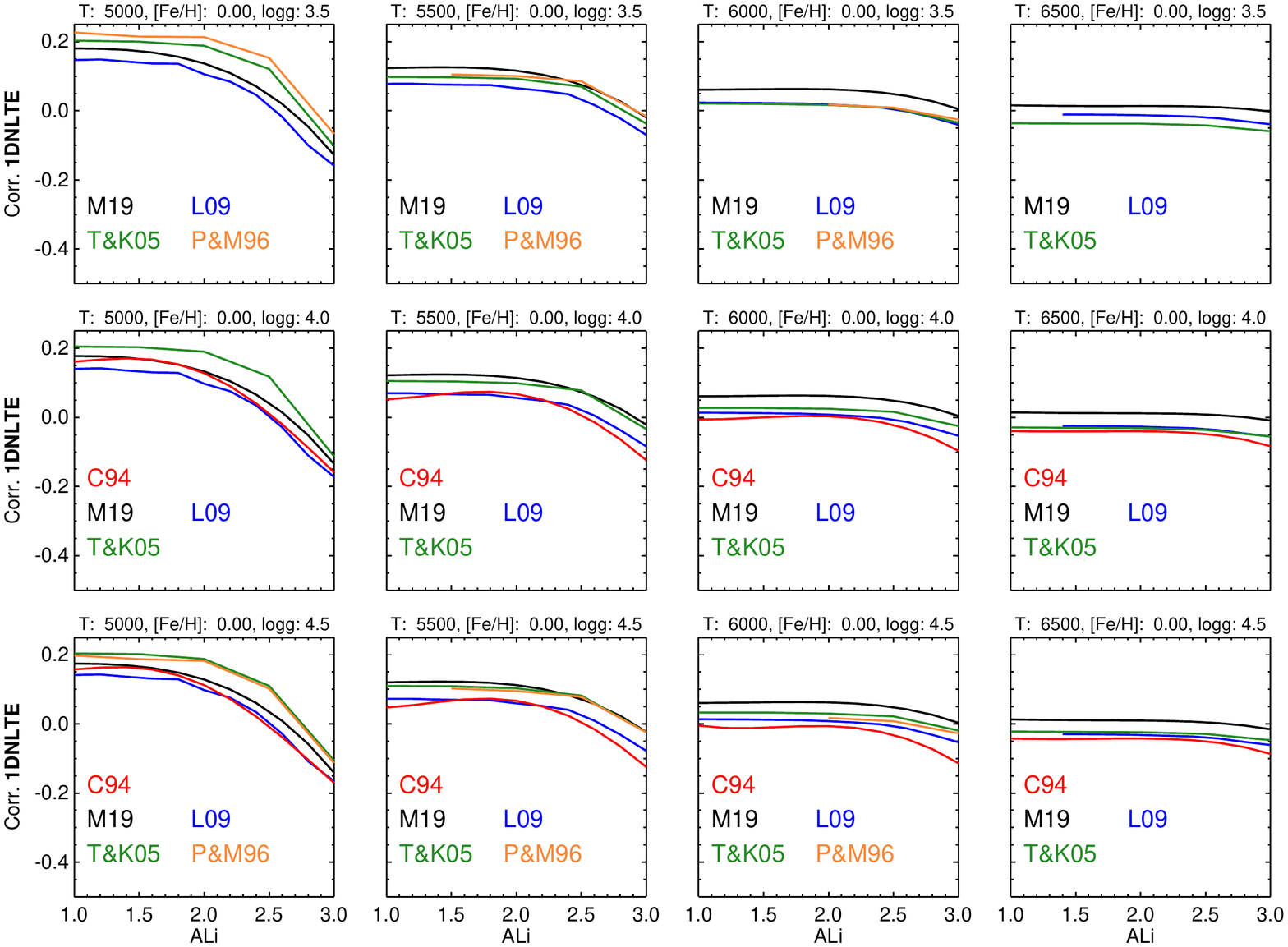}
	\caption{
          Comparison between the results of our fitting function
          \texttt{FFI} for the 1D\,NLTE lithium abundance corrections
          and those published by other authors for
          $\feh$\,=\,$0.0$, $\logg$\,=\,$3.5, 4.0,$ and $4.5$,
          $\teff$\,=\,$5000, 5500, 6000,$ and $6500$\,K.  The black lines
          labeled as M19 show our results, whereas the colored lines
          show the corrections of other authors:
          C94 \citep[][red]{carlsson94},
          P\&M96 \citep[][orange]{pavlenko96},
          T\&K05 \citep[][green]{takeda05}, and
          L09 \citep[][blue]{lind09}. }
	\label{authors_mm00}
\end{figure*}

\begin{figure*}[hbtp]
    \sidecaption
    \includegraphics[width=12cm,trim=1cm 1cm 1cm 1cm,clip]{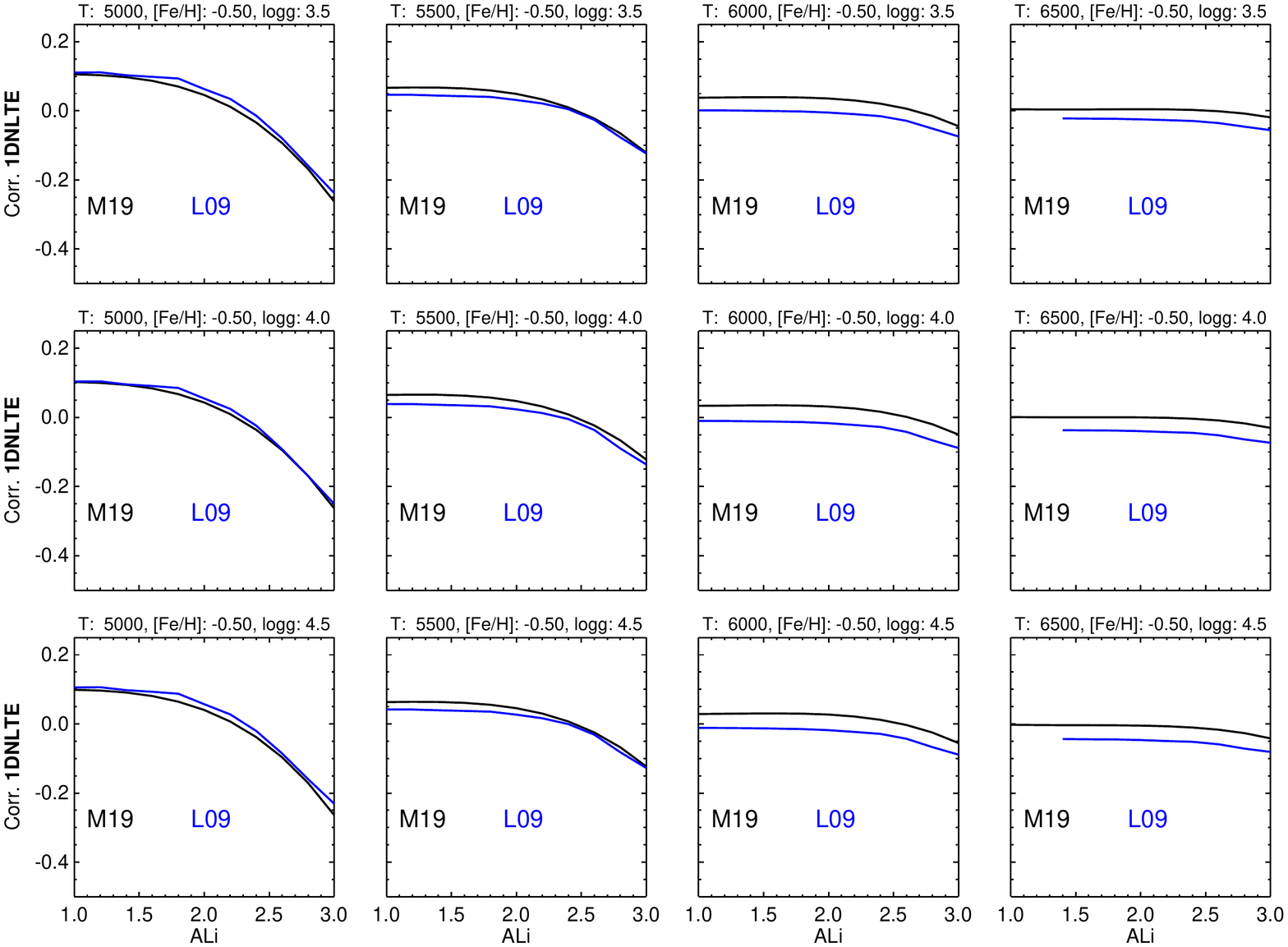}
	 \caption{Same as Figure \ref{authors_mm00} but for
                  $\feh=-0.5$. For this metallicity, a comparison is
                  only possible with the results of \citealt{lind09}. }
	\label{authors_mm05}
\end{figure*}

\begin{figure*}[hbtp]
    \sidecaption
    \includegraphics[width=12cm,trim=1cm 1cm 1cm 1cm,clip]{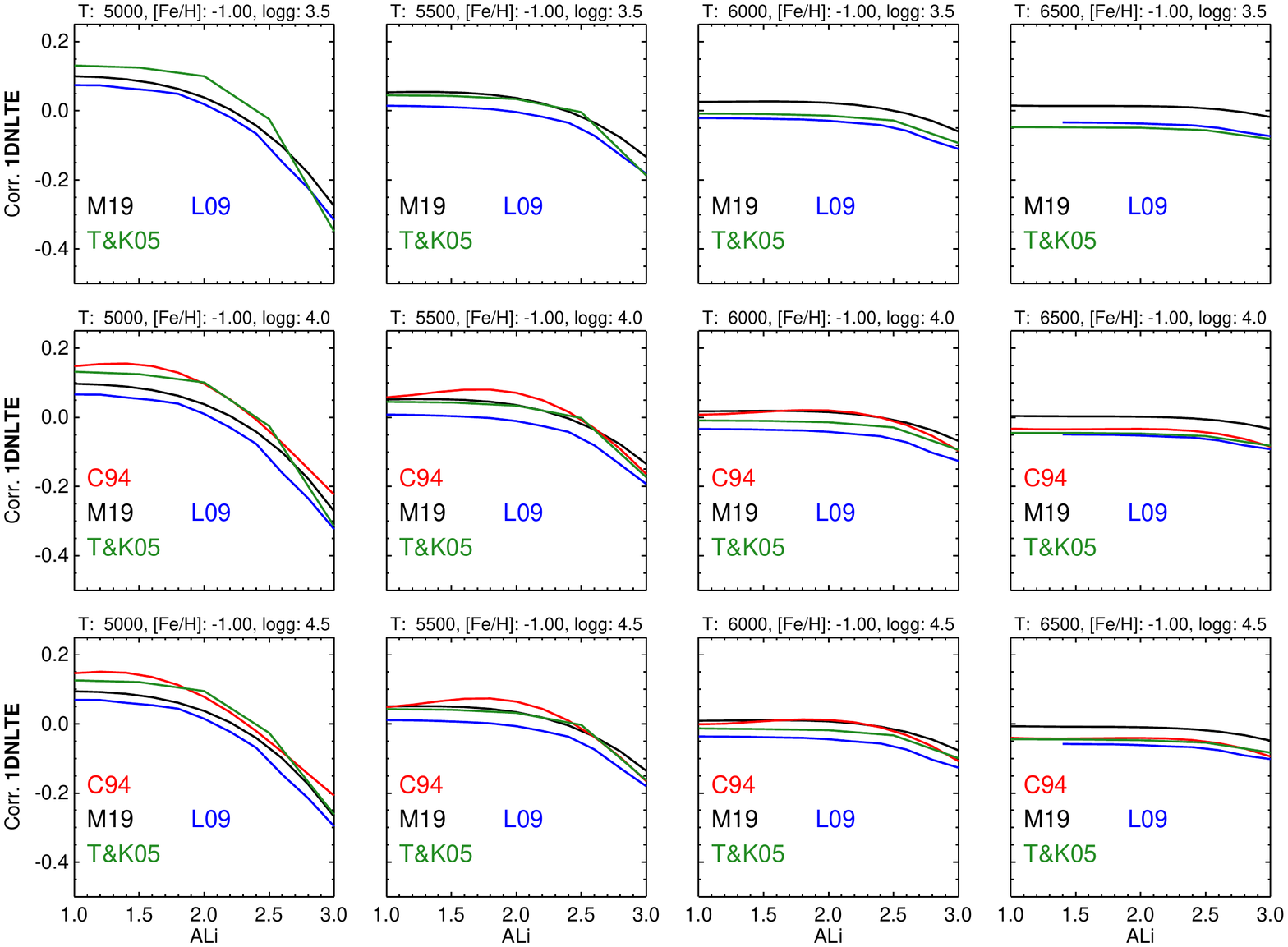}
	            \caption{Same as Figure \ref{authors_mm00} but for
                             $\feh=-1.0$. }
	\label{authors_mm10}
\end{figure*}

\begin{figure*}[hbtp]
    \sidecaption
    \includegraphics[width=12cm,trim=1cm 1cm 1cm 1cm,clip]{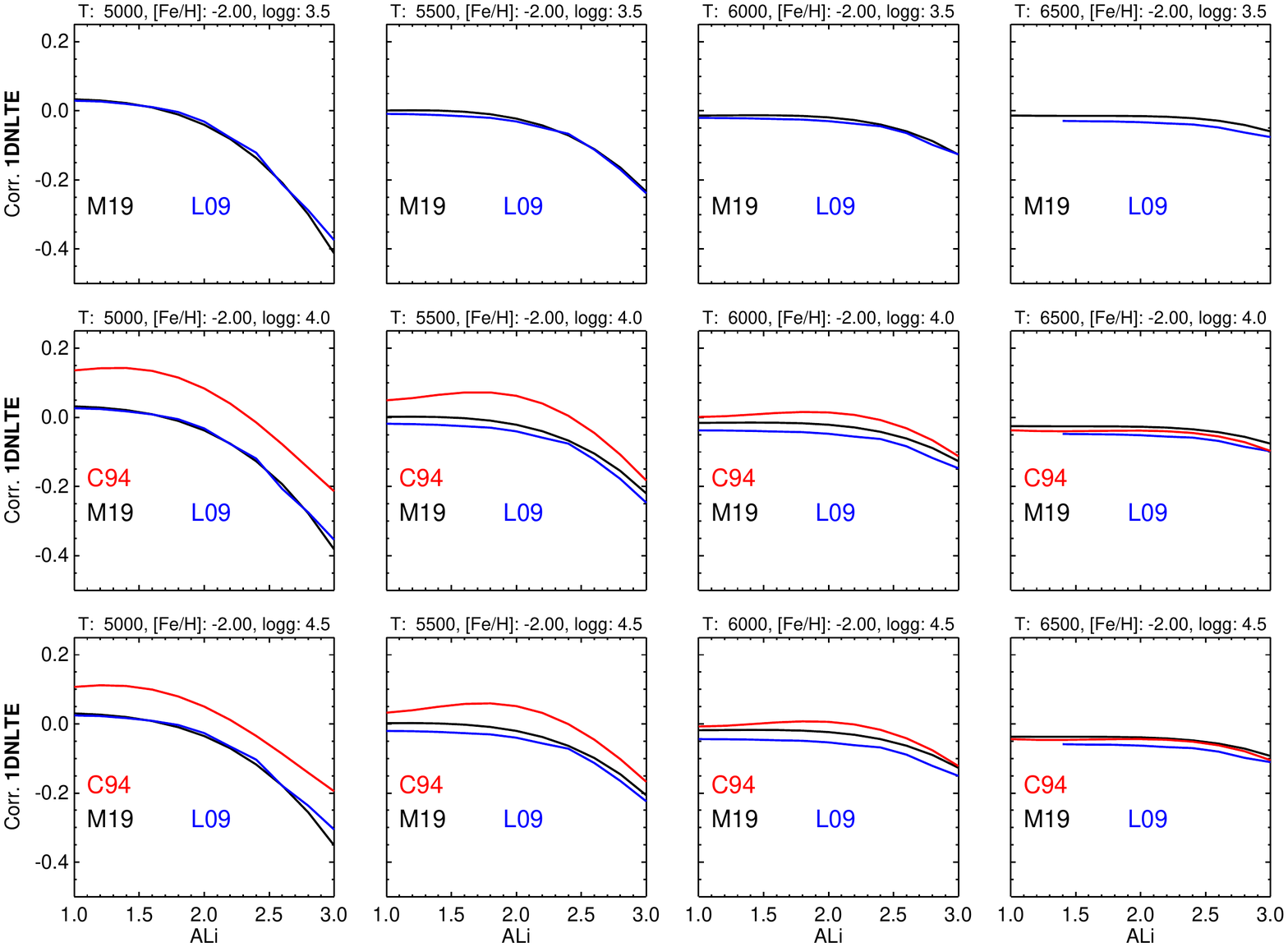}
	            \caption{Same as Figure \ref{authors_mm00} but for $\feh=-2.0$. }
	\label{authors_mm20}
\end{figure*}

\begin{figure*}[hbtp]
    \sidecaption
    \includegraphics[width=12cm,trim=1cm 1cm 1cm 1cm,clip]{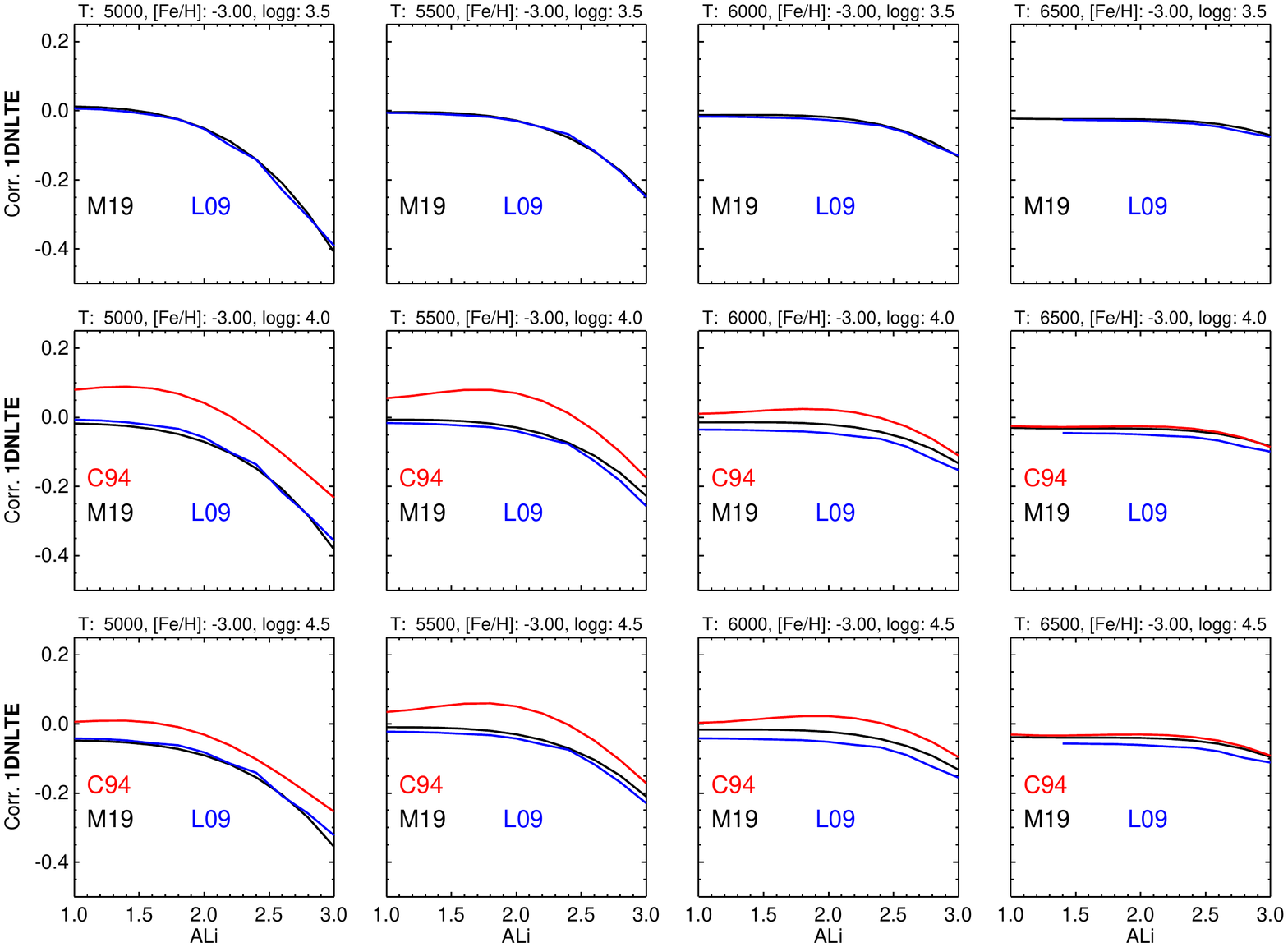}
	            \caption{Same as Figure \ref{authors_mm00} but for
                      $\feh=-3.0$. }
	\label{authors_mm30}
\end{figure*}

\subsection{1D\,NLTE lithium abundance corrections in the literature}
In the context of 1D\,NLTE lithium abundance corrections, there are a number of
other authors who contributed to this topic. We refer in this section to the
work by \cite{carlsson94}, \cite{pavlenko96}, \cite{takeda05} and more
recently \cite{lind09}. They all provided a series of 1D\,NLTE lithium
abundance corrections for the lithium resonance line at 670.8 nm, and in some
cases also for the subordinate line at 610.3 nm, for a diversity of stellar
parameters and lithium abundances. After briefly discussing their setups and
techniques, we show a comparison of the 1D\,NLTE lithium abundance corrections
between our results and those of the following authors.

\begin{itemize}[label={--}]
\item \cite{carlsson94}: their atmosphere grid contains one-dimensional LTE
  plane-parallel MARCS models \citep{gustafsson75} with \teff\ between 4500
  and 7500 K, $\mbox{\logg= 0.0, 2.0, 4.0}$ and 4.5 and
  $\feh=-3.0, -2.0, -1.0$, and $0.0$. The NLTE departure coefficients are
  computed adopting a Li model atom with 21 levels and 20 bound-bound
  transitions, allowing them to compute 1D\,NLTE lithium abundance corrections
  for both the resonance and the subordinate lithium lines. For the latter,
  however, they do not provide the values of the corrections since they are
  smaller and show significantly less variation across the grid than for
  the resonance line.  The 1D\,NLTE corrections for the \ion{Li}{i}
  $\lambda$670.8 nm line are given, for all the models of their grid, through
  a set of coefficients for a fourth-order polynomial fit to the values of
  their corrections as a function of $A$(Li). The assumed microturbulence
  is not specified.

\item \cite{pavlenko96}: the authors compute 1D\,NLTE and 1D\,LTE
  synthetic lithium line profiles and CoGs for three lithium lines: the
  resonance doublet ($\lambda$670.8 nm) and two subordinate lines
  ($\lambda$610.3 nm and $\lambda$812.6 nm). They provide 1D\,NLTE and 1D\,LTE
  equivalent widths in tabular form for a grid of \cite{kurucz1993a} ATLAS9
  model atmospheres for solar metallicity G-M dwarfs and subgiant stars with
  $3500 \leq T_{\rm eff} \leq 6000\, \mathrm{K}$ and $3.0 \leq \log\,g \leq
  4.5$ and for wide range of lithium abundances $A$(Li).
  Presumably, a fixed microturbulence parameter \micro=2.0\,\kms
  is adopted (details lacking). For the coolest models of their grid ($3500
  \leq T_{\rm eff} \leq 4000\, \mathrm{K}$), they find a very weak dependence
  of NLTE line profiles (and equivalent widths) on the surface gravity. In
  LTE, on the other hand, this dependence is quite strong, emphasizing the
  importance of using NLTE results in order to minimize the unrealistic
  effects of errors in \logg\, on lithium abundance determinations. For
  comparing these results to our corrections, we converted the LTE and NLTE
  equivalent widths of \cite{pavlenko96} to 1D\,NLTE abundance corrections (as
  a function of the 1D\,LTE abundance) by interpolation in their CoGs.

\item \cite{takeda05}: they employ an extensive grid of 100
  \cite{kurucz1993a} ATLAS9 model atmospheres to determine 1D\,NLTE lithium
  abundances from the lithium resonance doublet line for a sample
  of 160 F-K dwarfs and subgiant stars of the Galactic disc. The range of
  stellar parameters considered is quite large, covering
  $5000 \leq T_{\rm eff} \leq 7000\,\mathrm{K}$, $3.0 \leq \log\,g \leq 5.0$,
  and $-1.0 \leq \feh\leq +0.4$. A microturbulence parameter of
  \micro\,=\,$1.5$\,\kms\, has been assumed across the entire grid of models.
  They provide grids of theoretical equivalent widths and the corresponding
  NLTE corrections for $A$(Li)\,=\,$0.5, 1.0, 1.5, 2.0, 2.5, 3.0$ and $3.5$.

\item \cite{lind09}: this work represents the most comprehensive
  1D\,NLTE calculations of lithium abundance and corrections in late-type
  stars.  By using a 1D MARCS \citep{gustafsson08} model atmospheres grid
  spanning $\teff=[4000\ldots 8000]\, \mathrm{K}$, $\logg=[1.0\ldots 5.0]$
  and $\feh= [0.0\ldots -3.0]$, they computed synthetic line profiles for
  the lithium lines at $\lambda\,670.8$\,nm and $\lambda\,610.3$\,nm for
  $A\rm (Li)=[-0.30\ldots +4.20]$.  For models with $\logg\geq 3.0$ the
  spectra were computed adopting a
  microturbulence parameter \micro\,=\,$1.0$ and $2.0$\,\kms.  For models
  with $\logg\leq 3.0$ they adopted \micro\,=\,$2.0$ and $5.0$\,\kms. The
  LTE and NLTE equivalent widths were obtained by numerical integration
  over the line profiles, from which the two curves-of-growth can be
  constructed. The NLTE abundance corrections for each abundance were
  defined as the difference between the NLTE and the LTE lithium
  abundance that corresponds to the same equivalent width. This corresponds
  to the standard definition of NLTE abundance corrections
  and is consistent with the definition adopted in the present work
  (cf.\ Sect.\,\ref{sec:defcorr1}, \ref{sec:defcorr2}). 

  The model atom used for the calculation of the NLTE departure coefficients
  includes the same energy levels for neutral lithium as \cite{carlsson94},
  accounting for the hyperfine structure of \liseven\ (six
  components for the $\lambda$670.8\,nm line, three for the
  $\lambda$610.3\,nm line).
  The corrections are available either by means of tables or via web
  interface\footnote{\url{http://inspect.coolstars19.com}}
  that, similarly to our fitting function \texttt{FFI$_{\mathrm{1DNLTE}}$},
  allows a quick computation of the corrections as a function of the chosen
  input stellar parameters (\teff, \logg, \feh), lithium abundance $A$(Li)
  and, in their case, microturbulence parameter \micro.
\end{itemize}

In Figures \ref{authors_mm00} -- \ref{authors_mm30}, we show the comparison
between our results for the 1D\,NLTE $A$(Li) corrections (by applying
\texttt{FFI$_{\mathrm{1DNLTE}}$}) and those of the other authors described
previously, for metallicities $\feh= 0.0, -0.5, -1.0, -2.0$ and $-3.0$. Each
Figure consists in $4\times3$ panels where different effective temperatures and
gravities are shown in different rows and columns, respectively. Our results
are represented by black lines, whereas the colored lines show the corrections
according to the other authors, as explained by the legend of each panel.
The full comparison across our grid of stellar parameters could only be done
with \cite{lind09}. The other authors cover limited or different ranges of
metallicity, temperatures, and gravities and for this reason are missing in
some of the comparison plots.

Our 1D\,NLTE corrections are based on 1D\,\lhdm\ spectra computed
with the microturbulence parameter \micro\  as listed in Table
\ref{appendixA}. Since the microturbulence is an input parameter
of the IDL routine provided by \citet{lind09}%
\footnote{\url{http://www.mpia.de/homes/klind/index/INSPECT.html}},
we can ensure that the comparison if fully consistent regarding the adopted
value of \micro, which is especially relevant for high lithium abundances
where the lines are strong and potentially affected by small-scale velocity
fields in the stellar photosphere. For the comparison with the other
authors, full consistency regarding the choice of \micro\ is not enforced.

Figure \ref{authors_mm00} shows the 1D\,NLTE corrections for solar
metallicity. This plot provides a good overview of the general trend of the
corrections, since such trends are qualitatively maintained also for the other
\feh.  For the coolest models ($\teff=5000$ and 5500 K with $\logg$\,=\,$3.5$
and $4.0$) the curves drop steeply when the lithium resonance line saturates
at large $A$(Li) values.  On the other hand, for weaker lines (low $A$(Li))
the corrections become almost independent on the lithium abundance and assume
a constant value.  For $\teff=5000$ K, our fitting function yields a positive
1D\,NLTE correction of $+0.18$\,dex (black line in the upper left plot of
Figure \ref{authors_mm00}).  The location of this knee is moving towards
larger lithium abundance as the effective temperature increases, together with
a flattening of the overall trend. At high metallicities ($\feh=0.0, -0.5$),
the corrections are seen to be almost independent of \logg.

The global trend of our 1D\,NLTE lithium abundance corrections is found to be
in good agreement with the other works for every metallicity. Although some
scatter can be appreciated especially at $\teff=5000$ K, our results fit well
within the other curves, indicating general consistency. A systematic shift of
\textasciitilde 0.05 dex between our results and the others samples of
corrections can be noticed from $\teff=6000$\,K onwards and for metallicity
$0.0$ (Fig.\,\ref{authors_mm00}), $-0.5$ (Fig.\,\ref{authors_mm05}) and $-1.0$
(Fig.\,\ref{authors_mm10}).  The origin of this offset could be traced back to
the fact that different authors choose different model atmospheres.  A
different choice of the model atmospheres can indeed significantly affect the
derived abundance corrections. For example, \cite{pavlenko96} and
\cite{takeda05} employed the same ATLAS9 models, and their results correspond
quite well with each other (Fig.\,\ref{authors_mm00}, orange and green lines).
If we compare their results with \cite{lind09} who adopt MARCS model
atmospheres (blue line in Fig.\,\ref{authors_mm00}), we can observe some
offset in the corrections.  Additionally, even different versions of the same
type of models can lead to significant differences in the resulting abundance
corrections. This is the case for \cite{carlsson94} (red line in
Fig.\,\ref{authors_mm00}) and \cite{lind09} who both use MARCS models in which
the newer version employed by latter authors includes more line opacity.  As
explained by \cite{lind09}, this causes a steeper temperature gradient in the
upper part of the photosphere which, in turn, leads to a higher
over-ionization. For this reason, the use of newer MARCS model atmospheres
introduces a systematic shift in the overall trend of the abundance
corrections (towards more positive, less negative corrections with respect to
the results obtained with the old MARCS models), the size of which depends,
however, on the stellar parameters.

\subsection{Impact of different model atmospheres}
The effect of using MARCS models instead of 1D\,\lhdm\ models has been
investigated by \cite{harutyunyan18}.  They adopted the same 1D\,\lhdm\ model
atmospheres as in the present work and experienced a similar systematic offset
in their NLTE corrections when comparing their results with \cite{lind09}.

\begin{figure}[h]
    \centering
    \includegraphics[width=\linewidth,trim=2cm 2cm 2cm 2cm,clip]{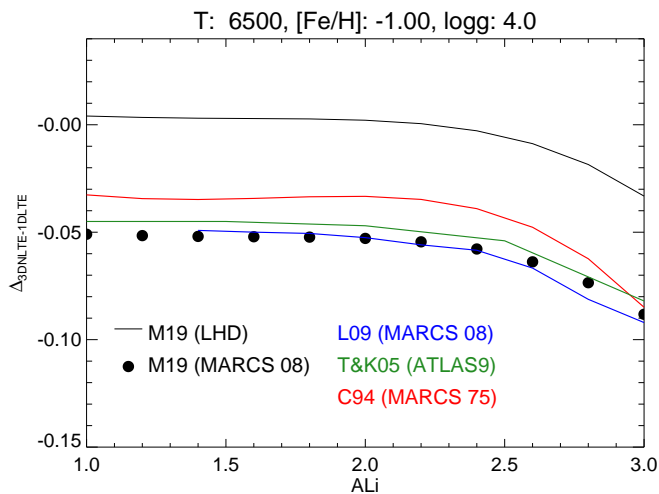}
    \caption[Effect of different model atmospheres on 1D\,NLTE $A$(Li)
      corrections]{Comparison between our 1D\,NLTE $A$(Li) corrections with
      the other authors for the model with \mbox{$\teff=6500$ K},
      \mbox{$\logg=4.0$} and \mbox{$\feh=-1.0$}. Our original corrections
      (M19, black solid line) have been shifted down (M19, black dots) by
      0.055 as found by \cite{harutyunyan18} to account for the difference
      between \lhdm\ and MARCS model atmospheres. The version of the models
      adopted by the different authors are indicated in the legend.}
    \label{marcs}
\end{figure}

To understand the reason for this offset, they recomputed the 1D corrections
by using one of the MARCS models of Lind et al. \citep{gustafsson08} with
parameters $\teff$\,=\,$6500$\,K, $\logg$\,=\,$4.0$, $\feh$\,=\,$-1.0$.
Assuming a lithium abundance of $A$(Li)\,=\,$2.0$, they found that the
resulting correction was $0.055$\,dex lower with respect to the value obtained
by means of the corresponding 1D\,\lhdm \ model. Applying the same MARCS--LHD
offset of $-0.055$\,dex over the whole range of $A$(Li) between $1.0$ and $3.0$,
their modified corrections were found to be in almost perfect agreement with
the work by Lind et al. (see Fig.\,\ref{marcs}).

The result of this experiment suggests that the main responsible for the
differences between our 1D\,NLTE abundance corrections and those of the
other authors is the different choice of the input model
atmospheres. In addition, other factors such as a different lithium model
atom and the treatment of the relevant physical processes in the calculation
of the NLTE departure coefficients (e.g., collisions with neutral hydrogen,
charge transfer reactions) may also play a role.

\section{Applications and discussion}
\label{sec:pplicability}
\subsection{NLTE corrections along Spite plateau}

Applying our fitting functions \texttt{FFI$_{\rm 3DNLTE}$} and
\texttt{FFI$_{\rm 1DNLTE}$}, we can study the influence of the
different line formation models on the distribution of stars
along the Spite plateau. For this purpose, we define a sample Spite
plateau stars with stellar parameters in the range
$5500$\,K\,$\le$\,\teff\,$\le$\,$6500$\,K ($\Delta=20$\,K),
$3.5$\,$\le$\,\logg\,$\le$\,$4.5$ ($\Delta=0.5$),
$-3.0$\,$\le$\,\feh\,$\le$\,$-1.5$  ($\Delta=0.025$), distributed
on a 3D rectangular grid. We assume that each of the $9333$ stars of the
sample has the same lithium abundance of $A$(Li)\,=\,$2.2$, such that, by
construction, they lie on a perfectly flat, infinitely thin Spite plateau
when analyzed in 3D\,NLTE.

\begin{figure}[htb]
    \centering
    \mbox{\includegraphics[width=\linewidth,trim=16 0 0 28,clip]{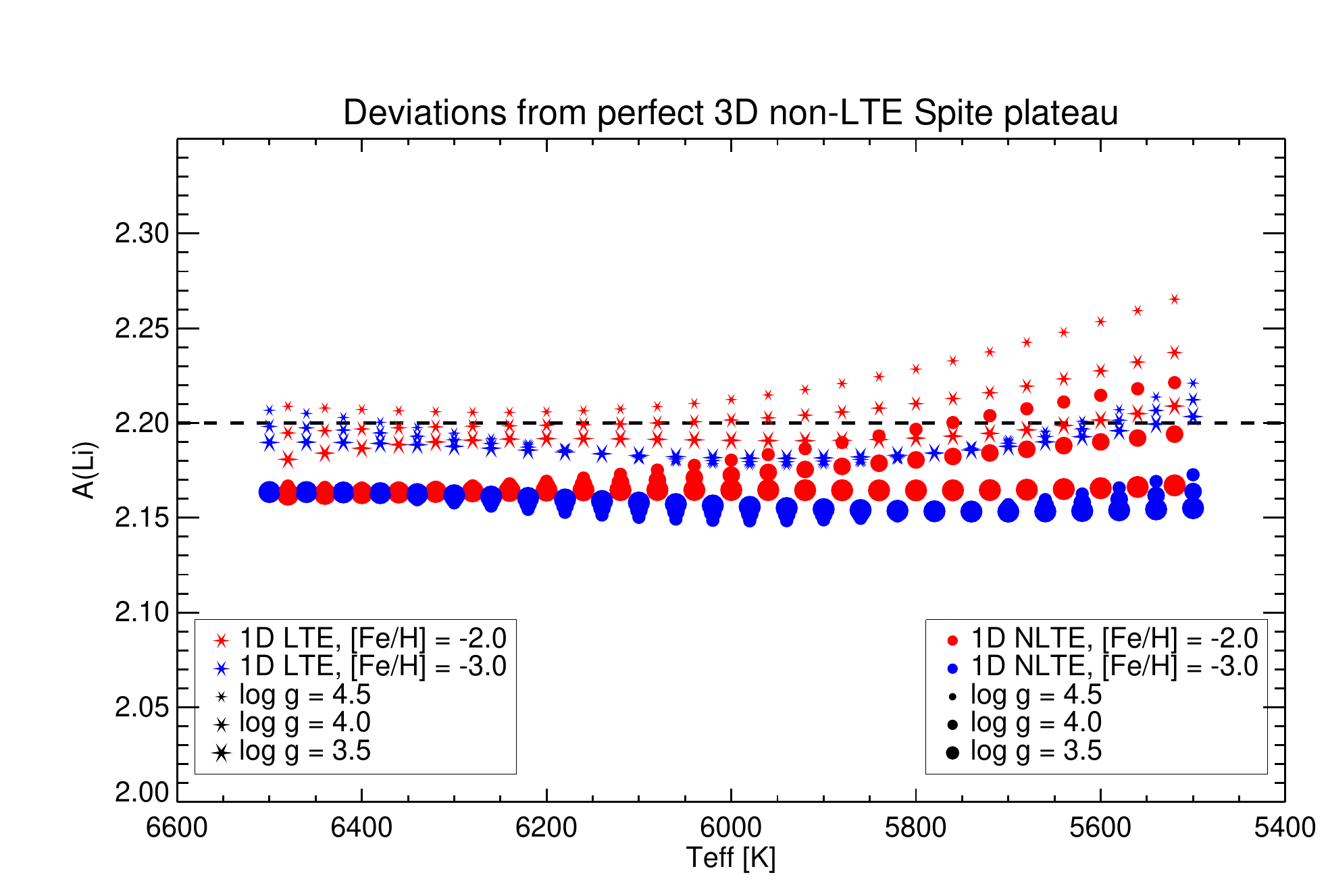}}
    \mbox{\includegraphics[width=\linewidth,trim=16 0 0 28,clip]{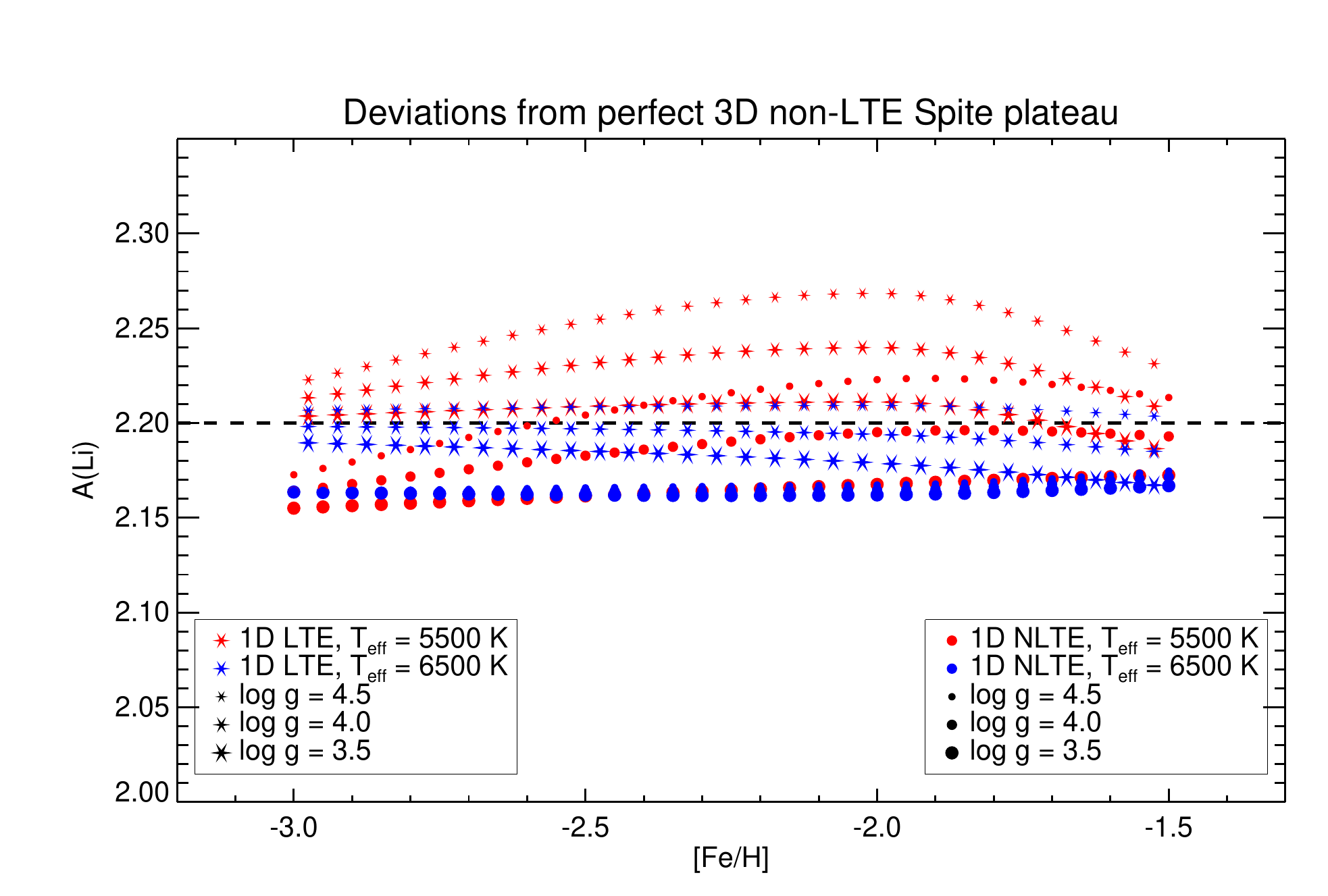}}
    \caption{Lithium abundance distribution for a sample of stars with stellar
      parameters in the range $5500$\,K\,$\le$\,\teff\,$\le$\,$6500$\,K,
      $3.5$\,$\le$\,logg\,$\le$\,$4.5$, $-3.0$\,$\le$\,\feh\,$\le$\,$-1.5$,
      plotted as a function of effective temperature (top) and metallicity
      (bottom). By construction, all stars lie on a perfectly flat, infinitely
      thin Spite plateau at $A$(Li)=$2.2$ when analyzed with 3D\,NLTE model
      spectra (dashed line). The corresponding 1D\,LTE (stars) and 1D\,NLTE
      (dots) abundances are derived by applying fitting functions
      \texttt{FFI$_{\rm 3DNLTE}$} and \texttt{FFI$_{\rm 1DNLTE}$},
      respectively. See legend for further details.}
    \label{spite_plateau}
\end{figure}

When instead analyzed in 1D\,LTE, the stars
are distributed in a lithium abundance range $2.167 \le A$(Li)\,$\le 2.268$ 
(see Fig.\,\ref{spite_plateau}, star symbols). The largest negative deviation
from the plateau of $-0.033$\, dex is found for stars with \teff\,=\,$6500$\,K,
\logg\,=\,$3.5$, \feh\,=\,$-1.5$, while the most positive deviation of
$+0.068$\,dex is found for stars with \teff\,=\,$5500$\,K, \logg\,=\,$4.5$,
\feh\,=\,$-2.0$. The mean LTE Li abundance of the sample is unchanged, 
$A$(Li)\,=\,$2.2006$, the standard deviation is as small as
$\sigma=0.0162$\,dex. On the other hand, the analysis of the sample
with 1D\,NLTE line formation is characterized by a systematic shift of
the Li abundance by $-0.03$\,dex to $A$(Li)\,=\,$2.17$ (see distribution
of filled dots in Fig.\,\ref{spite_plateau}). The total spread of the
abundances is $\sigma=0.0134$\,dex, similar to the 1D\,LTE case.

We conclude from this exercise that the error introduced by
analyzing the Spite plateau with 1D\,LTE models instead of employing
the full 3D\,NLTE machinery is small, at most $\pm 0.05$\,dex
in extreme cases. Our results also indicate that the 1D\,NLTE analysis
may introduce a small systematic offset of $\Delta A$(Li)\,$\approx -0.03$\,dex.
We see no evidence that 3D and / or NLTE effects might be responsible
for the ``meltdown'' of the Spite plateau at metallicities below
\feh\,$\approx$\,$-3.0$, although we cannot demonstrate this explicitly since,
strictly speaking, our fitting functions are not applicable to such low
metallicities.

The dependence of the corrections on \teff\ and \logg\ along an isochrone at
metallicity \feh\,$\approx$\,$-2.2$ is illustrated for three stars of globular
cluster \object{NGC\,6397} studied by \citet{nordlander12}, representing the
turn-off point (TOP), the subgiant branch (SGB), and the base of the red giant
branch (bRGB), respectively. The stellar parameters of these objects and their
Li abundance derived in 1D\,NLTE, taken from Table\,3 (temperature scale
``100s'') of \citet{nordlander12}, are reproduced in Cols.\,(1) -- (5) of
Table\,\ref{top-rgb}. With the help of our fitting functions \texttt{FFI$_{\rm
    1DNLTE}$} and \texttt{FFI$_{\rm 3DNLTE}$}, we can easily estimate the
1D\,LTE and 3D\,NLTE lithium abundance, respectively, for the given 1D\,NLTE
value. The results are collected in Cols.\,(6) -- (7) of
Table\,\ref{top-rgb}. We find that the 1D\,LTE abundance coincides with the
3D\,NLTE result for the TOP and (marginally) the SGB, but with the 1D\,NLTE
result for the bRGB.  Remarkably, the 3D\,NLTE $-$ 1D\,NLTE abundance
difference is $+0.035$\,dex for all three stars, such that all abundance
gradients with \teff\ remain perfectly unchanged. The small offset is not
affecting the conclusions of \citet{nordlander12} whatsoever.

\begin{table}[htbp]
\centering
\caption{Lithium abundances derived for three members of NGC\,6397 with
different line formation models.}
\label{top-rgb}
\begin{tabular}{lcccccc}
\hline\noalign{\smallskip}
Group &  \teff & \logg &  \feh & $A$(Li) & $A$(Li) & $A$(Li) \\
      &  [K]   &       &       & {\tiny 1D\,NLTE} & {\tiny 1D\,LTE} &
{\tiny 3D\,NLTE}\\
\hline\noalign{\smallskip}
TOP  &   $6354$ & $4.01$ & $-2.23$ &  $2.26$ &  $2.30$  &  $2.30$ \\
SGB  &   $5905$ & $3.71$ & $-2.19$ &  $2.39$ &  $2.44$  &  $2.43$ \\
bRGB &   $5556$ & $3.38$ & $-2.13$ &  $1.46$ &  $1.46$  &  $1.50$ \\
\hline
\end{tabular}
\end{table}

\subsection{Microturbulence dependence}
\label{sec:micro}
In this work, microturbulence is assumed to be a function of the fundamental
stellar parameters according to Eq.\,(\ref{microDutraFerreira}).  For this
reason, \micro\ is not considered as an independent parameter of our fitting
functions.

The microturbulence dependence given by Eq.\,(\ref{microDutraFerreira})
is based on \cobold~3D hydrodynamical simulations of the small-scale
velocity field in stellar atmospheres and gives a reasonable approximation to
empirical microturbulence trends measured in reals stars, as demonstrated
by \citet{dutraferreira16}.
It might nevertheless happen that the \micro\ parameter measured in individual
objects would exhibit major deviations from the adopted relation. In this
case, our fitting functions \texttt{FFII} ($A$(Li)\,$\rightarrow$\,EW) and
\texttt{FFIII} (EW\,$\rightarrow$\,$A$(Li)) would become unreliable for
stronger lines whose equivalent width is susceptible to the influence of
microturbulence. However, the abundance corrections are much less
sensitive to \micro\ \citep[see, e.g.,][]{lind09}, and the results obtained
from fitting function \texttt{FFI} should still be valid, as long as the
measured \micro\ is used for the determination of the 1D\,LTE lithium
abundance $A$(Li).

\subsection{Blend lines}
\label{sec:blends}
The fact that the fitting functions elaborated in this work are constructed
for the pure, unblended Li feature does not imply that they cannot be applied
to stars where some blend lines of other elements interfere with the lithium
doublet. In this case, the EW value of the lithium resonance doublet cannot be
measured directly from an observed spectrum but must be deduced by some
technique to disentangle the contributions of the interfering blend lines from
that of the pure lithium spectrum. This is not an uncommon procedure and there
are different methods to achieve such EW extraction.  For example, the solar
spectrum is characterized by a weak lithium feature immersed in several blend
lines. Nevertheless, \cite{brault75} succeeded in subtracting the spectrum of
the blend lines (mainly the laboratory spectrum of the CN molecular band) from
that of the Sun, thus isolating the lithium feature. Another technique to
derive the pure \ion{Li}{i} EW from a blended spectrum is to use the spectrum
of a reference star with similar stellar parameters, but being essentially
lithium-free, as a template which is subtracted from the original blended
spectrum to retrieve the de-blended pure lithium line profile
(e.g., \citealt{strassmeier15}, \citealt{soriano15}). A necessary condition for
this procedure to be valid is that all lines must be sufficiently weak, lying
on the linear part of the curve-of-growth.

In general, whenever a LTE lithium abundance can be estimated from an observed
spectrum, the 3D\,NLTE (or 1D\,NLTE) correction calculated with our fitting
functions can be applied.
    
\subsection{Lithium isotopic ratio}

The present investigation is based on synthetic spectra that were computed for
a fixed (meteoritic) lithium isotopic ratio of \iso=8.2\%, assuming that the
derived abundance corrections are insensitive to this choice. For
\feh\,=\,$0.0$ and $-2.0$, we have checked the validity of this assumption by
recomputing the full set of 1D\,\lhdm\ spectra with \iso=0.0, which is
probably a more realistic assumption since \lisix\ is expected to be
completely destroyed during the pre-main sequence evolution of the stars in
the parameter range under consideration.  We find that the corrections
computed with \iso=0.0 start to deviate significantly from the standard
corrections (\iso=8.2\%) only for the strongest lines (EW $\ga 250$\,m\AA) due
to saturation effects that lead to slightly more negative corrections. In the
most extreme case, the difference amounts to $-0.038$\,dex (\teff=$5000$\,K,
\logg=$3.5$, \feh=$-2.0$, $A$(Li)=$3.0$).

The fitting functions \texttt{FFI}, \texttt{FFII}, \texttt{FFIII} were
designed to fit the data computed with \iso=8.2\%. However, we found that they
can also be used to obtain the corrections valid for \iso=0.0 and the
corresponding conversion between EW and $A$(Li) by applying an empirical shift
of the input $A$(Li) (\texttt{FFI, FFII}) or the input EW (\texttt{FFIII}) as
follows:
\begin{eqnarray}
\label{iso_corr}
\Delta_0 &=&
\texttt{\rm FFI}\left(\teff,\logg,A\mathrm{(Li)+0.035}\right) \nonumber \\
\log_{10}(\mathrm{EW}_0) &=&
\texttt{FFII}\left(\teff,\logg,A\mathrm{(Li)+0.045}\right) - 0.045\nonumber \\
A_0\mathrm{(Li)} &=&
\texttt{FFIII}\left(\teff,\logg,1.109\times\mathrm{EW}\right)- 0.045\, ,
\end{eqnarray}
where quantities with subscript $0$ indicate the results adjusted for zero
\lisix\ contribution. With this modification, the fitting functions reproduce
the 1D data computed with \iso=0.0 with almost the same accuracy as the
original fitting functions reproduce the data computed with \iso=8.2\%. We
assume that the same manipulation works also for the intermediate
metallicities as well as for the 3D case.

\section{Conclusions}
\label{conclusions}

We extensively investigated the lithium line formation in solar-type FGK stars
of different metallicity from a purely theoretical perspective.  From the
CIFIST grid of 3D~\cobold\ model atmospheres, we utilized a large sub-set of
models and calculated 3D\,NLTE and 1D\,NLTE lithium abundance corrections from
synthetic curves-of-growth, spanning a wide range of stellar parameters and
lithium abundances.  This allowed us to study the magnitude of such 3D and
NLTE effects across the grid. The derived corrections can be applied to
improve the accuracy of 1D\,LTE abundance determinations for stellar
metallicities in the range $-3.0\le\feh\le0.0$.

We found these corrections to be rather small (but still significant) for
weaker lines (low $A\rm(Li)_{1DLTE}$ and higher \teff), where they are of the
order of $\pm0.05$ dex. Their magnitude increases substantially for
cooler temperatures and lower metallicities, especially for high lithium
abundances ($A\rm(Li)_{1DLTE}\geq 2.0$) where the corrections can reach the
$-0.30$ dex level. It is also worth noting that the abundance corrections
become systematically more negative towards the more metal-poor stars
(with a possible reversal of the trend near [Fe/H]\,=\,$-2.0$) for
which reliable lithium abundances are of particular interest.

Since 3D model atmosphere simulations and NLTE line formation calculations are
time consuming and require considerable computational resources, we first
condensed the results of our study into a set of tables that provide the
3D\,NLTE abundance corrections as a function of the stellar parameters and the
1D\,LTE lithium abundance $A^\ast$(Li) to which the correction shall be
applied. We then translated the tabulated corrections into analytical
functions that evaluate the 3D\,NLTE or 1D\,NLTE corrections at arbitrary
\teff, \logg, $A^\ast$(Li), separately for a fixed grid of metallicities \feh.
In a similar way, we also provide analytical fitting functions for directly
converting a given lithium abundance into an equivalent width or vice versa, a
given equivalent width into a lithium abundance (3D\,NLTE or 1D\,NLTE).  The
coefficients of these multi-dimensional polynomial functions are tabulated in
the appendix.

For further convenience, we have developed a Python
  script\footnote{\url{https://gitlab.aip.de/mst/Li_FF/}} that performs the
evaluation of all six fitting functions for given \teff, \logg, \feh,
and \iso, using the polynomial coefficients tabulated as a function of
\feh, applying the corrections for isotopic ratio if needed, and
subsequently interpolating to the desired \feh.

By means of the fitting functions developed in this work, the results of
complex 3D and NLTE calculations are readily accessible and quickly applicable
to large samples of stars.  In this way, our work contributes to improving the
accuracy of spectroscopic lithium abundance determinations across a wide range
of stellar metallicities, and may thus help to address the open questions
related to the lithium content in metal-poor and solar-like stellar
atmospheres.

Our results confirm that for metal-poor stars on the Spite plateau, 3D and
NLTE effects are by far too small to provide a solution for the cosmological
lithium problem, the persistent disagreement between the stellar plateau at
$A$(Li)\,$\approx 2.2$ and the primordial lithium abundance of
$A$(Li)\,$\approx 2.5$ inferred from WMAP and \textit{Planck} measurements of
the cosmic microwave background. At the same time, the metallicity dependence
of the corrections offers no explanation for the Li ``meltdown'' observed
below [Fe/H]\,$\approx$\,$-3.0$.

It has recently been claimed that stellar turbulent diffusion models
are able to explain the lithium depletion of the Spite plateau, although their
inclusion in stellar evolution models is not exempted from some
uncertainties. To better constrain the underlying physics, extremely precise
lithium abundance measurements along isochrones of metal-poor globular clusters
are necessary. Using our fitting functions for the lithium abundance
corrections, it is easy to show that differential 3D and NLTE effects between
stars at the turn-off point and the base of the red giant branch are less than
$0.04$\,dex.

\begin{acknowledgements}
We thank the State of Brandenburg (MWFK) and the German Federal Ministry 
of Education and Research (BMBF) for their continuous funding of basic 
solar-stellar research at the Leibniz Institute for Astrophysics Potsdam
(AIP). A.M. is grateful for financial support from the Leibniz Graduate
School for Quantitative Spectroscopy in Astrophysics, a joint project of
AIP and the Institute of Physics and  Astronomy  of  the  University  of
Potsdam. The critical comments of an anonymous referee helped to
improve the manuscript significantly.
\end{acknowledgements}

%
%
\bibliographystyle{aa}
\bibliography{references}


\begin{appendix} 
\section{Model atmospheres used}
\label{appendixA}  
In Table\,\ref{tab:Models} we compile the main properties of the full grid of
models used in this work. The name of the 3D model (col.\,$1$) is followed by
the number of opacity bins (col.\,$2$) and the number of representative
snapshots evaluated for the respective model (col.\,$3$). Columns ($4$)--($6$)
list the fundamental stellar parameters (\teff, \logg\ and \feh) which are
identical for 3D and 1D\,\lhdm\ models. The effective temperature reported in
col.\,($4$) is obtained by averaging the emergent radiative flux over the
sub-sample of the selected $N_{\rm snap}$ snapshots of each
\cobold\ simulation. Finally, col.\,($7$) shows the microturbulence parameter
\micro\ which is computed by means of Eq.\,(\ref{microDutraFerreira}) and is
adopted for the spectral line profile synthesis from the 1D\,\lhdm\ models.

\begin{table*}[htbp]
\caption{List of the main properties of the models that have been used for developing the fitting functions. 
The models that are missing in the CIFIST grid ($\teff=6500$ K, $\logg=3.5$)
are highlighted in bold face. For these parameters we still provide
values for the lithium abundance corrections by means of extrapolation from
the existing models.}
\centering
\begin{tabular}{lccccrc}
\hline\hline
Model  name     & Opacity & Number of  & \teff & \logg & [Fe/H]  & \micro\ \\
                & bins    & snapshots  & [K]   &  [dex]&  [dex]  &  [\kms]         \\\hline 
\texttt{d3t50g35mm00n01} & 5 & 19ss & 4920 & 3.5 &  0.0 & 0.90 \\
\texttt{d3t50g35mm05n01} & 6 & 20ss & 4960 & 3.5 & -0.5 & 0.92 \\ 
\texttt{d3t50g35mm10n01} & 6 & 20ss & 4930 & 3.5 & -1.0 & 0.90 \\ 
\texttt{d3t50g35mm20n01} & 6 & 20ss & 4980 & 3.5 & -2.0 & 0.93 \\ 
\texttt{d3t50g35mm30n01} & 6 & 20ss & 4980 & 3.5 & -3.0 & 0.93 \\ \hline
\texttt{d3t50g40mm00n01} & 6 & 20ss & 4950 & 4.0 &  0.0 & 0.82 \\ 
\texttt{d3t50g40mm05n01} & 6 & 20ss & 5000 & 4.0 & -0.5 & 0.84 \\ 
\texttt{d3t50g40mm10n01} & 6 & 20ss & 4990 & 4.0 & -1.0 & 0.84 \\ 
\texttt{d3t50g40mm20n01} & 6 & 20ss & 4960 & 4.0 & -2.0 & 0.83 \\ 
\texttt{d3t50g40mm30n02} & 6 & 20ss & 5160 & 4.0 & -3.0 & 0.89 \\ \hline
\texttt{d3t50g45mm00n04} & 5 & 20ss & 4980 & 4.5 &  0.0 & 0.82 \\ 
\texttt{d3t50g45mm05n01} & 6 & 20ss & 5040 & 4.5 & -0.5 & 0.83 \\ 
\texttt{d3t50g45mm10n03} & 6 & 19ss & 5060 & 4.5 & -1.0 & 0.84 \\ 
\texttt{d3t50g45mm20n03} & 6 & 19ss & 5010 & 4.5 & -2.0 & 0.83 \\ 
\texttt{d3t50g45mm30n03} & 6 & 20ss & 4990 & 4.5 & -3.0 & 0.82 \\ \hline
\texttt{d3t55g35mm00n01} & 5 & 18ss & 5430 & 3.5 &  0.0 & 1.13 \\ 
\texttt{d3t55g35mm05n01} & 6 & 20ss & 5470 & 3.5 & -0.5 & 1.15 \\ 
\texttt{d3t55g35mm10n01} & 6 & 19ss & 5480 & 3.5 & -1.0 & 1.16 \\ 
\texttt{d3t55g35mm20n01} & 6 & 20ss & 5500 & 3.5 & -2.0 & 1.17 \\ 
\texttt{d3t55g35mm30n01} & 6 & 20ss & 5540 & 3.5 & -3.0 & 1.18 \\ \hline
\texttt{d3t55g40mm00n01} & 5 & 20ss & 5480 & 4.0 &  0.0 & 0.99 \\ 
\texttt{d3t55g40mm05n01} & 6 & 19ss & 5530 & 4.0 & -0.5 & 1.01 \\ 
\texttt{d3t55g40mm10n01} & 6 & 20ss & 5530 & 4.0 & -1.0 & 1.01 \\ 
\texttt{d3t55g40mm20n01} & 6 & 20ss & 5470 & 4.0 & -2.0 & 0.99 \\ 
\texttt{d3t55g40mm30n01} & 6 & 20ss & 5480 & 4.0 & -3.0 & 0.99 \\ \hline
\texttt{d3t55g45mm00n01} & 5 & 20ss & 5490 & 4.5 &  0.0 & 0.91 \\ 
\texttt{d3t55g45mm05n01} & 6 & 20ss & 5460 & 4.5 & -0.5 & 0.91 \\ 
\texttt{d3t55g45mm10n01} & 6 & 20ss & 5470 & 4.5 & -1.0 & 0.91 \\ 
\texttt{d3t55g45mm20n01} & 6 & 20ss & 5480 & 4.5 & -2.0 & 0.91 \\ 
\texttt{d3t55g45mm30n01} & 6 & 20ss & 5490 & 4.5 & -3.0 & 0.91 \\ \hline
\texttt{d3t59g35mm00n01} & 5 & 20ss & 5880 & 3.5 &  0.0 & 1.34 \\ 
\texttt{d3t59g35mm05n02} & 6 & 20ss & 5760 & 3.5 & -0.5 & 1.29 \\ 
\texttt{d3t59g35mm10n01} & 6 & 20ss & 5890 & 3.5 & -1.0 & 1.34 \\ 
\texttt{d3t59g35mm20n01} & 6 & 20ss & 5860 & 3.5 & -2.0 & 1.33 \\ 
\texttt{d3t59g35mm30n01} & 6 & 20ss & 5870 & 3.5 & -3.0 & 1.34 \\ \hline
\texttt{d3t59g40mm00n01} & 5 & 18ss & 5930 & 4.0 &  0.0 & 1.13 \\ 
\texttt{d3t59g40mm05n01} & 6 & 20ss & 5920 & 4.0 & -0.5 & 1.13 \\ 
\texttt{d3t59g40mm10n02} & 6 & 20ss & 5850 & 4.0 & -1.0 & 1.11 \\ 
\texttt{d3t59g40mm20n02} & 6 & 20ss & 5860 & 4.0 & -2.0 & 1.11 \\ 
\texttt{d3t59g40mm30n02} & 6 & 20ss & 5850 & 4.0 & -3.0 & 1.11 \\ \hline
\texttt{d3t59g45mm00n01} & 5 & 19ss & 5870 & 4.5 &  0.0 & 0.98 \\ 
\texttt{d3t59g45mm05n01} & 6 & 20ss & 5900 & 4.5 & -0.5 & 0.98 \\ 
\texttt{d3t59g45mm10n01} & 6 & 08ss & 5920 & 4.5 & -1.0 & 0.99 \\ 
\texttt{d3t59g45mm20n01} & 6 & 18ss & 5920 & 4.5 & -2.0 & 0.99 \\ 
\texttt{d3t59g45mm30n01} & 6 & 19ss & 5920 & 4.5 & -3.0 & 0.99 \\ \hline
\texttt{d3t63g35mm00n01} & 5 & 21ss & 6130 & 3.5 &  0.0 & 1.45 \\ 
\texttt{d3t63g35mm05n02} & 6 & 20ss & 6150 & 3.5 & -0.5 & 1.46 \\ 
\texttt{d3t63g35mm10n01} & 6 & 20ss & 6210 & 3.5 & -1.0 & 1.49 \\ 
\texttt{d3t63g35mm20n01} & 6 & 20ss & 6290 & 3.5 & -2.0 & 1.53 \\ 
\texttt{d3t63g35mm30n01} & 6 & 20ss & 6310 & 3.5 & -3.0 & 1.54 \\ \hline
\texttt{d3t63g40mm00n01} & 5 & 20ss & 6230 & 4.0 &  0.0 & 1.23 \\ 
\texttt{d3t63g40mm05n01} & 6 & 20ss & 6250 & 4.0 & -0.5 & 1.24 \\ 
\texttt{d3t63g40mm10n01} & 6 & 20ss & 6260 & 4.0 & -1.0 & 1.24 \\ 
\texttt{d3t63g40mm20n01} & 6 & 16ss & 6280 & 4.0 & -2.0 & 1.24 \\ 
\texttt{d3t63g40mm30n01} & 6 & 20ss & 6270 & 4.0 & -3.0 & 1.24 \\ \hline

\end{tabular}
\label{tab:Models}
\end{table*}

\addtocounter{table}{-1}
\begin{table*}[htbp]
\caption{\ldots continued.}
\centering
\begin{tabular}{lccccrc}
\hline\hline
Model  name     & Opacity & Number of  & \teff & \logg & [Fe/H]  & \micro\ \\
                & bins    & snapshots  & [K]   &  [dex]&  [dex]  &  [\kms]         \\\hline 
\texttt{d3t63g45mm00n01} & 5 & 20ss & 6230 & 4.5 &  0.0 & 1.04 \\ 
\texttt{d3t63g45mm05n01} & 6 & 20ss & 6230 & 4.5 & -0.5 & 1.04 \\ 
\texttt{d3t63g45mm10n01} & 6 & 20ss & 6240 & 4.5 & -1.0 & 1.04 \\ 
\texttt{d3t63g45mm20n01} & 6 & 19ss & 6320 & 4.5 & -2.0 & 1.05 \\ 
\texttt{d3t63g45mm30n01} & 6 & 18ss & 6270 & 4.5 & -3.0 & 1.05 \\ \hline
\texttt{\bf d3t65g35mm00} & - & -    & 6500 & 3.5 &  0.0 & 1.62 \\ 
\texttt{\bf d3t65g35mm05} & - & -    & 6500 & 3.5 & -0.5 & 1.62 \\ 
\texttt{\bf d3t65g35mm10} & - & -    & 6500 & 3.5 & -1.0 & 1.62 \\ 
\texttt{\bf d3t65g35mm20} & - & -    & 6500 & 3.5 & -2.0 & 1.62 \\ 
\texttt{\bf d3t65g35mm30} & - & -    & 6500 & 3.5 & -3.0 & 1.62 \\ \hline
\texttt{d3t65g40mm00n01} & 5 & 20ss & 6490 & 4.0 &  0.0 & 1.31 \\ 
\texttt{d3t65g40mm05n01} & 6 & 20ss & 6520 & 4.0 & -0.5 & 1.32 \\ 
\texttt{d3t65g40mm10n01} & 6 & 20ss & 6500 & 4.0 & -1.0 & 1.31 \\ 
\texttt{d3t65g40mm20n01} & 6 & 19ss & 6530 & 4.0 & -2.0 & 1.32 \\ 
\texttt{d3t65g40mm30n01} & 6 & 20ss & 6410 & 4.0 & -3.0 & 1.29 \\ \hline 
\texttt{d3t65g45mm00n01} & 5 & 20ss & 6460 & 4.5 &  0.0 & 1.08 \\ 
\texttt{d3t65g45mm05n03} & 6 & 20ss & 6490 & 4.5 & -0.5 & 1.08 \\ 
\texttt{d3t65g45mm10n01} & 6 & 19ss & 6460 & 4.5 & -1.0 & 1.08 \\ 
\texttt{d3t65g45mm20n01} & 6 & 19ss & 6530 & 4.5 & -2.0 & 1.09 \\ 
\texttt{d3t65g45mm30n01} & 6 & 12ss & 6450 & 4.5 & -3.0 & 1.08 \\ \hline

\end{tabular}
\label{tab:Models2}
\end{table*}

\clearpage

\section{Coefficients}
\label{appendixB}

In this appendix we tabulate the numerical coefficients for the fitting
functions provided in this work.  In all of the following tables, the first
column contains the name of each numerical coefficient and in the next five
columns their numerical value for each one of the five metallicities of our
grid.

\subsection{Fitting functions \texttt{FFI}}
\begin{figure*}[h]
    \sidecaption
    \mbox{\includegraphics[trim=0 0 0 40,width=12cm]{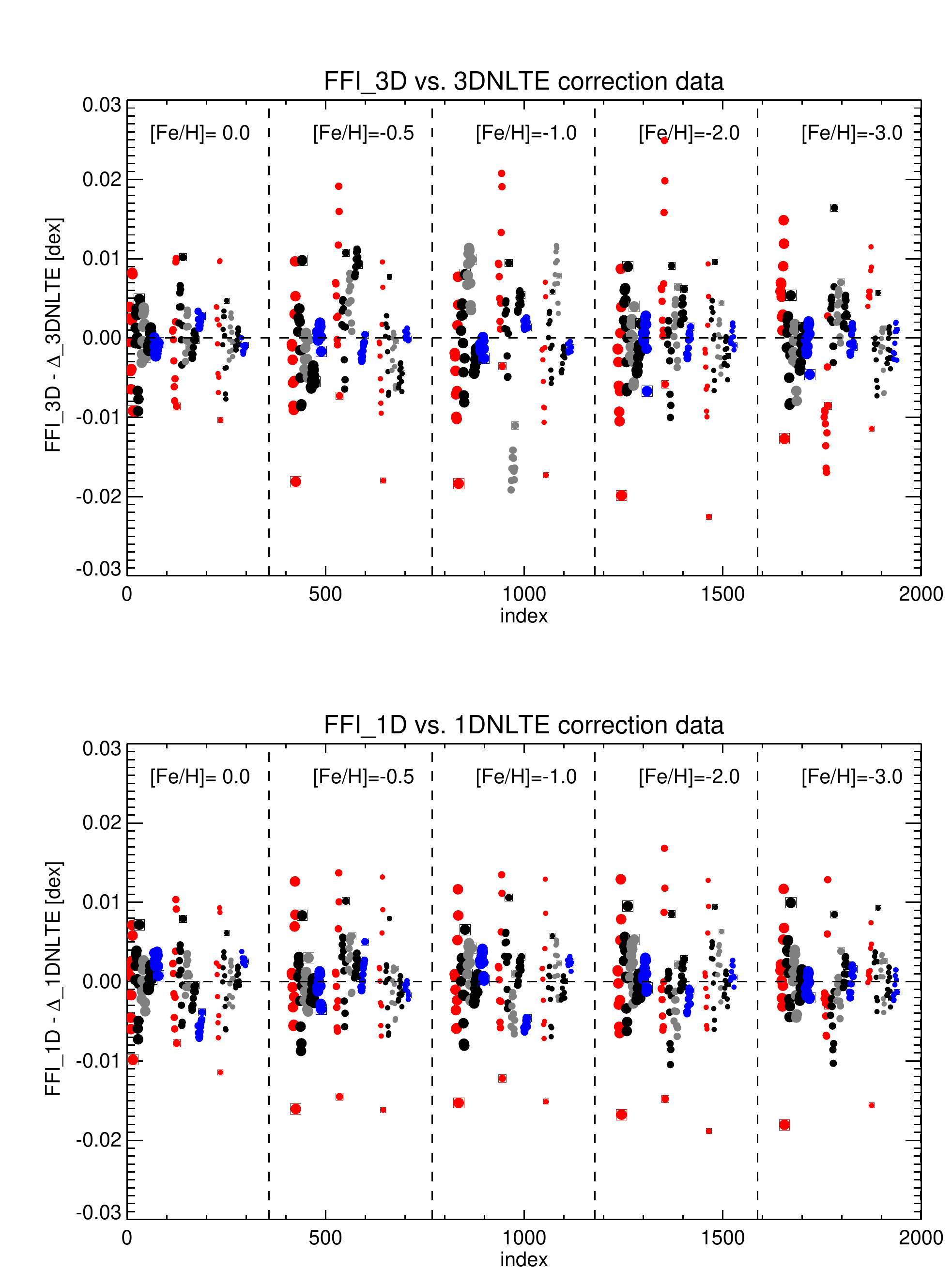}}
    \caption{Performance of the Li abundance correction fitting functions
      \texttt{FFI$_{\rm 3DNLTE}$} (top) and  \texttt{FFI$_{\rm 1DNLTE}$}
      (bottom). For each metallicity, a total of $165$ points indicates the
      difference (\texttt{FFI}\,$-$\,input correction) as a function of an
      index that runs over $A$(Li), \teff, and \logg. Large (left group),
      intermediate (middle group), and small dots (right group) correspond to
      \logg\,=\,$3.5$, $4.0$, and $4.5$, respectively. For each gravity,
      \teff\ increases from left to right ($5000$\,K: red, $5500$\,K: black,
      $5900$\,K: gray, $6300$\,K: black, $6500$\,K: blue).
      Each effective temperature is represented by $10$ dots indicating the
      results for \mbox{$1.0 \le A$(Li)\,$\le 2.8$} plus a square for
      $A$(Li)\,=\,$3.0$.}
    \label{ff1_errors}
\end{figure*}

\begin{figure*}[h]
    \sidecaption
    \mbox{\includegraphics[trim=0 0 0 40,width=12cm]{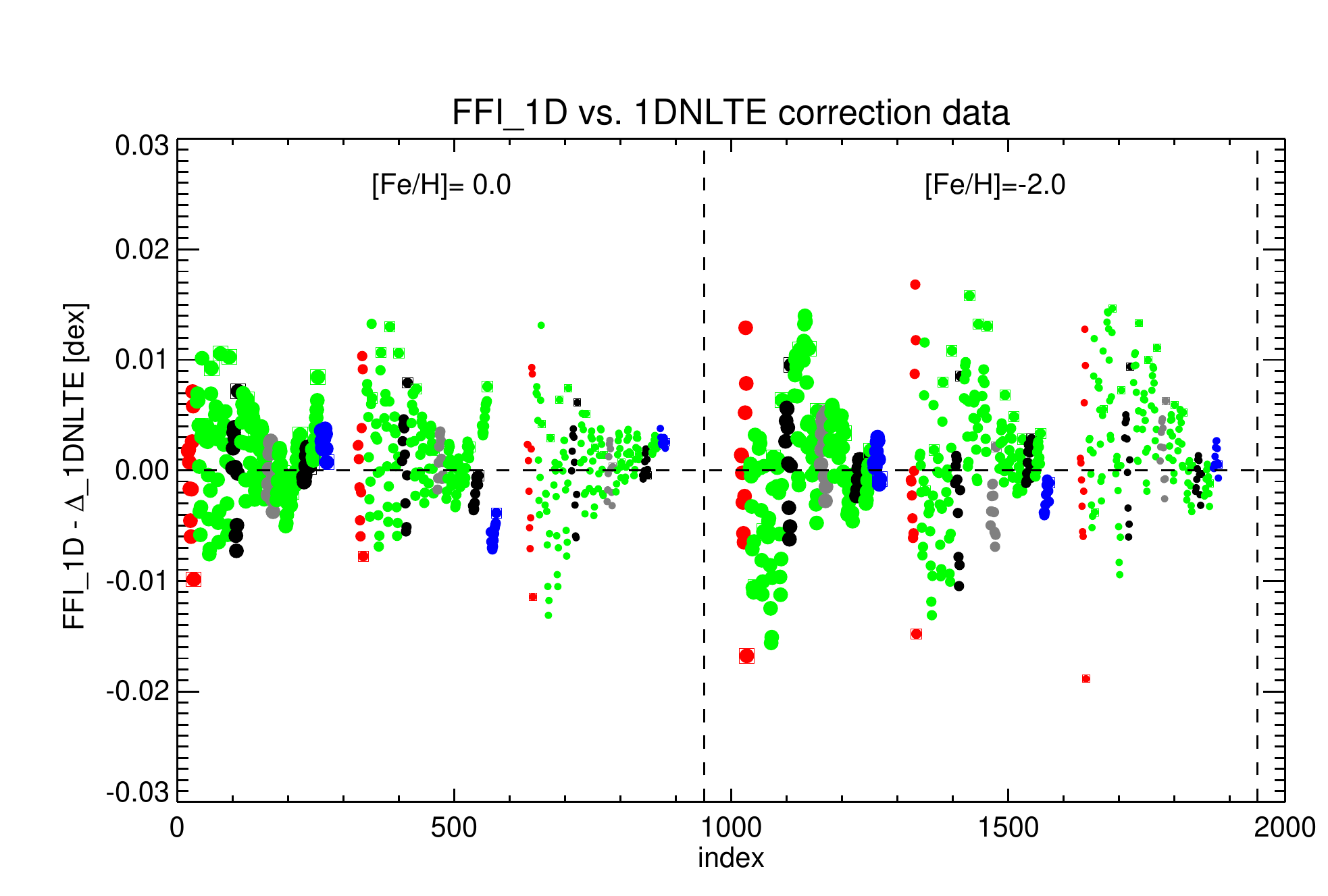}}
    \caption{Partial view of Fig.\,\ref{ff1_errors} (bottom), showing
       for metallicities [Fe/H]\,=\,$0.0$ and $-2.0$ the difference
       (\texttt{FFI}$_{\rm 1DNLTE}-$ input correction) as a function of
       spectrum number, now for a refined temperature grid with
       $\Delta$\teff\,=\,$100$\,K. For each metallicity, large,
       intermediate, and small dots correspond to \logg\,=\,$3.5$, $4.0$,
       and $4.5$, respectively. For each gravity, \teff\ increases from left
       to right. Color coding for the temperatures of the original grid is
       the same as in Fig.\,\ref{ff1_errors}; in addition, green dots
       represent the results for the intermediate 1D\,\lhdm\ test models.
       As before, each effective temperature is represented by $10$ dots for
       \mbox{$1.0 \le A$(Li)\,$\le 2.8$} plus a square for $A$(Li)\,=\,$3.0$.}
    \label{ff1_errors_lhd}
\end{figure*}

Tables \ref{coeff_3Dcorr} and \ref{coeff_1Dcorr} list the coefficients
$\mathrm{c}_{ijk}$ defined by Eq.\,(\ref{eqcorr3DNLTE}) for
\texttt{FFI$_{\rm 3DNLTE}$} and \texttt{FFI$_{\rm 1DNLTE}$}, respectively.
Figure\,\ref{ff1_errors} gives an overview of the fitting function
errors at the nodes of the grid for all metallicities, while
Figure\,\ref{ff1_errors_lhd} illustrates the interpolation errors
evaluated on a refined temperature grid of 1D\,\lhdm\ models for
[Fe/H]\,=\,$0.0$ and $-2.0$.

\begin{table}[htbp]
\centering
\caption[Coefficients $c_{ijk}$ for \texttt{FFI$_{\rm 3DNLTE}$} (3D\,NLTE
abundance corrections)]{Numerical coefficients $c_{ijk}$ of the fitting
function \texttt{FFI$_{\rm 3DNLTE}$} approximating the  3D\,NLTE lithium
abundance corrections $\Delta_{\rm 3DNLTE-1DLTE}$ (Eq.\,\ref{eqcorr3DNLTE})
for the five metallicities of our grid.}
\label{coeff_3Dcorr}
\renewcommand{\arraystretch}{1.2}
\resizebox{\linewidth}{!}{
\begin{tabular}{crrrrr}
\hline\hline
\multirow{2}{*}{$\mathrm{c}_{ijk}$}& \multicolumn{5}{c}{\feh}\\\cline{2-6} 
 &  0.0 dex   & -0.5 dex   & -1.0 dex   & -2.0 dex   &  -3.0 dex   \\
\hline
c$_{000}$ &  0.1934700 &  0.1137574 &  0.1059171 &  0.0110396 &  0.0302943 \\ 
c$_{100}$ & -0.0239075 & -0.0257894 & -0.0307899 & -0.0160289 & -0.0730075 \\ 
c$_{001}$ & -0.1479080 & -0.1209572 & -0.1251443 &  0.0061589 & -0.0496778 \\ 
c$_{101}$ &  0.1187087 & -0.0034667 &  0.1605346 & -0.2109836 &  0.1059637 \\ 
c$_{002}$ &  0.1429609 &  0.2182036 &  0.1630807 &  0.1021026 &  0.1900663 \\ 
c$_{102}$ & -0.3658550 &  0.0199786 & -0.4815992 &  0.3694110 &  0.0988528 \\ 
c$_{003}$ & -0.1536380 & -0.2343304 & -0.1063167 & -0.1538708 & -0.2032447 \\ 
c$_{103}$ &  0.3933365 &  0.0043123 &  0.4460331 & -0.1835116 & -0.2011892 \\ 
c$_{004}$ &  0.0504229 &  0.0804113 &  0.0225933 &  0.0569504 &  0.0615852 \\ 
c$_{104}$ & -0.1317260 & -0.0116685 & -0.1317514 &  0.0189035 &  0.0706434 \\ 
c$_{010}$ & -0.0112007 & -0.0095449 & -0.0071110 & -0.0101823 & -0.0140992 \\ 
c$_{110}$ &  0.0014934 &  0.0041522 & -0.0008530 &  0.0101311 &  0.0110694 \\ 
c$_{011}$ & -0.0157936 & -0.0057599 & -0.0049469 &  0.0077487 &  0.0031318 \\ 
c$_{111}$ & -0.0035633 & -0.0132778 & -0.0108938 & -0.0188577 &  0.0021892 \\ 
c$_{012}$ & -0.0029972 & -0.0097395 & -0.0218368 & -0.0330288 & -0.0016796 \\ 
c$_{112}$ &  0.0015286 &  0.0149844 &  0.0249090 &  0.0265209 &  0.0200220 \\ 
c$_{013}$ &  0.0226405 &  0.0190282 &  0.0262789 &  0.0303408 &  0.0146414 \\ 
c$_{113}$ & -0.0007266 & -0.0054095 & -0.0108755 & -0.0155570 & -0.0209639 \\ 
c$_{020}$ &  0.0228360 &  0.0220577 &  0.0198897 &  0.0230061 &  0.0223846 \\ 
c$_{120}$ &  0.0001464 & -0.0048820 & -0.0002827 & -0.0148570 & -0.0242930 \\ 
c$_{021}$ & -0.0015479 & -0.0023546 &  0.0026504 & -0.0048869 & -0.0188727 \\ 
c$_{121}$ &  0.0028341 &  0.0042002 & -0.0048818 &  0.0090188 & -0.0036151 \\ 
c$_{022}$ & -0.0304533 & -0.0255947 & -0.0278502 & -0.0282937 & -0.0237323 \\ 
c$_{122}$ & -0.0001805 & -0.0008105 &  0.0014137 &  0.0056108 &  0.0164049 \\ 
c$_{030}$ & -0.0047771 & -0.0070773 & -0.0069892 & -0.0073567 & -0.0025587 \\ 
c$_{130}$ & -0.0015685 &  0.0022109 &  0.0023593 &  0.0061664 &  0.0125796 \\ 
c$_{031}$ &  0.0190796 &  0.0192168 &  0.0190010 &  0.0229878 &  0.0236960 \\ 
c$_{131}$ & -0.0001362 & -0.0008104 &  0.0003031 & -0.0062793 & -0.0082827 \\ 
c$_{040}$ & -0.0040627 & -0.0040545 & -0.0040723 & -0.0051027 & -0.0060987 \\ 
c$_{140}$&  0.0002129 & -0.0004152 & -0.0006280 & -0.0000427 & -0.0006542 \\ \hline 
\end{tabular}
}
\end{table}

\begin{table}[htbp]
\centering
\caption[Coefficients $c_{ijk}$ for \texttt{FFI$_{\rm 1DNLTE}$} (1D\,NLTE
abundance corrections)]{Numerical coefficients $c_{ijk}$ for the fitting
function \texttt{FFI$_{\rm 1DNLTE}$} approximating the 1D\,NLTE lithium
abundance corrections $\Delta_{\rm 1DNLTE-1DLTE}$ (Eq.\,\ref{eqcorr3DNLTE})
for the five metallicities of our grid.}
\label{coeff_1Dcorr}
\renewcommand{\arraystretch}{1.2}
\resizebox{\linewidth}{!}{
\begin{tabular}{crrrrr}
\hline\hline
\multirow{2}{*}{$\mathrm{c}_{ijk}$}& \multicolumn{5}{c}{\feh}\\\cline{2-6} 
 &  0.0 dex   & -0.5 dex   & -1.0 dex   & -2.0 dex   &  -3.0 dex   \\
\hline
  c$_{000}$ &  0.1791557 &  0.1050792 &  0.0990588 &  0.0328342 &  0.0124536 \\
  c$_{100}$ & -0.0073790 & -0.0084633 & -0.0070312 & -0.0032602 & -0.0587628 \\
  c$_{001}$ & -0.1406743 & -0.1530184 & -0.2250787 & -0.1734033 & -0.0807081 \\
  c$_{101}$ &  0.0561255 &  0.0583358 &  0.1517873 &  0.0513971 &  0.2173629 \\
  c$_{002}$ &  0.1231245 &  0.2275670 &  0.4254515 &  0.3390962 &  0.1571438 \\
  c$_{102}$ & -0.1741239 & -0.1280650 & -0.4558604 & -0.1418215 & -0.2973344 \\
  c$_{003}$ & -0.1497231 & -0.1886872 & -0.3898447 & -0.2990925 & -0.1397715 \\
  c$_{103}$ &  0.1935423 &  0.0873244 &  0.4176210 &  0.1312618 &  0.1654042 \\
  c$_{004}$ &  0.0536468 &  0.0499225 &  0.1203871 &  0.0905645 &  0.0401296 \\
  c$_{104}$ & -0.0675082 & -0.0183153 & -0.1236082 & -0.0432368 & -0.0335796 \\
  c$_{010}$ & -0.0111539 & -0.0073209 & -0.0051128 & -0.0041143 & -0.0080132 \\
  c$_{110}$ &  0.0046478 &  0.0038263 &  0.0008493 &  0.0058993 &  0.0047513 \\
  c$_{011}$ & -0.0042783 &  0.0023326 &  0.0007908 & -0.0005678 & -0.0023488 \\
  c$_{111}$ & -0.0146549 & -0.0232104 & -0.0205851 & -0.0043562 &  0.0108387 \\
  c$_{012}$ & -0.0085580 & -0.0180569 & -0.0272235 & -0.0223643 &  0.0063592 \\
  c$_{112}$ &  0.0069384 &  0.0264662 &  0.0329991 &  0.0043666 &  0.0036971 \\
  c$_{013}$ &  0.0196931 &  0.0185953 &  0.0255633 &  0.0231517 &  0.0073388 \\
  c$_{113}$ &  0.0002411 & -0.0085541 & -0.0130252 & -0.0057535 & -0.0107872 \\
  c$_{020}$ &  0.0216719 &  0.0188139 &  0.0160754 &  0.0149103 &  0.0149583 \\
  c$_{120}$ & -0.0036719 & -0.0032942 & -0.0003223 & -0.0091198 & -0.0155050 \\
  c$_{021}$ & -0.0064836 & -0.0034779 &  0.0022292 &  0.0004144 & -0.0149522 \\
  c$_{121}$ &  0.0096988 &  0.0076570 &  0.0005769 &  0.0071321 & -0.0049674 \\
  c$_{022}$ & -0.0249244 & -0.0211086 & -0.0245816 & -0.0255060 & -0.0194343 \\
  c$_{122}$ & -0.0038088 & -0.0031284 & -0.0005112 &  0.0051264 &  0.0128407 \\
  c$_{030}$ & -0.0052306 & -0.0068116 & -0.0056458 & -0.0046855 & -0.0006729 \\
  c$_{130}$ &  0.0006315 &  0.0013540 &  0.0010385 &  0.0028883 &  0.0091173 \\
  c$_{031}$ &  0.0188565 &  0.0183322 &  0.0181051 &  0.0205227 &  0.0213070 \\
  c$_{131}$ & -0.0010174 & -0.0012430 & -0.0004732 & -0.0056278 & -0.0069360 \\
  c$_{040}$ & -0.0040438 & -0.0040839 & -0.0043396 & -0.0054409 & -0.0063454 \\
  c$_{140}$ & -0.0000474 & -0.0001415 & -0.0001646 &  0.0006245 & -0.0001072 \\
  \hline 
\end{tabular}
}
\end{table}


\clearpage

\subsection{Fitting functions \texttt{FFII}}
\begin{figure*}[h]
    \sidecaption
    \mbox{\includegraphics[trim=0 0 0 40,width=12cm]{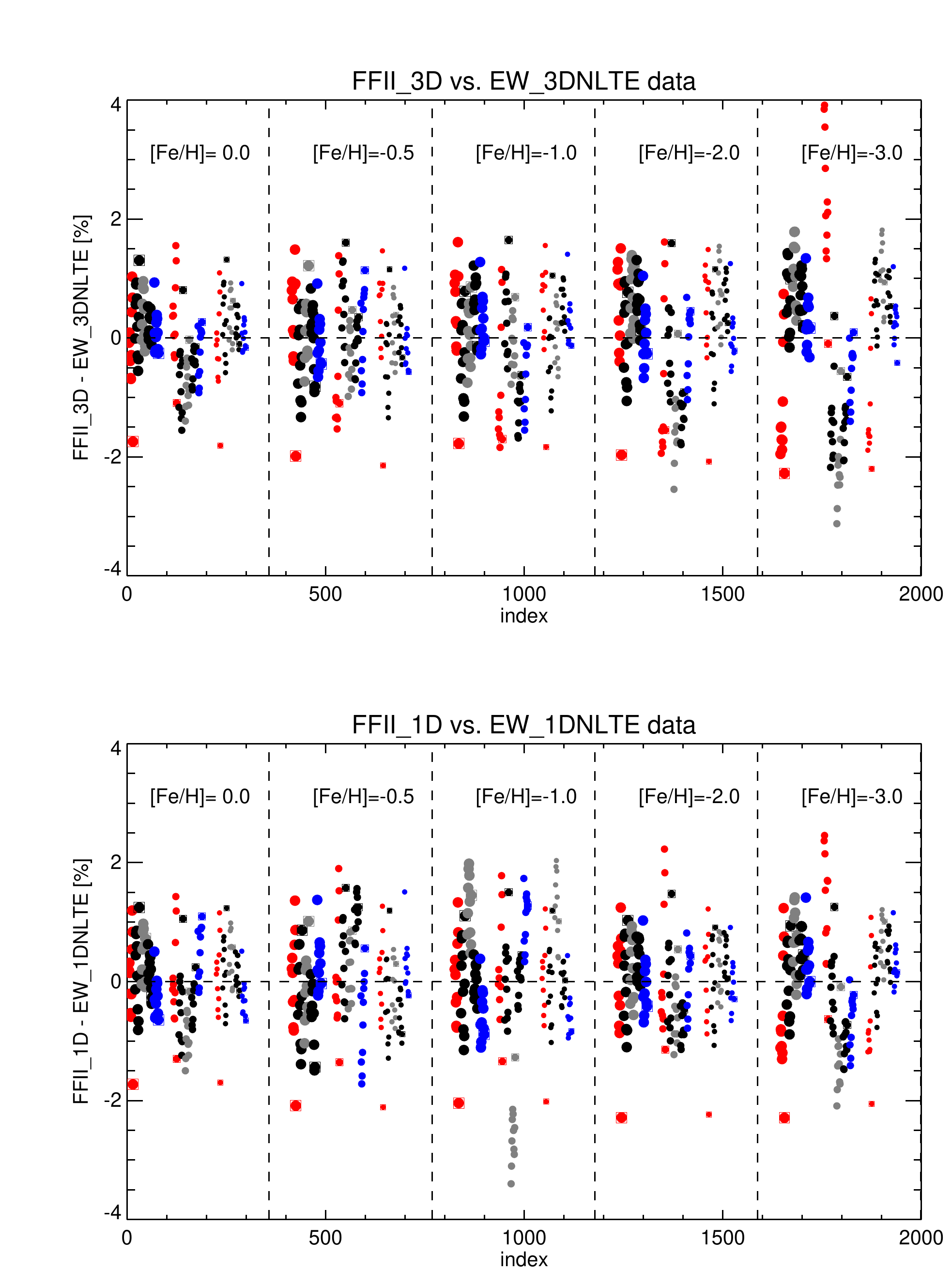}}
    \caption{Performance of the Li equivalent width fitting functions
      \texttt{FFII$_{\rm 3DNLTE}$} (top) and  \texttt{FFII$_{\rm 1DNLTE}$}
      (bottom). For each metallicity, a total of $165$ points indicates the
      relative difference (\texttt{FFII}\,$-$\,input EW) as a function of an
      index that runs over $A$(Li), \teff, and \logg. Large (left group),
      intermediate (middle group), and small dots (right group) correspond to
      \logg\,=\,$3.5$, $4.0$, and $4.5$, respectively. For each gravity,
      \teff\ increases from left to right ($5000$\,K: red,
      $5500$\,K: black, $5900$\,K: gray, $6300$\,K: black, $6500$\,K: blue).
      Each effective temperature is represented by $10$ dots indicating the
      results for \mbox{$1.0 \le A$(Li)\,$\le 2.8$} plus a square for
      $A$(Li)\,=\,$3.0$.}
    \label{ff2_errors}
\end{figure*}

\begin{figure*}[h]
    \sidecaption
    \mbox{\includegraphics[trim=0 0 0 40,width=12cm]{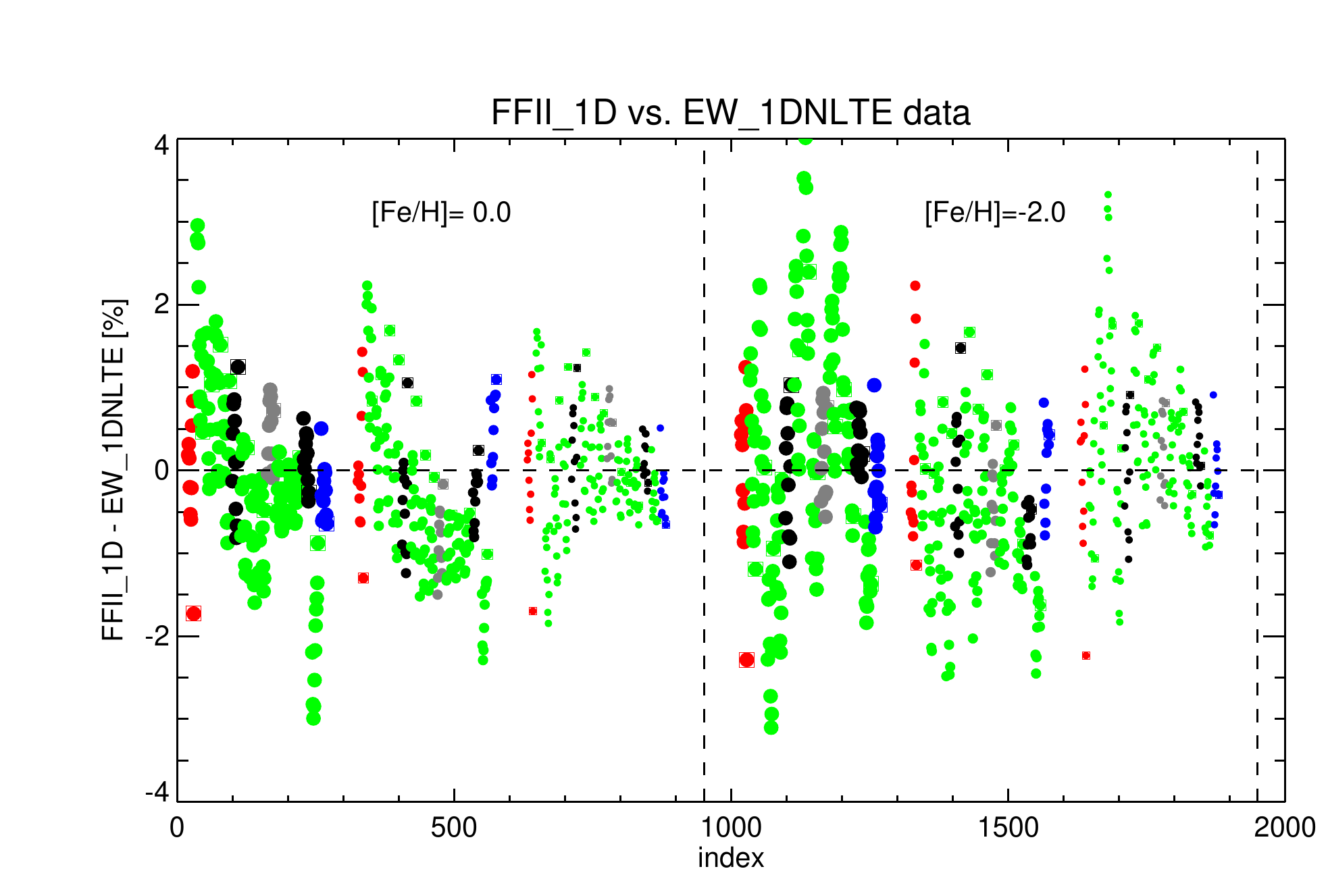}}
    \caption{Partial view of Fig.\,\ref{ff2_errors} (bottom), showing
       for metallicities [Fe/H]\,=\,$0.0$ and $-2.0$ the difference
       (\texttt{FFII}$_{\rm 1DNLTE}-$ input EW) as a function of
       spectrum number, now for a refined temperature grid with
       $\Delta$\teff\,=\,$100$\,K. For each metallicity, large,
       intermediate, and small dots correspond to \logg\,=\,$3.5$, $4.0$,
       and $4.5$, respectively. For each gravity, \teff\ increases from left
       to right. Color coding for the temperatures of the original grid is
       the same as in Fig.\,\ref{ff2_errors}; in addition, green dots
       represent the results for the intermediate 1D\,\lhdm\ test models.
       As before, each effective temperature is represented by $10$ dots for
       \mbox{$1.0 \le A$(Li)\,$\le 2.8$} plus a square for $A$(Li)\,=\,$3.0$.}
    \label{ff2_errors_lhd}
\end{figure*}

Tables \ref{FF_EW3D} and \ref{FF_EW1D} contain the coefficients
$\mathrm{w}_{ijk}$ defined by Eq.\,(\ref{eqFFII}) for the equivalent width
fitting functions \texttt{FFII}$_{\rm 3DNLTE}$ and \texttt{FFII}$_{\rm 1DNLTE}$,
respectively.
Figure\,\ref{ff2_errors} gives an overview of the fitting function
  errors at the nodes of the grid for all metallicities, while
  Figure\,\ref{ff2_errors_lhd} illustrates the interpolation errors
evaluated on a refined temperature grid of 1D\,\lhdm\ models for
[Fe/H]\,=\,$0.0$ and $-2.0$.

\begin{table}[htbp]
\centering
\caption[Coefficients $w_{ijk}$ for \texttt{FFII$_{A\rm (Li)-EW}$} (3D\,NLTE)]
        {Numerical  coefficients $w_{ijk}$ (Eq.\,\ref{eqFFII}) for the
         equivalent width fitting function \texttt{FFII}$_{\rm 3DNLTE}$,
         derived for the five metallicities of our grid.}
\label{FF_EW3D}
\renewcommand{\arraystretch}{1.2}
\resizebox{\linewidth}{!}{
\begin{tabular}{crrrrr}
\hline\hline
\multirow{2}{*}{$\mathrm{w}_{ijk}$}& \multicolumn{5}{c}{\feh}\\\cline{2-6} 
 &  0.0 dex   & -0.5 dex   & -1.0 dex   & -2.0 dex   &  -3.0 dex   \\
\hline
w$_{000}$ &  0.1475069 &  0.2245458 &  0.2239167 &  0.2849084 &  0.2068955 \\ 
w$_{100}$ &  0.0347089 &  0.0339855 &  0.0306456 &  0.0223937 &  0.0133552 \\ 
w$_{001}$ & -1.0208849 & -1.1469250 & -1.2259755 & -1.3272777 & -0.9932886 \\ 
w$_{101}$ & -0.0669732 &  0.1537401 &  0.2371681 &  0.3296509 &  0.1479368 \\ 
w$_{002}$ &  0.2054467 &  0.5791591 &  0.7219083 &  0.8228920 &  0.4085410 \\ 
w$_{102}$ &  0.2143639 & -0.2905753 & -0.4994409 & -0.6173923 & -0.3156809 \\ 
w$_{003}$ &  0.0014179 & -0.2833503 & -0.3976060 & -0.3689554 & -0.2145625 \\ 
w$_{103}$ & -0.2303063 &  0.1320859 &  0.3419425 &  0.3533569 &  0.2474548 \\ 
w$_{004}$ & -0.0188241 &  0.0516219 &  0.0878278 &  0.0580685 &  0.0553802 \\ 
w$_{104}$ &  0.0782668 & -0.0003866 & -0.0704236 & -0.0529642 & -0.0656191 \\ 
w$_{010}$ &  1.0254700 &  1.0536114 &  1.0482122 &  1.0525770 &  1.0448927 \\ 
w$_{110}$ & -0.0123804 & -0.0402674 & -0.0328853 & -0.0274988 & -0.0895899 \\ 
w$_{011}$ & -0.0370201 & -0.0563704 & -0.0354928 & -0.0487050 & -0.0955188 \\ 
w$_{111}$ & -0.0224963 & -0.0349607 & -0.0656007 & -0.0428780 &  0.0236516 \\ 
w$_{012}$ & -0.1005337 & -0.1450037 & -0.1672949 & -0.2049884 & -0.0942996 \\ 
w$_{112}$ &  0.0257704 &  0.0782432 &  0.0999685 &  0.0993408 &  0.0422533 \\ 
w$_{013}$ &  0.0385082 &  0.0430121 &  0.0515689 &  0.0606149 &  0.0203157 \\ 
w$_{113}$ & -0.0049530 & -0.0276824 & -0.0309441 & -0.0416862 & -0.0082573 \\ 
w$_{020}$ & -0.0012442 & -0.0111443 & -0.0123488 & -0.0087517 & -0.0016854 \\ 
w$_{120}$ &  0.0115008 &  0.0312578 &  0.0315381 &  0.0227707 &  0.0650281 \\ 
w$_{021}$ &  0.0727088 &  0.1046035 &  0.1053274 &  0.1329422 &  0.1065020 \\ 
w$_{121}$ &  0.0038124 & -0.0052565 & -0.0022758 & -0.0118978 & -0.0372886 \\ 
w$_{022}$ & -0.0223655 & -0.0130742 & -0.0143628 & -0.0106278 & -0.0077959 \\ 
w$_{122}$ & -0.0049537 & -0.0047638 & -0.0063133 &  0.0020711 & -0.0033688 \\ 
w$_{030}$ & -0.0159346 & -0.0212027 & -0.0199292 & -0.0273466 & -0.0233474 \\ 
w$_{130}$ & -0.0035786 & -0.0086870 & -0.0090606 & -0.0054171 & -0.0138632 \\ 
w$_{031}$ &  0.0101170 &  0.0031635 &  0.0027428 & -0.0033878 & -0.0005013 \\ 
w$_{131}$ &  0.0004436 &  0.0013816 &  0.0015143 &  0.0004107 &  0.0057651 \\ 
w$_{040}$ & -0.0033853 & -0.0021711 & -0.0022738 & -0.0007230 & -0.0017665 \\ 
w$_{140}$ &  0.0002804 &  0.0008126 &  0.0008187 &  0.0006392 &  0.0011240 \\ \hline
\end{tabular}
}
\end{table}

\begin{table}[htbp]
\centering
\caption[Coefficients $w_{ijk}$ for \texttt{FFII$_{A\rm (Li)-EW}$} (1D\,NLTE)]
        {Numerical  coefficients $w_{ijk}$ (Eq.\,\ref{eqFFII}) for the
         equivalent width fitting function \texttt{FFII}$_{\rm 1DNLTE}$,
         derived for the five metallicities of our grid.}
\label{FF_EW1D}
\renewcommand{\arraystretch}{1.2}
\resizebox{\linewidth}{!}{
\begin{tabular}{crrrrr}
\hline\hline
\multirow{2}{*}{$\mathrm{w}_{ijk}$}& \multicolumn{5}{c}{\feh}\\\cline{2-6} 
 &  0.0 dex   & -0.5 dex   & -1.0 dex   & -2.0 dex   &  -3.0 dex   \\
\hline
w$_{000}$ &  0.1668157 &  0.2358674 &  0.2338765 &  0.2696325 &  0.2404062 \\ 
w$_{100}$ &  0.0149293 &  0.0173628 &  0.0081412 &  0.0048003 & -0.0302344 \\ 
w$_{001}$ & -1.0165459 & -1.1335444 & -1.0981356 & -1.1334040 & -0.9817465 \\ 
w$_{101}$ & -0.0210521 &  0.1219691 &  0.2055447 &  0.0515904 &  0.0651075 \\ 
w$_{002}$ &  0.1444629 &  0.6690733 &  0.3493729 &  0.5578882 &  0.4516058 \\  
w$_{102}$ &  0.1364904 & -0.2768881 & -0.3623220 & -0.0860166 &  0.0226130 \\ 
w$_{003}$ &  0.1038219 & -0.4558280 &  0.0130863 & -0.2255391 & -0.2897598 \\ 
w$_{103}$ & -0.1815287 &  0.2097250 &  0.1752576 &  0.0647830 & -0.0531376 \\ 
w$_{004}$ & -0.0652711 &  0.1265842 & -0.0550262 &  0.0308042 &  0.0826828 \\ 
w$_{104}$ &  0.0725737 & -0.0500634 & -0.0108072 & -0.0134427 &  0.0153335 \\ 
w$_{010}$ &  1.0132868 &  1.0417293 &  1.0374104 &  1.0327493 &  1.0081008 \\ 
w$_{110}$ & -0.0064728 & -0.0379439 & -0.0345743 & -0.0132608 & -0.0198091 \\ 
w$_{011}$ & -0.0207368 & -0.0674934 & -0.0327851 & -0.0506050 & -0.0586258 \\ 
w$_{111}$ & -0.0421332 & -0.0226979 & -0.0642451 & -0.0323514 & -0.0026935 \\ 
w$_{012}$ & -0.1076595 & -0.1195333 & -0.1635718 & -0.1613043 & -0.0994343 \\ 
w$_{112}$ &  0.0420931 &  0.0555183 &  0.1068789 &  0.0442803 &  0.0442668 \\ 
w$_{013}$ &  0.0443194 &  0.0354617 &  0.0530645 &  0.0442895 &  0.0199720 \\ 
w$_{113}$ & -0.0087720 & -0.0160399 & -0.0309735 & -0.0087144 & -0.0101419 \\ 
w$_{020}$ &  0.0064495 & -0.0006828 & -0.0036903 &  0.0062150 &  0.0227895 \\ 
w$_{120}$ &  0.0083023 &  0.0282699 &  0.0317697 &  0.0112594 &  0.0099326 \\ 
w$_{021}$ &  0.0627791 &  0.0999068 &  0.0990063 &  0.1104659 &  0.0848678 \\ 
w$_{121}$ &  0.0091549 & -0.0062510 & -0.0082175 & -0.0011886 & -0.0238849 \\ 
w$_{022}$ & -0.0242035 & -0.0155056 & -0.0155906 & -0.0093547 & -0.0059664 \\ 
w$_{122}$ & -0.0067343 & -0.0054689 & -0.0079125 & -0.0053423 & -0.0008217 \\ 
w$_{030}$ & -0.0166254 & -0.0235642 & -0.0215853 & -0.0282541 & -0.0286804 \\ 
w$_{130}$ & -0.0027657 & -0.0069611 & -0.0073040 & -0.0009272 &  0.0041802 \\ 
w$_{031}$ &  0.0127171 &  0.0048428 &  0.0042163 &  0.0004064 &  0.0033798 \\ 
w$_{131}$ & -0.0003911 &  0.0015833 &  0.0030302 &  0.0002764 &  0.0023297 \\ 
w$_{040}$ & -0.0036484 & -0.0021369 & -0.0023126 & -0.0012000 & -0.0016354 \\ 
w$_{140}$ &  0.0002877 &  0.0005860 &  0.0004177 &  0.0000390 & -0.0008902 \\ \hline
\end{tabular}
}
\end{table}


\clearpage

\subsection{Fitting functions \texttt{FFIII}}
\begin{figure*}[h]
    \sidecaption
    \mbox{\includegraphics[trim=0 0 0 40,width=12cm]{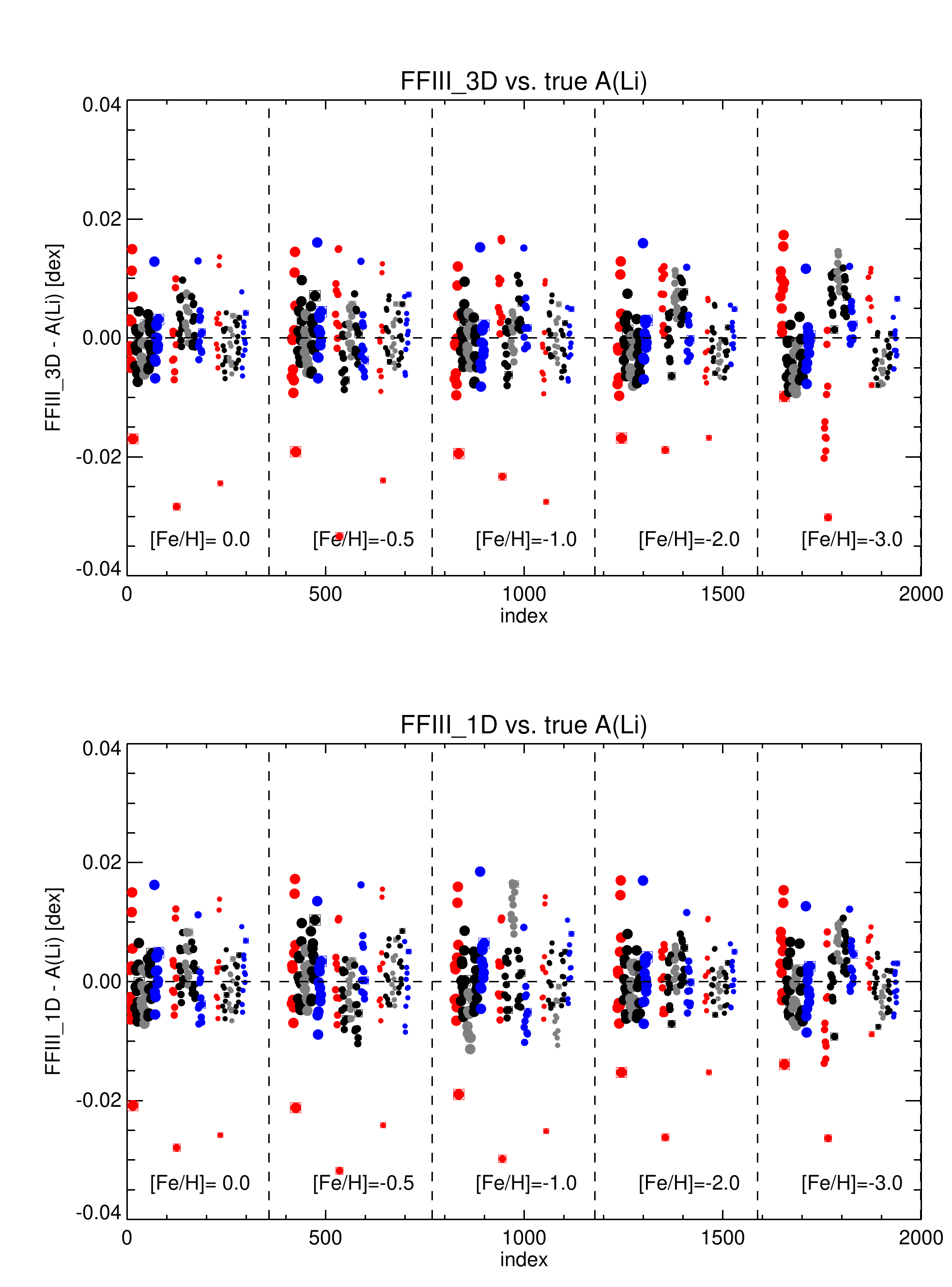}}
    \caption{Performance of the NLTE Li abundance fitting functions
      \texttt{FFIII$_{\rm 3DNLTE}$} (top) and \texttt{FFIII$_{\rm 1DNLTE}$}
      (bottom). For each metallicity, a total of $165$ points indicates the
      difference (\texttt{FFIII}\,$-$\,input abundance) as a function of an
      index that runs over $A$(Li), \teff, and \logg. Large (left group),
      intermediate (middle group), and small dots (right group) correspond to
      \logg\,=\,$3.5$, $4.0$, and $4.5$, respectively. For each gravity,
      \teff\ increases from left to right ($5000$\,K: red,
      $5500$\,K: black, $5900$\,K: gray, $6300$\,K: black, $6500$\,K: blue).
      Each effective temperature is represented by $10$ dots indicating the
      results for \mbox{$1.0 \le A$(Li)\,$\le 2.8$} plus a square for
      $A$(Li)\,=\,$3.0$.}
    \label{ff3_errors}
\end{figure*}

\begin{figure*}[h]
    \sidecaption
    \mbox{\includegraphics[trim=0 0 0 40,width=12cm]{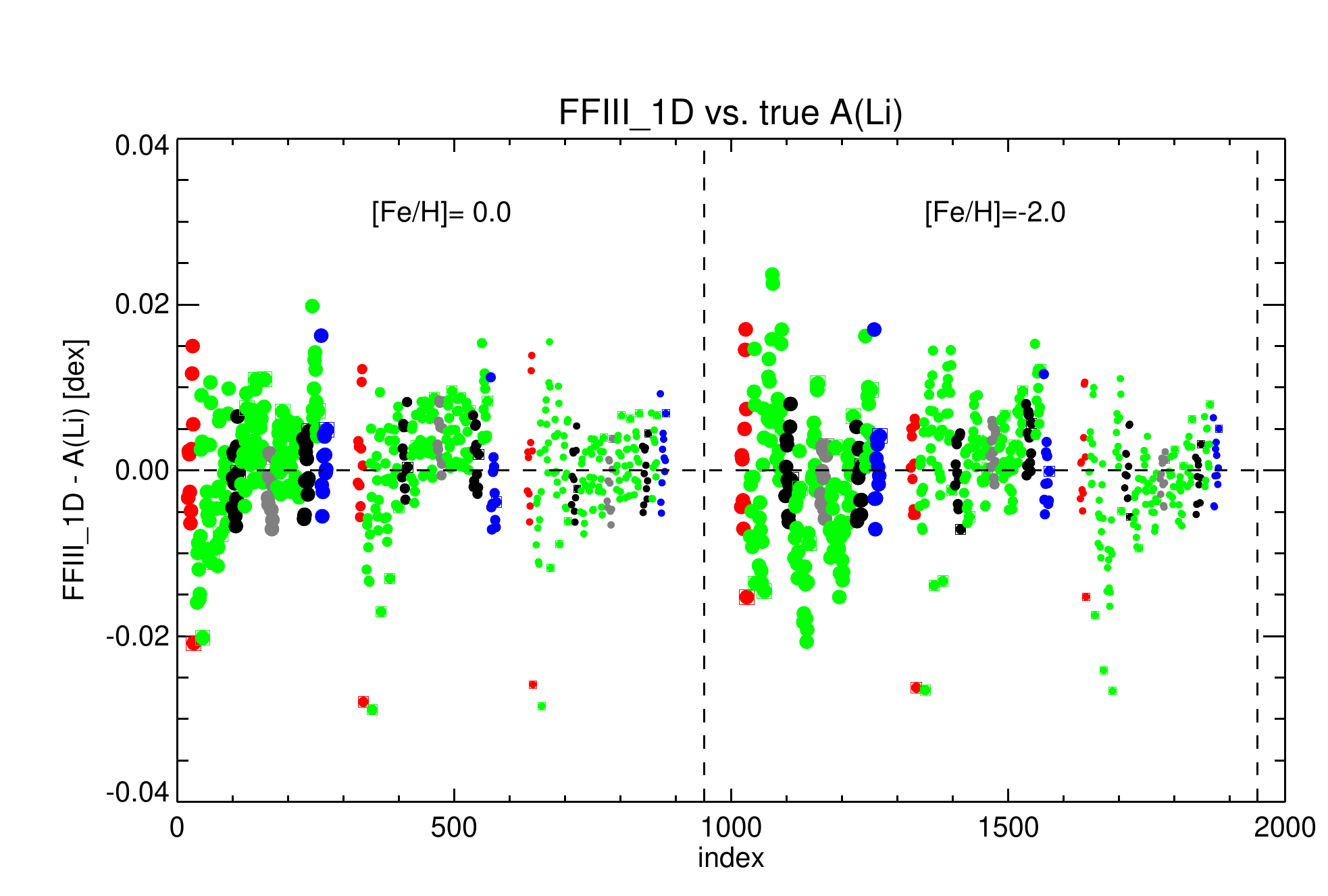}}
    \caption{Partial view of Fig.\,\ref{ff3_errors} (bottom), showing
       for metallicities [Fe/H]\,=\,$0.0$ and $-2.0$ the difference
       (\texttt{FFIII}$_{\rm 1DNLTE}-$ input abundance) as a function of
       spectrum number, now for a refined temperature grid with
       $\Delta$\teff\,=\,$100$\,K. For each metallicity, large,
       intermediate, and small dots correspond to \logg\,=\,$3.5$, $4.0$,
       and $4.5$, respectively. For each gravity, \teff\ increases from left
       to right. Color coding for the temperatures of the original grid is
       the same as in Fig.\,\ref{ff3_errors}; in addition, green dots
       represent the results for the intermediate 1D\,\lhdm\ test models.
       As before, each effective temperature is represented by $10$ dots for
       \mbox{$1.0 \le A$(Li)\,$\le 2.8$} plus a square for $A$(Li)\,=\,$3.0$.}
    \label{ff3_errors_lhd}
\end{figure*}

In the context of the ``inverse'' fitting function  \texttt{FFIII},
the coefficients $p_m$ and $q_m$ needed for the normalization of the input
equivalent width according to Eq.\,(\ref{eqScalingEW}), are listed in
Table\,\ref{coeffParabolas}. The coefficients $\mathrm{a}_{ijk}$ for the
related fitting function (Eq.\,\ref{eqFFIII}) are presented in
Table\,\ref{coeffFFIII3D} for the 3D case and in Table \ref{coeffFFIII1D}
for the 1D case.
Figure\,\ref{ff3_errors} gives an overview of the fitting function errors at
the nodes of the grid for all metallicities, while
Figure\,\ref{ff3_errors_lhd} illustrates the interpolation errors evaluated on
a refined temperature grid of 1D\,\lhdm\ models for [Fe/H]\,=\,$0.0$ and
$-2.0$.

\begin{table}[htbp]
\centering
\caption[Coefficients for the parabolic functions $p(z)$ and $q(z)$]{Numerical
  coefficients for the parabolic functions $p(z)$ and $q(z)$ for the scaling of
  the EW values (\mbox{Eq.\,\ref{eqScalingEW}}) for the five metallicities
  of our grid.}
\label{coeffParabolas}
\renewcommand{\arraystretch}{1.2}
\resizebox{\linewidth}{!}{
\begin{tabular}{crrrrrrrrrr}
\hline\hline
      & \multicolumn{5}{c}{3D\,NLTE}   \\                                            
      &  0.0 dex   & -0.5 dex   & -1.0 dex   & -2.0 dex   & -3.0 dex   \\ \hline
$p_0$ &  1.1666957 &  1.2504734 &  1.2464725 &  1.3043333 &  1.2126223 \\
$p_1$ & -0.9744305 & -0.9486030 & -0.9792976 & -1.0051502 & -0.8792871 \\
$p_2$ &  0.0998992 &  0.0860085 &  0.1103587 &  0.1234108 &  0.0831803 \\
$q_0$ &  2.5298152 &  2.5573504 &  2.5622675 &  2.6001840 &  2.5832374 \\ 
$q_1$ & -0.3196788 & -0.2843448 & -0.3114402 & -0.3440792 & -0.3490530 \\
$q_2$ & -0.0940990 & -0.0961721 & -0.0804780 & -0.0614842 & -0.0479861 \\ \hline
      & \multicolumn{5}{c}{1D\,NLTE} \\
      &  0.0 dex   & -0.5 dex   & -1.0 dex   & -2.0 dex   & -3.0 dex   \\ \hline  
$p_0$ &  1.1717118 &  1.2508765 &  1.2407464 &  1.2765080 &  1.2212788 \\
$p_1$ & -0.9636974 & -0.9391062 & -0.9457412 & -0.9397265 & -0.8184223 \\ 
$p_2$ &  0.0969080 &  0.0888007 &  0.1022377 &  0.1074416 &  0.0538786 \\
$q_0$ &  2.5373471 &  2.5629222 &  2.5642192 &  2.5933402 &  2.5898440 \\
$q_1$ & -0.3065712 & -0.2770424 & -0.2907207 & -0.3123179 & -0.3037169 \\
$q_2$ & -0.0994211 & -0.0945872 & -0.0840242 & -0.0640284 & -0.0647640 \\ \hline
\end{tabular}
}
\end{table}

\begin{table}[htbp]
\centering
\caption[Coefficients $a_{ijk}$ for \texttt{FFIII$_{\mathrm{EW}-A\rm (Li)}$}
  (3D\,NLTE)]
  {Numerical coefficients $a_{ijk}$ for the fitting function 
  \texttt{FFIII$_{\rm 3DNLTE}$} (Eq.\,\ref{eqFFIII}) for the five metallicities of
   our grid.}
\label{coeffFFIII3D}
\renewcommand{\arraystretch}{1.2}
\resizebox{\linewidth}{!}{
\begin{tabular}{crrrrr}
\hline\hline
\multirow{2}{*}{$\mathrm{a}_{ijk}$}& \multicolumn{5}{c}{\feh}\\\cline{2-6} 
 &  0.0 dex   & -0.5 dex   & -1.0 dex   & -2.0 dex   &  -3.0 dex   \\
\hline
a$_{000}$ &  1.0129539 &  1.0048051 &  1.0073185 &  1.0018589 &  0.9843768 \\ 
a$_{100}$ & -0.0320762 & -0.0181588 & -0.0225066 & -0.0135131 &  0.0258085 \\ 
a$_{001}$ & -0.0047474 &  0.1561165 &  0.1868200 &  0.2595673 &  0.1117859 \\ 
a$_{101}$ &  0.0966525 & -0.1225815 & -0.1843958 & -0.2946401 & -0.1551465 \\ 
a$_{002}$ &  0.0107189 & -0.3957869 & -0.4996260 & -0.5756683 & -0.2655631 \\ 
a$_{102}$ & -0.2652379 &  0.2545796 &  0.4620044 &  0.6000681 &  0.3281206 \\ 
a$_{003}$ & -0.0300389 &  0.3105966 &  0.4199568 &  0.4078356 &  0.2412711 \\ 
a$_{103}$ &  0.2674269 & -0.1445944 & -0.3654113 & -0.4031672 & -0.2896941 \\ 
a$_{004}$ &  0.0199194 & -0.0706069 & -0.1063437 & -0.0853998 & -0.0668322 \\ 
a$_{104}$ & -0.0896722 &  0.0127179 &  0.0855543 &  0.0816975 &  0.0796513 \\ 
a$_{010}$ &  1.4826018 &  1.4363913 &  1.4479415 &  1.4363663 &  1.4851382 \\ 
a$_{110}$ & -0.0122440 & -0.0120726 & -0.0224080 & -0.0332608 & -0.0266341 \\ 
a$_{011}$ &  0.4809866 &  0.4890558 &  0.4584515 &  0.4533793 &  0.3895906 \\ 
a$_{111}$ &  0.0315554 &  0.0315003 &  0.0671112 &  0.0233328 &  0.0026447 \\ 
a$_{012}$ & -0.1093419 & -0.1443315 & -0.0745938 & -0.0858786 & -0.1282260 \\ 
a$_{112}$ & -0.0582297 & -0.0676528 & -0.1250455 & -0.0366054 &  0.0103118 \\ 
a$_{013}$ & -0.0811257 & -0.0608838 & -0.0967849 & -0.0887680 & -0.0404274 \\ 
a$_{113}$ &  0.0315529 &  0.0419535 &  0.0693063 &  0.0455033 &  0.0214958 \\ 
a$_{020}$ &  0.0397534 &  0.0658588 &  0.0650056 &  0.0664652 &  0.0979962 \\ 
a$_{120}$ &  0.0310409 &  0.0362215 &  0.0456162 &  0.0816874 &  0.0243971 \\ 
a$_{021}$ &  0.2289434 &  0.3330445 &  0.2619056 &  0.2722017 &  0.3058020 \\ 
a$_{121}$ &  0.0119482 & -0.0044280 &  0.0319514 & -0.0529941 & -0.0511880 \\ 
a$_{022}$ &  0.2664408 &  0.2528331 &  0.2841167 &  0.2869350 &  0.2237575 \\ 
a$_{122}$ & -0.0300937 & -0.0414550 & -0.0792486 & -0.1155353 & -0.1034352 \\ 
a$_{030}$ & -0.3033826 & -0.4262138 & -0.3866560 & -0.3930803 & -0.4543772 \\ 
a$_{130}$ & -0.0669515 & -0.0577996 & -0.0822305 & -0.0379099 &  0.0486115 \\ 
a$_{031}$ & -0.7817028 & -0.8713536 & -0.8517458 & -0.8782503 & -0.7977752 \\ 
a$_{131}$ &  0.0597030 &  0.1003228 &  0.1256164 &  0.2541338 &  0.2421975 \\ 
a$_{040}$ &  0.7897918 &  0.9366162 &  0.9056341 &  0.9497166 &  0.9366128 \\ 
a$_{140}$ & -0.0023798 & -0.0321545 & -0.0377751 & -0.1493477 & -0.1960495 \\ \hline
\end{tabular}
}
\end{table}

\begin{table}[htbp]
\centering
\caption[Coefficients $a_{ijk}$ for \texttt{FFIII$_{\mathrm{EW}-A\rm (Li)}$}
  (1D\,NLTE)]
  {Numerical coefficients $a_{ijk}$ for the fitting function 
  \texttt{FFIII$_{\rm 1DNLTE}$} (Eq.\,\ref{eqFFIII}) for the five metallicities of
   our grid.}
\label{coeffFFIII1D}
\renewcommand{\arraystretch}{1.2}
\resizebox{\linewidth}{!}{
\begin{tabular}{crrrrr}
\hline\hline
\multirow{2}{*}{$\mathrm{a}_{ijk}$}& \multicolumn{5}{c}{\feh}\\\cline{2-6} 
 &  0.0 dex   & -0.5 dex   & -1.0 dex   & -2.0 dex   &  -3.0 dex   \\
\hline
a$_{000}$ &  1.0034880 &  0.9969588 &  0.9946962 &  0.9947520 &  0.9761783 \\ 
a$_{100}$ & -0.0149131 & -0.0012568 &  0.0012896 & -0.0014573 &  0.0395208 \\ 
a$_{001}$ & -0.0070700 &  0.1701429 &  0.0837970 &  0.1427269 &  0.1456125 \\ 
a$_{101}$ &  0.0600870 & -0.1039627 & -0.1483714 & -0.0204912 & -0.0444751 \\ 
a$_{002}$ &  0.0778053 & -0.5129522 & -0.1001108 & -0.3366641 & -0.3424794 \\ 
a$_{102}$ & -0.1876247 &  0.2598852 &  0.3102002 &  0.0590798 & -0.0621787 \\ 
a$_{003}$ & -0.1373179 &  0.4948993 & -0.0283389 &  0.2445354 &  0.3244218 \\ 
a$_{103}$ &  0.2092790 & -0.2249752 & -0.1888596 & -0.0636695 &  0.0678418 \\ 
a$_{004}$ &  0.0658496 & -0.1461374 &  0.0473187 & -0.0465606 & -0.0964052 \\ 
a$_{104}$ & -0.0800654 &  0.0586372 &  0.0229167 &  0.0126802 & -0.0208056 \\ 
a$_{010}$ &  1.4863001 &  1.4397544 &  1.4525720 &  1.4504639 &  1.4938299 \\ 
a$_{110}$ & -0.0129561 & -0.0107809 & -0.0196197 & -0.0334373 & -0.0392963 \\ 
a$_{011}$ &  0.4811911 &  0.4921130 &  0.4573048 &  0.4333948 &  0.3481754 \\ 
a$_{111}$ &  0.0435972 &  0.0324621 &  0.0619665 &  0.0383781 &  0.0388136 \\ 
a$_{012}$ & -0.1143308 & -0.1531218 & -0.0943043 & -0.1027095 & -0.1124205 \\ 
a$_{112}$ & -0.0699413 & -0.0597135 & -0.1132561 & -0.0532867 & -0.0459816 \\ 
a$_{013}$ & -0.0831284 & -0.0613116 & -0.0878985 & -0.0772232 & -0.0415169 \\ 
a$_{113}$ &  0.0356960 &  0.0364476 &  0.0626087 &  0.0514719 &  0.0511127 \\ 
a$_{020}$ &  0.0323323 &  0.0641480 &  0.0701729 &  0.0757278 &  0.0785333 \\ 
a$_{120}$ &  0.0297583 &  0.0277220 &  0.0287614 &  0.0668523 &  0.0775074 \\ 
a$_{021}$ &  0.2677503 &  0.3584981 &  0.3004327 &  0.3354400 &  0.4029507 \\ 
a$_{121}$ & -0.0101038 & -0.0243102 &  0.0291790 & -0.0641447 & -0.0807435 \\ 
a$_{022}$ &  0.2705392 &  0.2603878 &  0.2783551 &  0.2747025 &  0.2155212 \\ 
a$_{122}$ & -0.0249536 & -0.0318485 & -0.0715133 & -0.1116670 & -0.1230486 \\ 
a$_{030}$ & -0.3389264 & -0.4639441 & -0.4374807 & -0.4779056 & -0.5137561 \\ 
a$_{130}$ & -0.0418873 & -0.0263805 & -0.0514110 & -0.0003870 &  0.0050233 \\ 
a$_{031}$ & -0.8119198 & -0.8934028 & -0.8646667 & -0.8982424 & -0.8600781 \\ 
a$_{131}$ &  0.0660941 &  0.0973589 &  0.1099220 &  0.2498231 &  0.2818613 \\ 
a$_{040}$ &  0.8224663 &  0.9672014 &  0.9390500 &  1.0030741 &  1.0015136 \\ 
a$_{140}$ & -0.0164435 & -0.0449689 & -0.0440915 & -0.1620551 & -0.1865475 \\ \hline
\end{tabular}
}
\end{table}


\end{appendix}
\end{document}